\documentclass[twocol]{ametsocV6}
\usepackage{moreverb}
\usepackage{graphicx,overpic,natbib}
\usepackage{bm,url,color,colortbl,xcolor,soul}
\usepackage{url}
\usepackage{multirow}
\usepackage{booktabs} 
\usepackage{float}
\usepackage{wrapfig}
\usepackage{amsmath}
\usepackage{multirow}
\newcommand\BibTeX{{\rmfamily B\kern-.05em \textsc{i\kern-.025em b}\kern-.08em
T\kern-.1667em\lower.7ex\hbox{E}\kern-.125emX}}
\newcommand{\Fig}[1]{Figure~\ref{#1}}
\newcommand{\Tab}[1]{Table~\ref{#1}}

\newcommand{\EQ}{\begin{equation}}
\newcommand{\EN}{\end{equation}}
\newcommand{\Eq}[1]{Equation~(\ref{#1})}
\newcommand{\Eqs}[2]{Equations~(\ref{#1}) and~(\ref{#2})}
\newcommand{\blue}[1]{\textcolor{black}{#1}}

\graphicspath{{./fig/}}

\bibpunct{(}{)}{;}{a}{}{,}

\title{Large-eddy simulations of marine boundary-layer clouds associated with
cold air outbreaks during the ACTIVATE campaign– part 1: Case setup and
sensitivities to large-scale forcings \footnotemark[2]}
\authors{
Xiang-Yu Li\aff{a}\correspondingauthor{Xiang-Yu Li, xiangyu.li@pnnl.gov},
Hailong Wang\aff{a}\correspondingauthor{Hailong Wang, hailong.wang@pnnl.gov},
Jingyi Chen\aff{a},
Satoshi Endo\aff{e},
Geet George\aff{f},
Brian Cairns\aff{b}, Seethala Chellappan\aff{c},
Xubin Zeng\aff{j},
Simon Kirschler\aff{g},
Christiane Voigt\aff{g},
Armin Sorooshian\aff{i,j},
Ewan Crosbie\aff{d},
Gao Chen\aff{d},
Richard Anthony Ferrare\aff{d},
William I. Gustafson Jr.\aff{a},
Johnathan W Hair\aff{d},
Mary M Kleb\aff{d},
Hongyu Liu\aff{l},
Richard Moore\aff{d},
David Painemal\aff{h},
Claire Robinson\aff{h},
Amy Jo Scarino\aff{d},
Michael Shook\aff{d}, Taylor J Shingler\aff{d},
Kenneth Lee Thornhill\aff{d},
Florian Tornow\aff{b,k},
Heng Xiao\aff{a},
Luke D Ziemba\aff{d},
Paquita Zuidema\aff{c}
}
\affiliation{
\aff{a}Pacific Northwest National Laboratory \\
\aff{b}NASA Goddard Institute for Space Studies, New York, NY, United States \\
\aff{c} Rosenstiel School of Marine and Atmospheric Science,
University of Miami, Miami, United States \\
\aff{d}NASA Langley Research Center, Hampton, VA, United States \\
\aff{e}Brookhaven National Laboratory, Upton, NY, United States \\
\aff{f}Max Planck Institute for Meteorology, Hamburg, Germany \\
\aff{g}Deutsches Zentrum für Luft- und Raumfahrt (DLR), Oberpfaffenhofen, Germany \\
\aff{h}Science Systems and Applications, Inc. Hampton, Hampton, VA, United States \\
\aff{i}University of Arizona, Department of Chemical
and Environmental Engineering, Tucson, AZ, United States \\
\aff{j}University of Arizona, Department of Hydrology and Atmospheric Sciences, Tucson, AZ, United States \\
\aff{k}Columbia University of New York, Center for Climate Systems Research,
Earth Institute, New York, NY, United States \\
\aff{l}National Institute of Aerospace, Hampton, VA, United States
}

\abstract{
Large-eddy simulation (LES) is able to capture key boundary-layer (BL)
turbulence and cloud processes.  Yet, large-scale forcing and surface turbulent
fluxes of sensible and latent heat are often poorly prescribed for LES
simulations.  We derive these quantities from measurements and reanalysis
obtained for two cold air outbreak (CAO) events during Phase I of the Aerosol
Cloud meTeorology Interactions oVer the western ATlantic Experiment (ACTIVATE)
in February-March 2020.  We study the two contrasting CAO cases by performing
LES and test the sensitivity of BL structure and clouds to
large-scale forcings and turbulent heat fluxes.  Profiles of atmospheric state
and large-scale divergence and surface turbulent heat fluxes obtained from the
reanalysis data ERA5 agree reasonably well with those derived from ACTIVATE
field measurements for both cases at the sampling time and location.  Therefore,
we adopt the time evolving heat fluxes, wind and advective tendencies profiles
from ERA5 reanalysis data to drive the LES. We find that
large-scale thermodynamic advective tendencies and wind relaxations are
important for the LES to capture the evolving observed BL meteorological states
characterized by the hourly ERA5 reanalysis data and validated by the
observations.  We show that the divergence (or vertical velocity) is
important in regulating the BL growth driven by surface heat fluxes in LES
simulations.  The evolution of liquid water path is largely affected by the
evolution of surface heat fluxes. The liquid water path simulated in LES agrees
reasonably well with the ACTIVATE measurements.  This study paves the path to
investigate aerosol-cloud-meteorology interactions using LES informed and
evaluated by ACTIVATE field measurements.}

\begin{document}
\maketitle

\clearpage

\section{Introduction}

When viewed from space, about $70\%$ of Earth's surface is covered
by clouds \citep{Schneider17}.
Clouds, the regulator of the radiative
heating and cooling of the planet \citep{RAMANATHAN57},
represent a major complication in the current modeling of
the climate system \citep{Schneider17,2013_Stevens,Bony17}.
One of the most challenging problems of cloud--climate interactions is to
understand how cloud microphysical processes affect
atmospheric water and radiation budgets,
such as how precipitation efficiency affects
radiative properties of stratocumulous clouds \citep{boucher2013clouds}.

The Western North Atlantic Ocean (WNAO) region has attracted decades of
atmospheric research due to the complex atmospheric system
\citep{painemaloverview}, pollution outflow from North America
\citep{corraloverview}, and accessibility by aircraft and ships.
However, the subject of Aerosol-Cloud-Interaction (ACI) is the least investigated among all
the field campaign measurements over the WNAO \citep{sorooshian2020atmospheric}
partly because of the complicated chemical, physical, and dynamical processes in
this region.  ACI involves processes from the formation of nm-sized aerosols to
the life cycle of kilometer-sized clouds, which covers a scale range of about
$10^{12}$.  Such a scale separation coupled with turbulence poses great
challenge for both measurements and numerical modeling.  The spatial
distribution of aerosols and the ambient humidity fields determine the formation
of cloud droplets and ice crystals and their size distribution
\citep{shaw_2003}.  Precipitation and radiative properties of clouds are altered
by the size distribution of particles. The Aerosol Cloud meTeorology
Interactions oVer the western ATlantic Experiment (ACTIVATE) field campaign aims
to tackle ACI by performing comprehensive measurements of cloud macro/micro
properties and atmospheric states using two aircraft simultaneously, which can
be used to evaluate and constrain atmospheric models
\citep{sorooshian2019aerosol, sorooshian2020atmospheric}.

Large eddies of $\mathcal{O}(10^2-10^3\, \rm{m})$ in the planetary boundary layer
are important for turbulent mixing, heat/moisture transport, and cloud formation.
Large-eddy simulation (LES) has been widely used to model marine boundary-layer
clouds and ACI \citep{bretherton1999gcss, stevens2002effects, brown2002large,
ackerman2009large,wang2009evaluation, wang2009modeling, Roode2019, brilouet2020organized}.
LES resolves the intermediate and large
turbulent eddies (sub-inertial range of turbulence)
and parameterizes smaller scales 
using well-established parameterization schemes.
LES has advantages over cloud
resolving models and beyond as it can resolve the BL
eddies and over direct numerical simulations
\citep{li2018effect, li2018cloud, li2020condensational}
since it is able to simulate mesoscale cloud organizations in a sufficiently large domain.
One of the most challenging problems of using idealized LES
with doubly periodic boundary conditions to represent evolving clouds
is realistically configuring large-scale forcings (e.g., horizontal advection tendencies, divergence (D) of
flow, and surface heat fluxes) that determine the
spatiotemporal variation of large-scale ambient conditions for the cloud system.
The ``large-scale'' here refers to scales of $50-500 \, \rm{km}$
\citep{bony2019measuring}.
There are different ways to construct large-scale
forcings for typical LES domains.
For example, $D$ profiles, horizontal advection tendencies and heat fluxes can
be obtained from analysis/reanalysis products and other numerical models, as
well as measurements of variables used for the calculation.
To validate $D$ and heat fluxes obtained from numerical models,
observations such as sounding profiles of atmospheric state
and surface temperature measurements are needed.
\citet{bretherton1999gcss} forced single-column models
and two-dimensional eddy-resolving models using time-varying
boundary conditions from reanalysis data and found that these models predict the observed evolution of
boundary layer well. Similar forcing was applied to LES
in \citet{van2013gass}.
\citet{neggers2012continuous} drove LES using time-varying large-scale
forcing from general circulation models and argued that such a forcing
strategy can reproduce large-scale meteorological states and preserve
small-scale cloud physics.
\citet{endo2015racoro} constructed continuous large-scale
and surface forcings from reanalysis data to successfully
simulate continental boundary layer clouds
during the Routine Atmospheric Radiation Measurement Aerial Facility
Clouds with Low Optical Water Depths Optical Radiative Observations (RACORO)
campaign using an idealized LES model.
The application of such large-scale forcing schemes
for marine stratocumulus clouds in the WNAO region, however, has not been reported,
which is a focus of this study.

Marine stratocumulus clouds associated with cold air outbreaks (CAOs)
with mesoscale (scales larger than a few kilometers) fluctuations
are challenging to represent in climate models.
CAO occurs when cold air mass moves over a warm sea
surface, creating strong convection analogous to Rayleigh-B\'enard convection \citep{agee1987mesoscale}.
CAO events are characterized by stronger surface latent heat fluxes
of $\mathcal{O}(10^2-10^3\, \rm{W\, m}^{-2})$\citep{papritz2015climatology}
and subsidence of up to an order of magnitude,
compared to non-CAO cases due to the large temperature difference \citep{agee1987mesoscale}. This can
makes it difficult to simulate convection and clouds associated with CAOs.
The ratio of buoyancy force to shear (Richardson number), precipitation, and
entrainment contribute to the topological cloud structure of CAOs (e.g., cloud streets)
\citep{Roode2019}.
The cloud roll structure occurring
during CAOs was first simulated by \citet{liu2004high} using
a cloud-resolving model able to capture the transition
of clouds from two-dimensional roll structure to three-dimensional
closed cells. \citet{gryschka2005roll} performed the first
LES to simulate CAO cloud roll structures.
They found that roll structures are sensitive
to the spatial resolution of LES.
\citet{tomassini2017g} investigated how well
a CAO event over the North Atlantic Ocean
is represented in global models
as compared to LES.
They found that the global models employed in their study
underestimate the amount of cloud liquid water compared
to LES results.
More recently, \citet{Roode2019} performed an LES
intercomparison study of a CAO case observed during the CONSTRAIN
campaign. They found that the evolution of the
stratocumulus cloud deck and the timing of its
breakup differ significantly among seven LES models
and attributed this discrepancy to the inconsistency of microphysics
parameterizations between different LES models.
\citet{tornow2021preconditioning} investigated
a marine CAO case in the northwestern Atlantic and found that
frozen hydrometeors accelerate the transition of cloud decks into
broken cloud streets.
Here we aim to examine the roles of large-scale forcings and aerosols in
affecting the evolution of WNAO marine boundary-layer meteorology and clouds
associated with CAO using LES constrained by in-situ and remote sensing
measurements in a two-part serial study. The first part focuses on quantifying
sensitivities of meteorology and clouds to large-scale forcings and turbulent
surface fluxes. The second part will focus on characterizing cloud properties
and aerosol-cloud-meteorology interactions.

In this first part of two companion studies,
we first introduce two CAO cases sampled during the 2020 winter
deployment of ACTIVATE and describe the numerical experiment setup
for idealized LES to model the two cases.
Then we use divergence profiles and surface heat fluxes derived from 
ACTIVATE dropsondes and sea surface temperature (SST) measurements to first evaluate these
quantities from ERA5 reanalysis data.
We further examine the sensitivities
of LES results to surface heat fluxes and 
large-scale thermodynamic advective tendencies.
We adopt the same LES model and
large-scale forcing scheme as in \citet{endo2015racoro}.

\section{Observations, reanalysis data, and LES numerical experiment setup}

\subsection{ACTIVATE campaign}

The ACTIVATE field campaign
aims to collect sufficient measurements to understand interactions of
marine boundary-layer clouds with meteorological conditions and
aerosol particles, which eventually leads to improved physical understanding of
cloud micro/macro processes and reduced uncertainty
in their representation in global climate models.
A total of 150 coordinated flights with two
air-borne platforms is planned for three years
(2020-2022) over the western North
Atlantic Ocean ($25^\circ-50^\circ$N, $60^\circ-85^\circ$W) to characterize
aerosol-cloud-meteorology interactions in a systematic
and simultaneous manner \citep{sorooshian2019aerosol}.
This is being achieved by flying two aircraft simultaneously at
different altitudes.
The low-flying HU-25 Falcon
measures in-situ trace gases, aerosol, clouds, precipitation,
and meteorological properties below, in, and above clouds.
The higher-flying King Air above clouds
simultaneously acquires remote retrievals of aerosols
and clouds while launching dropsondes.

\begin{figure*}[t!]\begin{center}
\includegraphics[width=\textwidth]{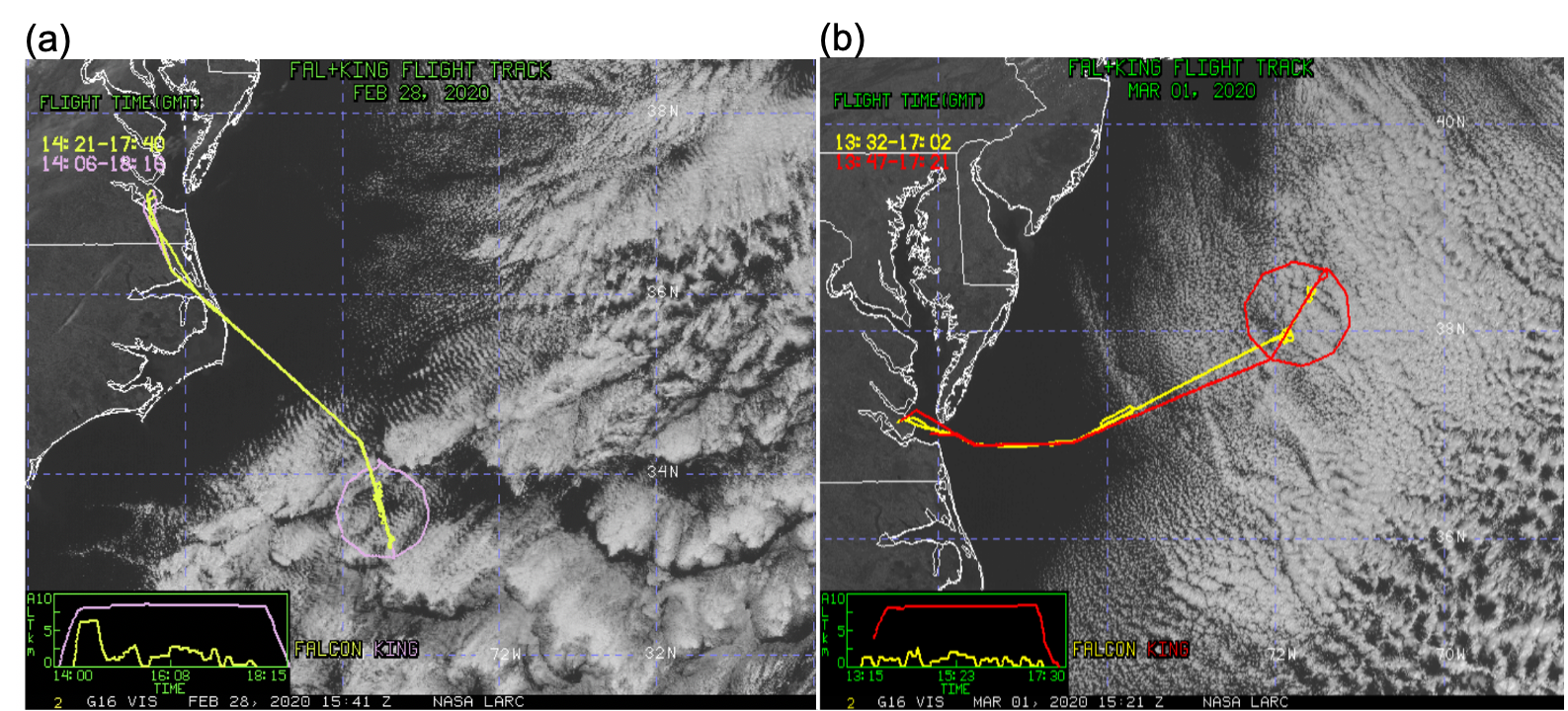}
\end{center}
\caption{Visible images for (a): 28 February 2020
and (b): 1 March 2020 cases from GOES-16 over the ACTIVATE
measurement region. Lower-left panels represent the flight
altitude as a function of UTC time for the HU-25 Falcon
(low-flying aircraft) and King Air (high-flying aircraft).
}
\label{visible}
\end{figure*}

\Fig{visible} shows flight tracks of King Air and HU-25
Falcon and visible images from GOES-16 satellite during the two CAO process-study
cases over the WNAO region on 28 February 2020 and 1 March 2020, corresponding
to Research Flight $\#10$ and $\#13$ (hereafter RF10 and RF13),
respectively.
11 dropsondes (model Vaisala NRD41)
were released from the King Air.
Each of them provided vertical profiles of
air pressure ${\rm p}$, temperature $T$, relative humidity RH,
and horizontal velocities $u$ and $v$
with a vertical resolution of 5-10 m and a
resolution (with associated uncertainty) of
0.01 hPa ($\pm 0.5$ hPa), $0.01^\circ C$ ($\pm 0.2^\circ C$),
$0.01\%$ ($\pm 0.3\%$), and $0.01\, {\rm m\, s}^{-1}$
($\pm 0.5\, {\rm m\, s}^{-1}$) \citep{drop}, respectively.
The King Air flew in a circular pattern with a diameter of about $152\, {\rm km}$
to cover the largest enclosed area for dropsonde measurements and to avoid
sharp turns. Such a flight pattern for the dropsonde measurements
was first proposed by \citet{lenschow1999measuring}.
This strategy has been used in other campaigns
to measure the large-scale divergence $D$,
such as Elucidating the Role of Cloud–Circulation Coupling in Climate
\citep{bony2019measuring}, Atlantic Tradewind Ocean-Atmosphere
Mesoscale Interaction Campaign \citep{quinn2021measurements}, 
and Next-Generation Aircraft Remote Sensing for Validation (NARVAL2) airborne
field campaign \citep{stevens2019high}.
Dropsondes were released at a height of about 8 km.
We interpolate the measured data evenly with a
vertical spacing of 10 m for further analysis.
Two contrasting CAO cases were observed over the WNAO
region on February 28 (RF10, dropsonde-circle center at $33.66^\circ$N, $286.69^\circ$E)
and March 1 (RF13, dropsonde-circle center at $38.01^\circ$N, $288.36^\circ$E)
as shown in \Fig{drop_map}.
\Tab{tab:drop} summarizes
the start/end time of dropsonde measurements,
the location, $10$-m wind speed $U_{10\rm m}$, 
$q_{v, \rm{10m}}$, $T_{\rm{10m}}$, and ERA5 SST at the center of dropsonde circle
for the February 28 and March 1 cases, respectively.

\begin{table*}[t!]
\caption{This table lists the start-end time of dropsonde measurements,
the location, $10$-m wind speed $U_{10\rm m}$, 
$q_{v, \rm{10m}}$, and $T_{\rm{10m}}$ at the center of dropsonde circle
for the 28 February 2020 and 1 March 2020 cases.
The corresponding SST from ERA5 reanalysis
is also documented.
} 
\centering
\setlength{\tabcolsep}{1pt}
\begin{tabular}{|c|c|c|c|c|c|c|c|c|}
\hline
\multirow{2}{*}{Case} & \multirow{2}{*}{Start-end times (UTC)} & \multirow{2}{*}{Lat ($^\circ$)}  & \multirow{2}{*}{Lon ($^\circ$)} &
\multirow{2}{*}{$U_{10\rm{m}}\,(\rm{m\, s}^{-1})$} & \multirow{2}{*}{$q_{v,10\rm{m}}$ ($\rm{g\, kg}^{-1}$)} & \multirow{2}{*}{$T_{10\rm{m}}\,(K)$} & \multicolumn{2}{c|}{SST ($K$)} \\
\cline{8-9}
& & & & & & & MW-IR & ERA5 \\
\cline{8-9}
\hline
0228 & 1540-1645 & 33.66 & 286.69 & 5.94 & 4.64 & 286.75& 293.25 & 293.60 \\  
0301 & 1451-1547 & 38.01 & 288.36 & 11.74 & 2.79 & 275.71 & 286.84 & 286.88 \\
\hline
\multicolumn{9}{p{0.3\textwidth}}{}
\end{tabular}
\label{tab:drop}
\end{table*}

Dropsonde measurements are used to
characterize the meteorological conditions and derive
large-scale divergence and surface heat fluxes for both cases.
Cloud droplets and ice crystals were observed for both cases.
The mean number concentration of cloud droplets obtained from
Fast Cloud Droplet Probe (FCDP, equipped on HU-25 Falcon)
measurement \citep{taylor2019aerosol, knop2021comparison} is about
$\langle N_c \rangle= 650\, {\rm cm}^{-3}$ for the February 28
case and $\langle N_c \rangle = 450\, {\rm cm}^{-3}$ for the March 1 case.
These values are acquired by averaging in-cloud FCDP measurement with
a lower cutoff of liquid water path of $0.02\, {\rm g\, kg}^{-1}$
and effective diameter of $3.5\, \mu{\rm m}$ (FCDP
covers a diameter range of 3.0 to 50.0 $\mu{\rm m}$).
There were also detailed measurements of aerosol particles including mass and
number concentration, composition, size distribution, hygroscopicity, and
optical properties.  Given the focus of this study, we only use the mean cloud
drop number in our LES sensitivity simulations on meteorological conditions and
large-scale forcings.  Liquid water path is retrieved from Research Scanning
Polarimeter (RSP) \citep{alexandrov2012accuracy, alexandrov2018retrievals}.
Given the instantaneous field of view of 14 mrad, typical cloud tops (about 2
km), and a flight altitude of the King Air during ACTIVATE ($8\sim9$ km), the nadir
pixel size of the RSP is approximately 100 m. To compare with the LES
with a 300 m horizontal grid spacing, we average the RSP sampling every 3 s,
given that the moving speed of King Air is about $100 \, \rm{m\, s}^{-1}$.
Fast in-situ 3-D wind measurements were performed
with an uncertainty of $5\%$ and a sampling frequency of 20 Hz.
The static air temperature was measured with an uncertainty of $5\%$
and a sampling frequency of 1 Hz.
The water vapor volume mixing ratio in ppmv was measured
by Diode Laser Hygrometer with an uncertainty of $5\%$ and
a sampling frequency of 1 Hz.

\subsection{ERA5 and MERRA-2 reanalysis data}

The ERA5 reanalysis data are generated using the fifth generation of European
Centre for Medium-Range Weather Forecasts’s Integrated Forecast System
\citep{hersbach2018era5}.  We use the ERA5 hourly data at a horizontal
resolution of 31 km. For three-dimensional fields, there are 137 model levels up
to a height of 80 km. Since ERA5 only provide $\bar{D}$ in datasets with
specified pressure levels, we use pressure-level data for the comparison of
$\bar{D}$ and the corresponding large-scale vertical velocity $w$.  The ERA5
large-scale forcings for the LES are obtained at the model levels instead of the
pressure levels because the model-level data
have a finer vertical mesh-size and can better characterize
the inversion layer. The ambient meteorological conditions for a given
location during a CAO usually evolve quickly due to strong winds and large
surface heat fluxes under winter mid-latitude weather disturbances.  Since the
measurement time window for the two CAO events is about one hour during the
ACTIVATE field campaign, we are not able to use the measurements directly to
drive the LES for many hours.  We validate the ERA5 reanalysis data
against the limited field measurements and then use the evolving forcing
conditions from ERA5 reanalysis data to drive the LES.

The Modern-Era Retrospective analysis for Research and
Applications version 2 (MERRA-2) \citep{merra-2} is also used to compare
with the ERA5 reanalysis and dropsonde measurements.
The MERRA-2 reanalysis data are generated using the Goddard Earth Observation
System version 5 (GEOS-5) with its Data Assimilation System version 5.12.4 \citep{gelaro2017modern}.
MERRA-2 has a horizontal resolution of $0.5^\circ \times 0.625^\circ$
with 72 model levels, from which the 3-hourly datasets
at 42 pressure-levels are interpolated. It also provides
1-hourly two-dimensional datasets.
We note that dropsonde measurements made during the
ACTIVATE campaign have {\it not} been assimilated in either the ERA5
or MERRA-2 reanalysis used in this study.
This allows us to validate meteorological states from LES and the reanalysis
against the dropsonde measurements. 

\subsection{Satellite measurements}

We use daily sea surface temperature (SST) retrieved from microwave and infrared
based satellite measurements (MW-IR SST) produced by Remote Sensing Systems
\citep{satellite}.  The SST product has a horizontal grid spacing of 9 km.  This
resolution is 3 times higher than the SST from ERA5 reanalysis data.

\subsection{LES numerical experiment setup}
\label{sec:ns}

We use the Weather Research and Forecasting (WRF) model \citep{skamarock2019description}
in the idealized LES mode (WRF-LES)  \citep{wang2009modeling}
to simulate the two CAO cases and test the sensitivities of the marine BL
and clouds to large-scale forcing and heat fluxes.
Doubly periodic boundary conditions are employed in horizontal directions.
The horizontal resolution is set to $dx=dy=300 \,\rm{m}$
with 200 lateral grid cells,
which results in a horizontal domain size of $L_x=L_y=60\,\rm{km}$.
The domain height is $z_{\rm top}=7\,\rm{km}$ with 153 vertical $\eta$-layers
($\eta = (p-p_{\rm T})/(p_{\rm S}-p_{\rm T})$ with
$p_{\rm S}$ and $p_{\rm T}$ the pressure at the
bottom and top of the model domain, respectively), which results
in a vertical mesh-size of about $33\, \rm{m}$ in the boundary layer.
The horizontal resolution of 300 m is quite coarse for LES
but it has proven to be able to simulate the formation and
evolution of cloud cellular structures in marine stratocumulus \citep{wang2009modeling}.
The periodic boundary condition in horizontal directions
is ideal for isolating main governing factors for cloud processes
and has been widely used for LES with lateral domain size even larger than
$60\, \rm{km}$ \citep{seifert2015large, bretherton2017understanding}.
The time step is set to $\Delta t = 3\, {\rm s}$ in all simulations.
Simulations are initiated at 06:00 UTC 
to allow sufficient model spin-up time before the WRF-LES results are
evaluated against measurements taken during 16:00-17:00 UTC on
February 28 and 15:00-16:00 UTC on March 1.

The two-moment Morrison cloud microphysics scheme
\citep{morrison2009impact} is used. 
In this part of the study, a constant number concentration of
cloud droplets derived from in-situ measurements
during the ACTIVATE campaign is prescribed in the
Morrison scheme to stay focused on cloud-meteorology interactions.
Both shortwave and longwave radiative schemes are originally from
the NCAR Community
Atmosphere Model (CAM 3.0), which were used in previous WRF-LES studies,
such as \citet{wang2009evaluation} and \citet{wang2009modeling}.
Surface heat fluxes and SST are all prescribed in the model as
the boundary conditions at the sea surface.

LES with horizontally uniform initial conditions and periodic boundary
conditions cannot predict changes in atmospheric state at scales larger than its
domain size. This is particularly true for rapidly evolving CAOs,
with a baroclinic structure and the resulting vertical wind-profiles that
cannot be properly simulated by LES
\citep{gryschka2014impact}. To circumvent this problem, we apply relaxation to horizontal wind
components $u$ and $v$ and advective tendencies to potential temperature
$\theta$ and water vapor mixing ratio $q_v$ as forcing terms in the prognostic
equations.  We adopt the same large-scale forcing and relaxation schemes as in
\citet{endo2015racoro}.
To derive the large-scale forcings,  we simplify
the governing equation of $\theta$ and $q_v$ by removing sink and source terms as,
\EQ
\label{eq:dT/dt}
\frac{\partial \theta}{\partial t} = -{\bm u}\cdot {\bm \nabla} \theta=
-u\frac{\partial \theta}{\partial x}-v\frac{\partial \theta}{\partial y}
-w\frac{\partial \theta}{\partial z},
\EN
\EQ
\label{eq:dqv/dt}
\frac{\partial q_v}{\partial t} = - {\bm u}\cdot {\bm \nabla} q_v=
-u\frac{\partial q_v}{\partial x}-v\frac{\partial q_v}{\partial y}
-w\frac{\partial q_v}{\partial z}.
\EN
Applying Reynolds decomposition to \Eqs{eq:dT/dt}{eq:dqv/dt}
and ignoring the perturbation terms $\overline{{\bm u}^\prime\cdot {\bm \nabla}\theta^\prime}$ and
$\overline{{\bm u}^\prime\cdot {\bm \nabla}q_v^\prime}$,
we obtain temporal tendencies at large scales,
\EQ
\label{eq:dT/dt2}
\frac{\partial \bar{\theta}}{\partial t} =\left[
-\bar{u}\frac{\partial \bar{\theta}}{\partial x}
-\bar{v}\frac{\partial \bar{\theta}}{\partial y}
-\bar{w}\frac{\partial \bar{\theta}}{\partial z}\right]_{\rm{ERA5}},
\EN
and
\EQ
\label{eq:dqv/dt2}
\frac{\partial \bar{q}_v}{\partial t}=\left[
-\bar{u}\frac{\partial \bar{q}_v}{\partial x}
-\bar{v}\frac{\partial \bar{q}_v}{\partial y}
-\bar{w}\frac{\partial \bar{q}_v}{\partial z}\right]_{\rm{ERA5}},
\EN
where the overbar denotes a large-scale mean.
The large-scale horizontal advective tendencies of $\bar{\theta}$
and $\bar{q}_v$ are given by the first two terms of
r.h.s of \Eq{eq:dT/dt2} and \Eq{eq:dqv/dt2}, respectively.
The third term in the r.h.s of \Eq{eq:dT/dt2} and \Eq{eq:dqv/dt2}
represents the large-scale vertical advective tendencies.
These large-scale advective tendencies are obtained from the
hourly ERA5 reanalysis data and applied to each grid cell.
Ignoring the perturbation terms
$\overline{{\bm u}^\prime\cdot {\bm \nabla}\theta^\prime}$ and
$\overline{{\bm u}^\prime\cdot {\bm \nabla}q_v^\prime}$
is for the practical reason that
ERA5 reanalysis data do not resolve intermediate scales
for our LES due to the relatively coarse mesh size of 31 km.
In addition, such a forcing scheme was also adopted
in previous studies, such as \citet{siebesma1995evaluation, endo2015racoro}.
\citet{van2019investigating} took the contribution of intermediate
scales into the large-scale forcing using $0.1^\circ\times 0.1^\circ$
mesh-sized forcing data.
However, in the present study, we aim to clearly define the scales to
be included as the large-scale contribution.
The horizontal wind components $u$ and $v$
are applied with a relaxation strategy
(i.e., nudging LES domain-average winds to a reference state),
as also used in previous LES studies
\citep{wang2008modeling, endo2015racoro}, defined by  
\EQ
\left.\frac{\partial u}{\partial t}\right\vert_{\rm R}=
\frac{u_{\rm ERA5}-\langle u \rangle}{\tau},
\label{eq:rlx_u}
\EN
\EQ
\left.\frac{\partial v}{\partial t}\right\vert_{\rm R}=
\frac{v_{\rm ERA5}-\langle v \rangle}{\tau},
\label{eq:rlx_v}
\EN
where $\langle \rangle$ denotes average over
the domain of WRF-LES, $\tau$ is the relaxation time scale,
which is set to be 1 hour in this study. The subscript ``R'' denotes
the relaxation adjustment to the horizontal wind components.
The grid-scale wind is determined by
\Eqs{eq:rlx_u}{eq:rlx_v} and pressure gradients as the Coriolis force is set to zero
in our LES. Overall, the large-scale forcing applied
to LES is homogeneous horizontally.

We acknowledge that applying relaxation of wind to WRF-LES
lacks physical judgment as also addressed in \citet{endo2015racoro}.
However, LES of horizontal winds with relaxation adjustments are
found to be comparable with the reanalysis and observational data.
This is not new and has been used in the single column
model \citep{randall1999alternative} and many LES works
\citep{neggers2012continuous, heinze2017evaluation}
in the meteorology community.
Even though the simulation domain is stationary 
and a horizontal periodic boundary condition is used,
the WRF-LES is set to take the cold air advection within CAO
into account through the large-scale advective tendencies
and wind relaxation described by
\Eqs{eq:dT/dt2}{eq:dqv/dt2} and \Eqs{eq:rlx_u}{eq:rlx_v},
respectively.

We also test the sensitivities of WRF-LES results
to prescribed surface heat fluxes obtained
from ERA5 reanalysis data.
\Tab{tab:param} lists parameters examined in the sensitivity tests.

\begin{table*}[t!]
\caption{List of WRF-LES with different
forcings. $\rm{SHF(t)_I}$ and $\rm{LHF(t)_I}$
denote sensible and latent heat fluxes calculated
interactively in WRF-LES.
} 
\centering
\setlength{\tabcolsep}{1pt}
\begin{tabular}{|c|c|c|c|c|c|c|c|}
\hline
Simulation & $\frac{\partial \bar{\theta}}{\partial t}$ $\&$ $\frac{\partial \bar{q}_v}{\partial t}$ & $u \& v$ relaxation  & SHF ($\rm{W\, m}^{-2}$) & LHF ($\rm{W\, m}^{-2}$) &$\bar{D}$& $N_c$ [$\rm{cm}^{-3}$] & dx (m) \\
\hline
0228A & Yes & Yes & 79.91  & 305.02 & Yes & 650 & 300 \\ 
0228B & No & Yes & 79.91  & 305.02 & Yes & 650  & 300\\  
0228C & Yes & No & 79.91  & 305.02 & Yes & 650  & 300\\  
0228D & No & No & 79.91  & 305.02  & Yes & 650  & 300\\  
0228E & Yes & Yes & SHF(t) & LHF(t) & Yes & 650 & 300 \\ 
0228F & Yes & Yes & $\rm{SHF(t)_I}$ & $\rm{LHF(t)_I}$ & Yes & 650  & 300\\ 
0228G & Yes & Yes & SHF(t) & LHF(t) & Yes & 650 & 100 \\ 
0301A & Yes & Yes & 231.76 & 382.18 & Yes & 450  & 300\\ 
0301B & Yes & Yes & 231.76 & LHF(t) & Yes & 450  & 300\\  
0301C & Yes & Yes & SHF(t) & 382.18 & Yes & 450  & 300\\  
0301D & Yes & Yes & 231.76 & 382.18 & No & 450  & 300\\  
0301E & Yes & Yes & SHF(t) & LHF(t) & Yes & 450  & 300\\  
0301F & Yes & Yes & $\rm{SHF(t)_I}$ & $\rm{LHF(t)_I}$ & Yes & 450  & 300\\  
0301G & Yes & Yes & SHF(t) & LHF(t) & Yes & 450  & 100\\  
\hline
\multicolumn{8}{p{0.3\textwidth}}{}
\end{tabular}
\label{tab:param}
\end{table*}

\section{Meteorological conditions and forcings for the two cases}

\begin{figure*}[t!]\begin{center}
\includegraphics[width=\textwidth]{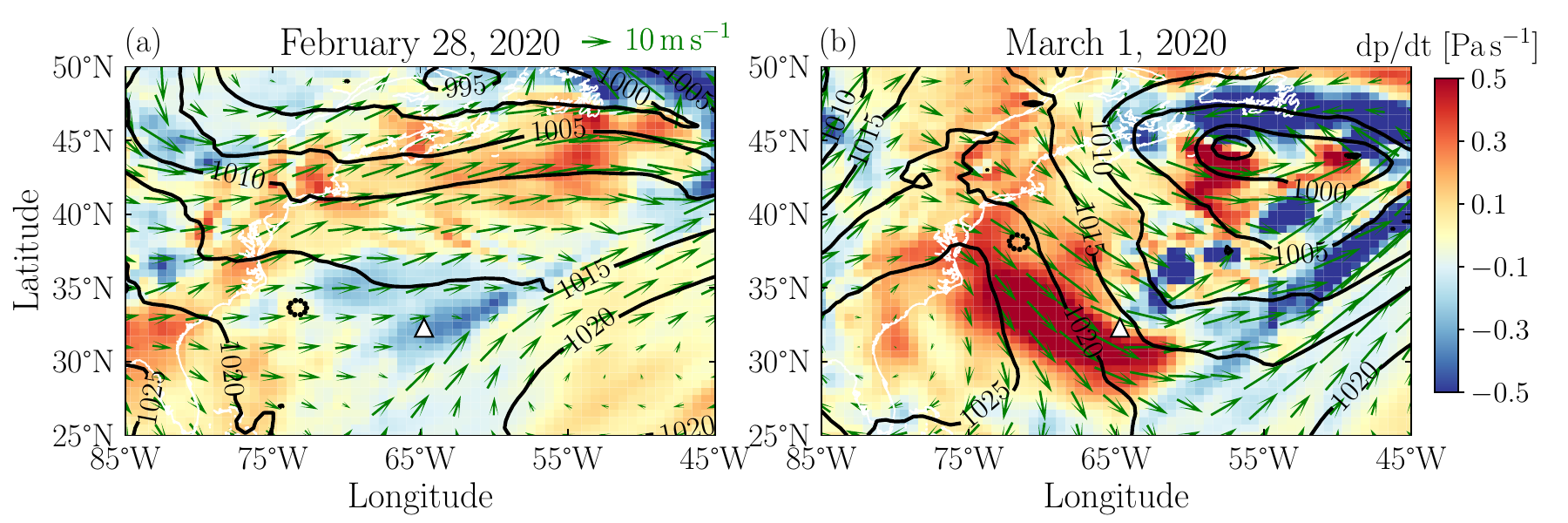}
\end{center}
\caption{Synoptic weather maps for the (a) February 28
and (b) March 1 cases. The colored contours, green arrows, and black
contours represent MERRA-2 omega velocity $\rm dp/dt$ at 600 hPa
(positive indicates downward), averaged winds ($\rm{m\, s^{-1}}$)
at 900 hPa, and sea level pressure at 18:00 UTC over the ACTIVATE
measurement region, respectively. \blue{Solid black circles
represent the location of dropsondes.} Triangles represent
the location of Bermuda Island. The \blue{white} lines indicate coastlines.
The length of green arrows is proportional to the magnitude of wind speeds.
The benchmark length represents $10\,\rm{m\,s^{-1}}$ wind speeds.
The instantaneous fields at 18:00 UTC is plotted to match the time of
MERRA-2 and dropsonde measurements.
}
\label{synoptic_maps}
\end{figure*}

\Fig{synoptic_maps} shows synoptic weather maps at 18:00 UTC from MERRA-2
for the February 28 and March 1 cases over the ACTIVATE measurement region. A low
pressure system at the upper-left domain on February 28 moved to the southeast
on March 1 with an anticyclone development along the coast.  The Februray 28 case is
featured by synoptic-scale ascending motion (negtitive omega velocity $\rm{dp/dt}$) and westerly
winds over the sampling domain.
The March 1 case features a subsidence region
(positive omega velocity $\rm{dp/dt}$) east of
the coastal anticyclone and dominant northwesterly winds
west of $60^\circ$W.

\subsection{Dropsonde measurements and derived divergence}
\label{sec:div}

\begin{figure*}[t!]\begin{center}
\includegraphics[width=0.48\textwidth]{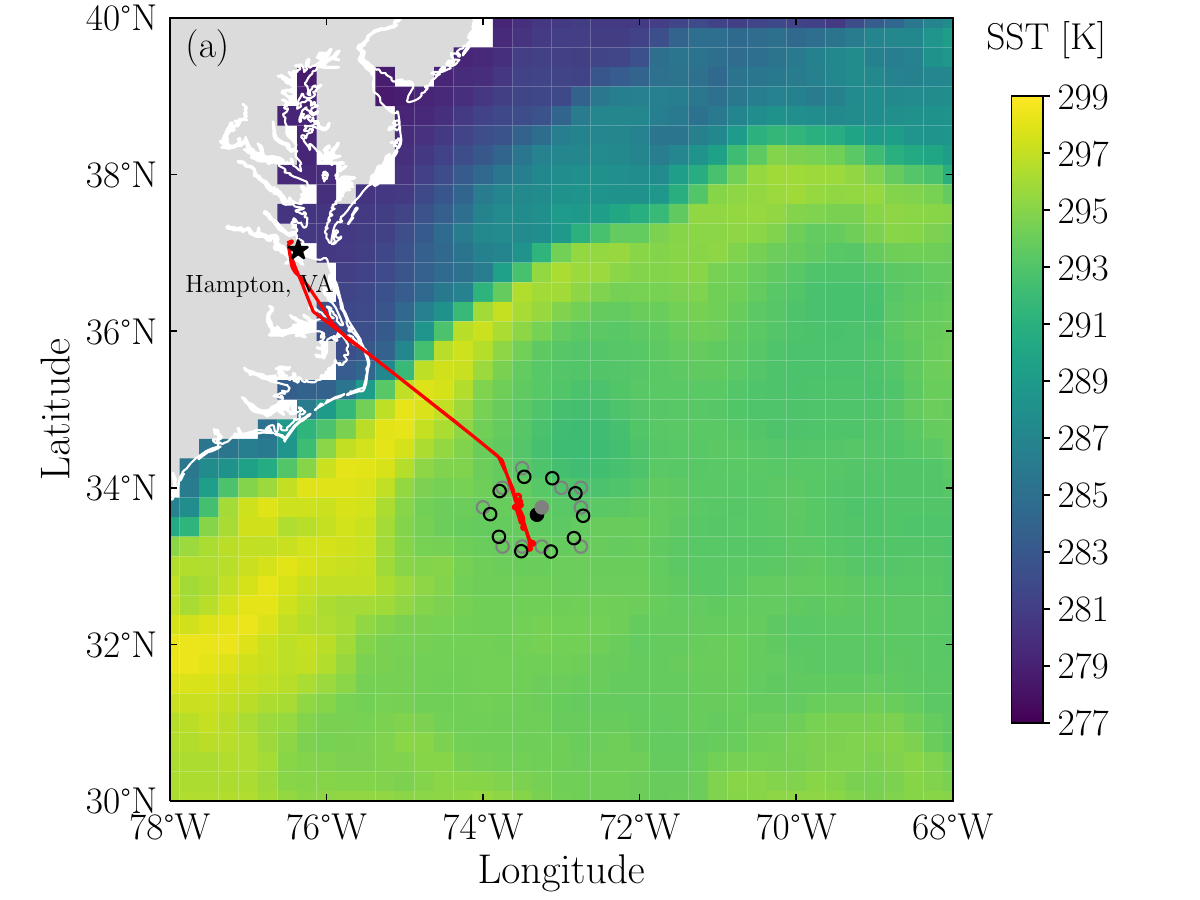}
\includegraphics[width=0.48\textwidth]{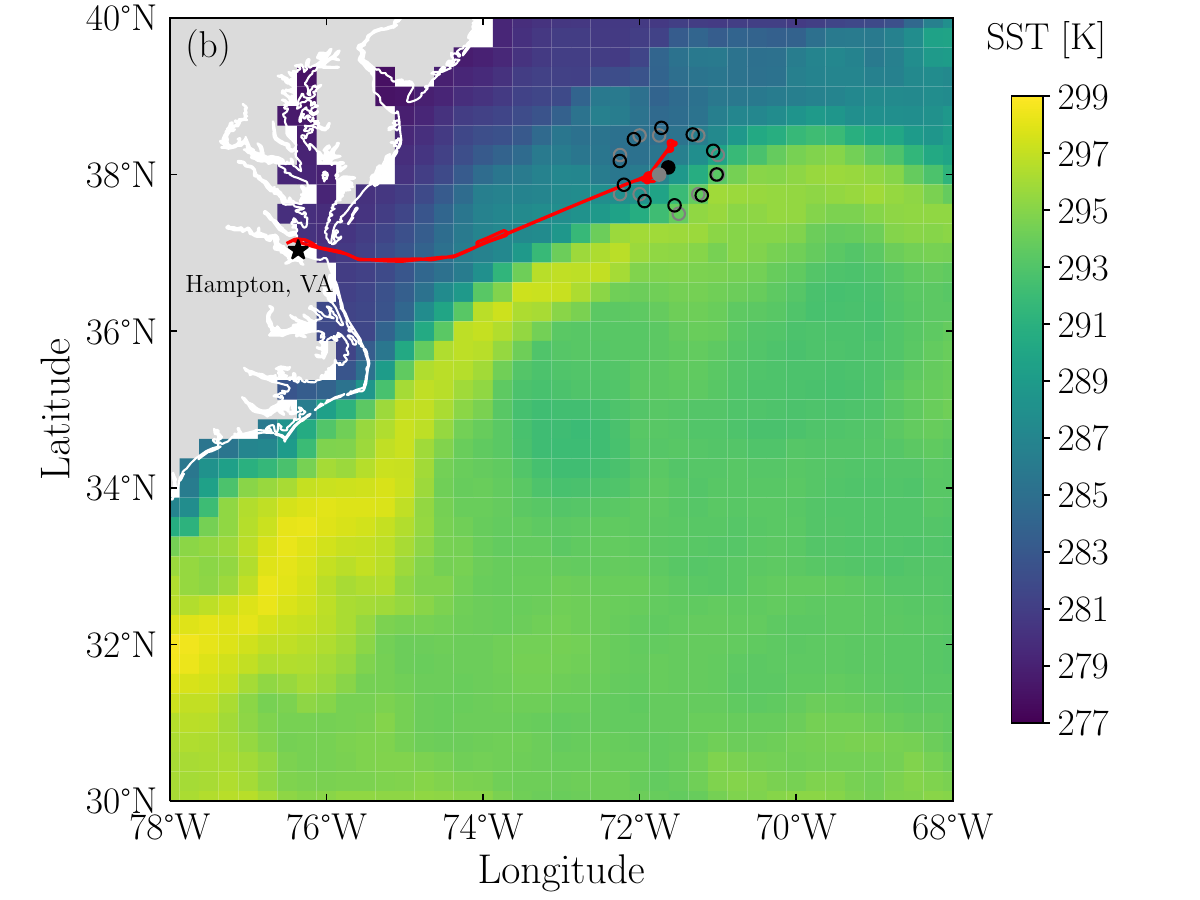}
\end{center}
\caption{Location of dropsondes for the (a): February 28
and (b): March 1 cases. Black open symbols
are the location of dropsondes at the surface and the gray ones are mapped
locations from the ERA5 reanalysis data.
Solid dots represent the center of the dropsonde circle.
The contour map shows ERA5 SST in the measurement region.
The black star represent the location of Hampton, VA
on this map. The \blue{red} curve shows the flight path.
\blue{The white lines and gray area indicate  coastlines and the land, respectively.}
}
\label{drop_map}
\end{figure*}

\Fig{drop_map} shows the location of individual dropsondes and the center of
dropsonde circle on an ERA5 SST map for both cases. The nearest ERA5 grid points
to the dropsondes are also shown in gray open symbols which are used to obtain
the SST for the corresponding dropsondes. Clearly, the SST is much warmer over
the circle on February 28 than March 1.

\begin{figure*}[t!]\begin{center}
\includegraphics[width=\textwidth]{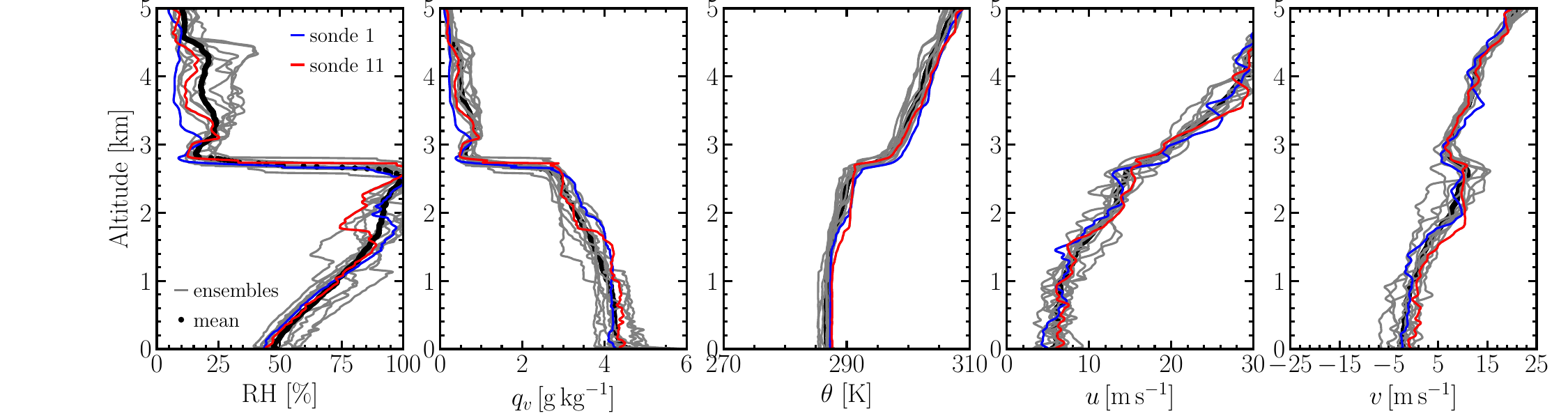}
\includegraphics[width=\textwidth]{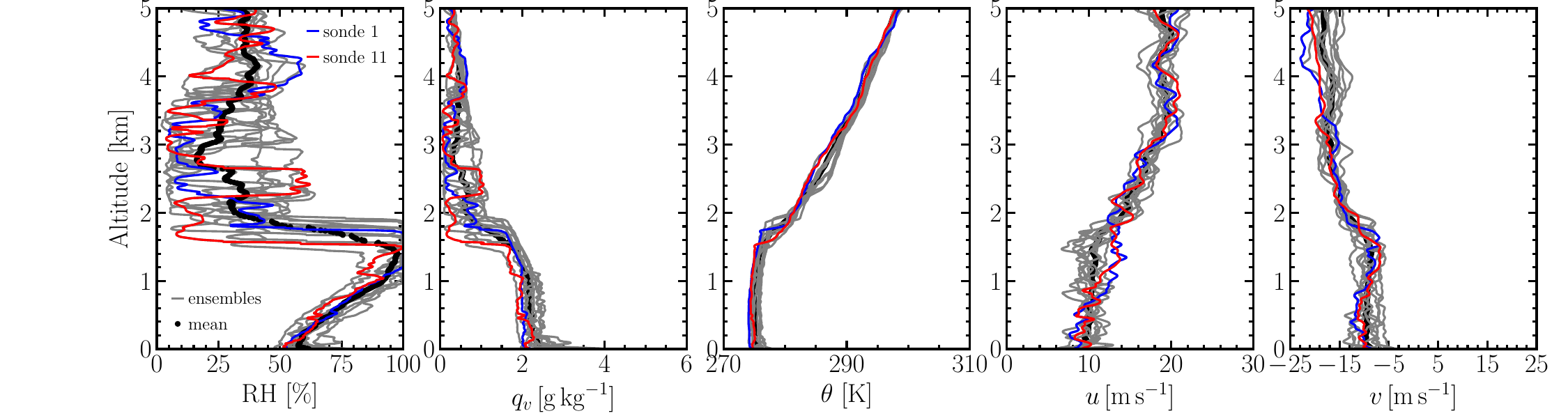}
\end{center}
\caption{Profiles from dropsonde measurements for the February 28 (upper panel)
and March 1 cases (lower panel). The gray lines represent vertical
profiles measured from 11 dropsondes and the thick
black lines represent the corresponding mean profile.
The blue and red curves represent the first and last
dropsonde, respectively, released at about the same location but one hour apart.
}
\label{drop_mapping_profiles}
\end{figure*}

\Fig{drop_mapping_profiles} shows the vertical profiles of RH, $q_v$, $\theta$,
$u$, and $v$ from dropsonde measurements for the two cases. The February 28 case
(RF10) is characterized by a deeper boundary layer with a depth of about $2.8\,
{\rm km}$ and a drier free troposphere compared to the March 1 case (RF13).
Individual RH and $q_v$ profiles show more fluctuations from the mean in the
free troposphere on March 1 than the February 28 case.  The boundary layer for
the March 1 case is shallower.  The magnitude of $u$ and $v$ increases rapidly
with height above the boundary layer, which is more profound on February 28,
showing a strong wind shear.  The meteorological states evolve substantially
during the one-hour sampling time period of both cases, as indicated by the
contrast between the first dropsonde (blue curve) and the last one (red curve)
that were released roughly at the same location.  The boundary layer became
deeper (shallower) with time on February 28 (March 1). 

The vertical velocity of airflow regulates the atmospheric water distribution
but is difficult to measure \citep{bony2019measuring}.  The continuity equation
of nearly incompressible airflow with velocity $\bm{u}=\bm{u}(u,v,w)$ is given
by
\begin{equation}
  {\bm \nabla}\cdot {\bm u}=\frac{\partial u}{\partial x}+\frac{\partial v}{\partial y}+\frac{\partial w}{\partial z}=0.
\label{eq:con}
\end{equation}
According to \Eq{eq:con}, the vertical velocity $w$ can be expressed as
\begin{equation}
w(z)=-\int_0^z\left(\frac{\partial u}{\partial x}+\frac{\partial v}{\partial y}\right){\rm d}z^\prime.
\end{equation}
Divergence is defined as
\begin{equation}
D=\frac{\partial u}{\partial x}+\frac{\partial v}{\partial y}.
\label{eq:d}
\end{equation}
Thus, \Eq{eq:d} can be written as
\begin{equation}
\label{eq:w}
w(z)=-\int_0^zD{\rm d}z^\prime.
\end{equation}
Therefore, $w$ can be indirectly obtained
from the measured horizontal wind components.
In the atmospheric boundary layer, motion of airflow is conventionally
decomposed to large and small scales. By applying Reynolds decomposition to
\Eq{eq:con}, we obtain the large-scale divergence,

\begin{equation}
  \bar{D}=\frac{\partial \bar{u}}{\partial x}+\frac{\partial \bar{v}}{\partial y}.
\label{eq:da}
\end{equation}

We follow the procedure described by \citet{lenschow2007divergence}
to calculate the divergence from dropsonde measurements,
details of which are given in Appendix~\ref{app:D}.
Since we use the linear regression method to estimate
$\bar{D}$, the standard error $\sigma_\epsilon$ can
be estimated.
The large-scale divergence $\bar{D}$ enters the third term in the
r.h.s of \Eq{eq:dT/dt2} and \Eq{eq:dqv/dt2} through \Eq{eq:w}.
Thus the effect of $\bar{D}$ is taken into account
via vertical motion, $w$, in the vertical component of the advective tendency.

\begin{figure*}[t!]\begin{center}
\includegraphics[width=0.48\textwidth]{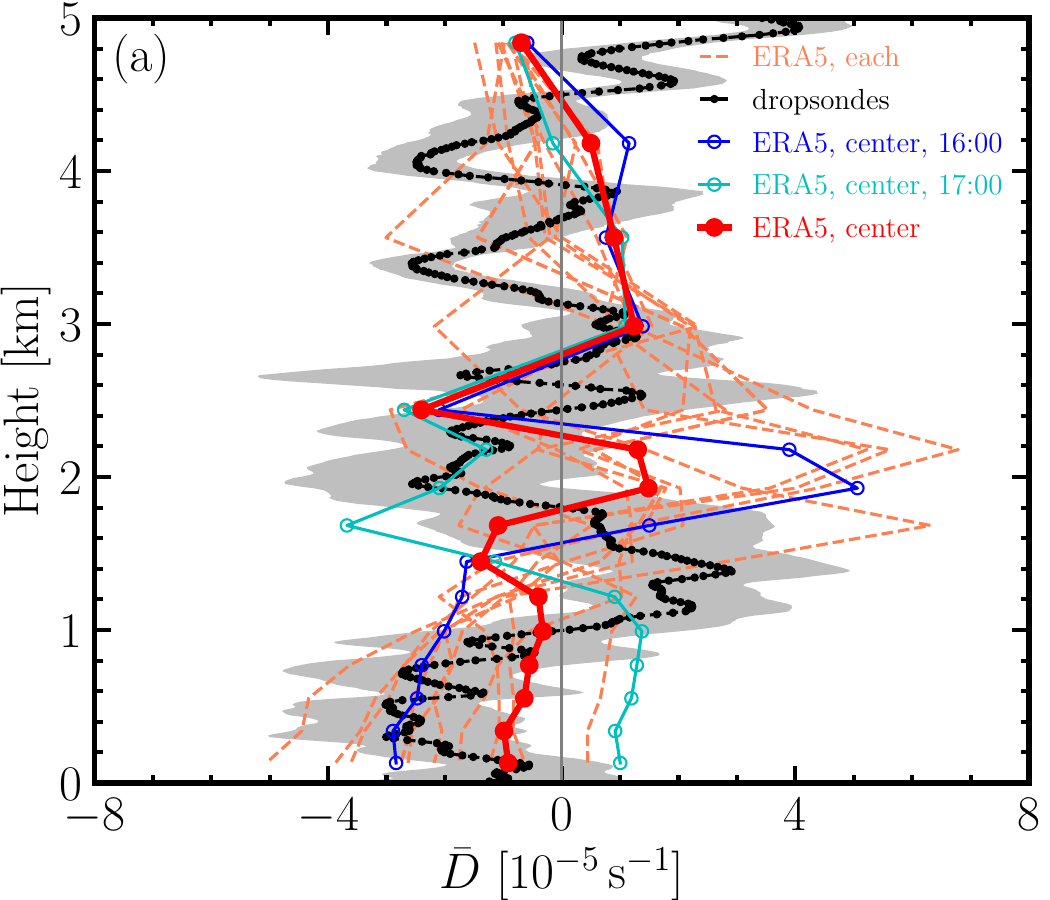}
\includegraphics[width=0.48\textwidth]{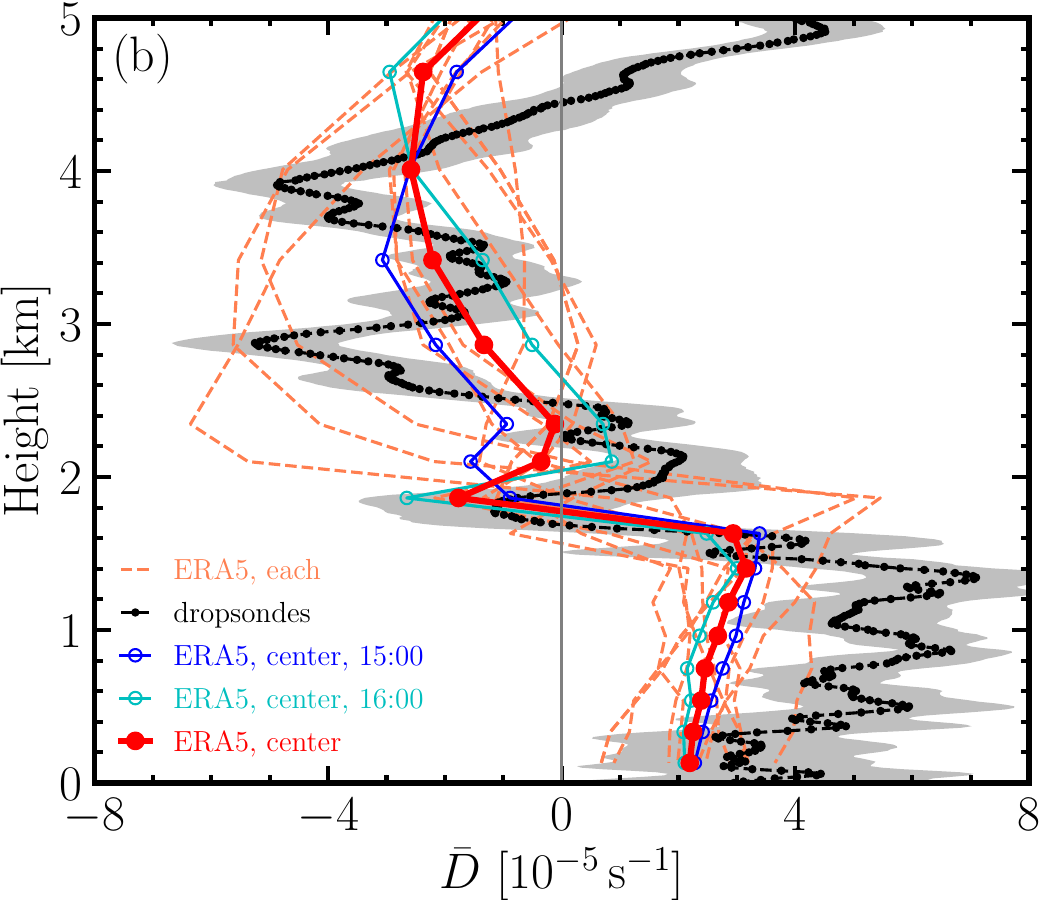}
\end{center}
\caption{Comparison of $\bar{D}$ profiles estimated from dropsonde measurements
with the one from ERA5 reanalysis data for the (a): February 28 and (b):
March 1 cases.  Black dotted lines represent $\bar{D}$ estimated from 10
dropsondes with $\pm\sigma_\epsilon$ uncertainty (gray shaded area).  Blue lines
represent ERA5-$\bar{D}$ profiles at 16:00 UTC for the February 28 and
17:00 UTC for the March 1 cases at the center of dropsondes. Cyan lines
represent ERA5-$\bar{D}$ profiles at 17:00 UTC for the February 28 and
16:00 UTC for the March 1 cases at the center of dropsondes. Red solid-dotted
lines represent $\bar{D}$ from ERA5 reanalysis data averaged during the
measurement time (between blue lines and cyan lines). The dashed coral-colored
lines represent ERA5 reanalysis data at the location of individual dropsondes
averaged during the measurement time for each case.
}
\label{era5_div_0228}
\end{figure*}

We compare $\bar{D}$ estimated from the dropsonde measurements with the one
obtained from ERA5 reanalysis data for both cases as shown in
\Fig{era5_div_0228}.  ERA5-$\bar{D}$ profile (red curve) averaged between 16:00
UTC and 17:00 UTC (two vertical profiles) at the dropsonde center is able to
capture the sign of $\bar{D}$ vertical-structure measured by dropsonde (black
curve) for the February 28 case (\Fig{era5_div_0228}(a)) within 1 km above the
surface. However, it differs from the dropsonde measurements in both signs and
magnitude above 1 km, which requires further investigation.  More
interestingly, ERA5-$\bar{D}$ profile evolves from convergence to divergence
within the boundary layer from 16:00 UTC (blue curve) to 17:00 UTC (cyan curve).
We also examine the ERA5 $\bar{D}$ averaged between 16:00 UTC and 17:00 UTC at
the location of each dropsonde as shown by coral-colored dashed lines in
\Fig{era5_div_0228}, which exhibits large spatial fluctuations. 
This demonstrates strong spatial inhomogeneity in
large-scale vertical motions over this area on February 28,
as also indicated by the sensitivity of $\bar{D}$ to different
dropsonde subsets as shown in \Fig{drop_div0228}.
The amplitude of the mean and largest value of dropsonde-$\bar{D}$ for the
February 28 case is $\langle|\bar{D}|\rangle=1.38\times10^{-5}\, {\rm s}^{-1}$
and $|\bar{D}|_{\rm max}=4.06\times10^{-5}\, {\rm s}^{-1}$, respectively.  For
the March 1, they are $\langle|\bar{D}|\rangle=2.99\times10^{-5}\, {\rm s}^{-1}$
and $|\bar{D}|_{\rm max}=7.09\times10^{-5}\, {\rm s}^{-1}$.  These values are
about one order of magnitude larger than the ones from non-CAO marine BL cloud
regimes.  The mean value of $\bar{D}$ inferred from the ensemble of radiosondes
during the Atlantic Trade-Wind Experiment \citep{augstein1973mass} and the
Barbados Oceanographic and Meteorological Experiment
\citep{holland1973measurements} is about $10^{-6}\, {\rm s}^{-1}$.  Similar
values were used in the case studies during the VAMOS
Ocean-Cloud-Atmosphere-Land Study \citep{rahn2010marine, wang2010modelling} and
Second Dynamics and Chemistry of Marine Stratocumulus field study (DYCOMS II)
field campaign \citep{wang2009modeling}. $\bar{D}$ estimated during the
NARVAL2 is about $10^{-5}\, \rm{s^{-1}}$ \citep{bony2019measuring}.

\Fig{era5_div_0228}(b) shows the same comparison but with the dropsonde
measurements conducted on March 1.
In this case, $\bar{D}$ obtained from ERA5
reanalysis data (red solid-dotted line) at the center
of dropsonde circle is able to capture the general vertical structure
of $\bar{D}$ estimated from dropsonde measurements.
Similar to the February 28 case, there is a strong spatiotemporal
variation in the ERA5 $\bar{D}$.

\begin{figure*}[t!]\begin{center}
\includegraphics[width=\textwidth]{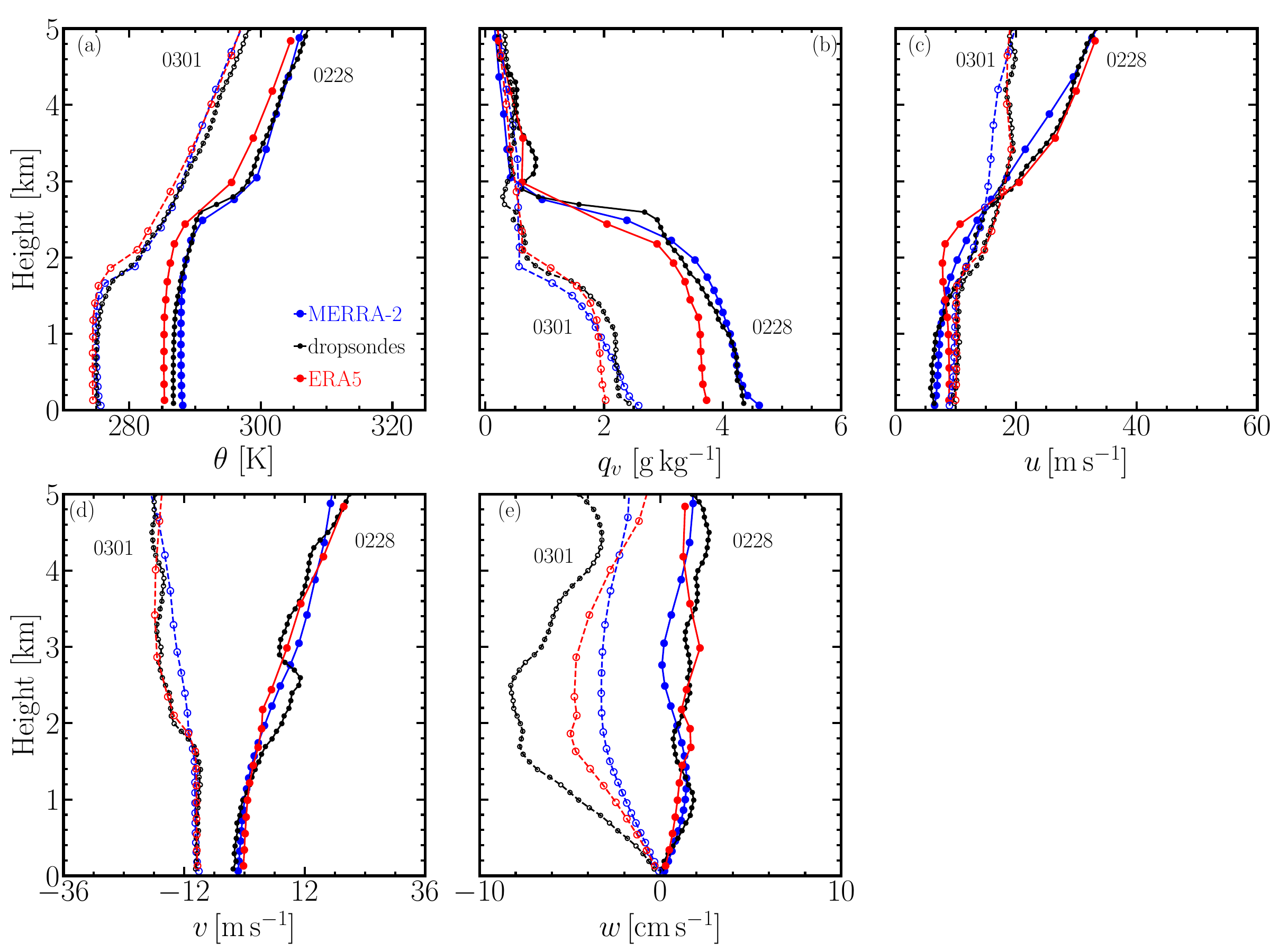}
\end{center}
\caption{Comparison among dropsonde measurements (black curves), ERA5 (red curves),
and MERRA-2 (blue curves) at the dropsonde center for both cases.
Solid symbols represent the
February 28 case (ERA5 data averaged between 16:00 and 17:00 UTC and MERRA-2
averaged between 15:00 and 18:00 UTC) and the open circles mark the March 1 case
(ERA5 averaged between 15:00 and 16:00 UTC and MERRA-2 at 15:00 UTC).
Large-scale vertical velocity $w$ from ERA5 corresponds to ERA5-$\bar{D}$ in \Fig{era5_div_0228}.
}
\label{w_comp}
\end{figure*}

We further compare the large-scale vertical velocity $w$ (subsidence) with
MERRA-2 reanalysis data as shown in \Fig{w_comp}(e) for both cases.  For the
February 28 case, the $w$ profile from MERRA-2  is averaged between 15:00 UTC
and 18:00 UTC and the one from ERA5 is averaged between 16:00 UTC and 17:00 UTC
to better match the dropsonde sampling time. Both the ERA5 and MERRA-2 can
reasonably capture the vertical profile of $w$ when compared with dropsonde
measurements for this case.  For the March 1 case, the $w$ profile from MERRA-2
reanalysis data at 15:00 UTC is used to compare with dropsonde measurements
while the one from ERA5 reanalysis data is averaged between 15:00 UTC and 16:00
UTC.  The ERA5 reanalysis data agree with the dropsonde measurements in the sign
but underestimate the magnitude.  The MERRA-2 does not capture the structure and
magnitude of the vertical profile of $w$ well.  Comparison of $\theta$, $q_v$,
$u$, and $v$ profiles is also shown in \Fig{w_comp}(a)--(d).  MERRA-2  shows a
slightly warmer boundary layer for the February 28 case while the ERA5  shows a
colder one.  Both MERRA-2 and ERA5 data capture the $\theta$ profile well for
the March 1 case. ERA5 yields a drier ($q_v$ profiles) boundary layer while
MERRA-2 capture the $q_v$ well compared to the dropsonde measurements for both
cases.  The $u$ and $v$ profiles within the boundary layer are represented well
by MERRA-2 and ERA5 data for both cases, given the large spread among the
individual dropsondes for the circled area (see \Fig{drop_mapping_profiles}).
The ERA5 captures those profiles above the boundary layer better than the
MERRA-2.  Overall, comparing to MERRA-2, ERA5 profiles are more consistent with
the dropsonde measurements, as also shown in \citet{Seethala21} for the broader
WNAO region. 

Since we aim to use the divergence as part of the large-scale
forcings to drive WRF-LES, the agreement of $\bar{D}$
(and the corresponding $w$) from ERA5
with the estimates from dropsondes for the March 1 case
affords confidence to use the hourly ERA5 divergence
to test the sensitivity of WRF-LES to time-varying large-scale forcings.

\subsection{Surface heat fluxes}
\label{sec:hf}

Turbulent sensible and latent heat fluxes at the surface
are important flux boundary conditions to
drive LES of boundary layer clouds.
They are responsible for the heat and moisture exchange between
the ocean and atmosphere.
Surface heat fluxes are challenging to measure and
estimate due to the nonlinear processes involved.
Therefore the so-called
bulk aerodynamics parameterization has been used
to estimate surface heat fluxes. Bulk aerodynamics
algorithms parameterize the turbulence instability
as well as the roughness length of the wind speed,
temperature and the water vapor mixing ratio \citep{zeng1998intercomparison}.
The surface sensible heat flux (SHF) and latent heat flux (LHF) are
given by \citep{smith1988coefficients}
\EQ
{\rm SHF}=C_T\rho c_{\rm p}U(T_{\rm s}-\theta),
\label{eq:ce}
\EN
\EQ
{\rm LHF}=C_EL_v\rho U(q_{\rm s}-q_v),
\label{eq:ct}
\EN
where $\rho$ is the air mass density,
$c_{\rm p}$ is the specific heat of air, 
$U$ is the mean wind speed
at a particular reference height, $T_{\rm s}$ is the SST,
$\theta$ is the potential temperature at the reference
height, $q_{\rm s}$ is the saturated
water vapor mixing ratio, $q_v$ is the water vapor mixing ratio,
and $L_v$ is the latent heat of evaporation.
The sensible heat flux coefficient $C_T$ and evaporation
coefficient $C_E$ are determined by the empirical
Monin–Obukhov (MO) similarity theory.
In this study, we adopt the bulk aerodynamics algorithms
developed by \citet{zeng1998intercomparison} to calculate
heat fluxes for comparison between ERA5 and dropsonde
measurements,
which we refer to as ``Z98'' hereafter.
The Z98 algorithm calculates $C_T$ and
$C_E$ based on instability analysis.
\Eq{eq:ce} and \Eq{eq:ct} are used to calculate
SHF and LHF after $C_T$ and $C_E$ are obtained.
The input parameters of the Z98 algorithm are
$T_{\rm s}$, 10-m (reference height)
wind speed $U_{10\rm m}$, temperature
$T_{10\rm m}$ (to calculate $\theta_{10 \rm m}$),
and the water vapor mixing ratio $q_{v, 10\rm m}$.

\begin{figure}[t!]\begin{center}
\includegraphics[width=0.48\textwidth]{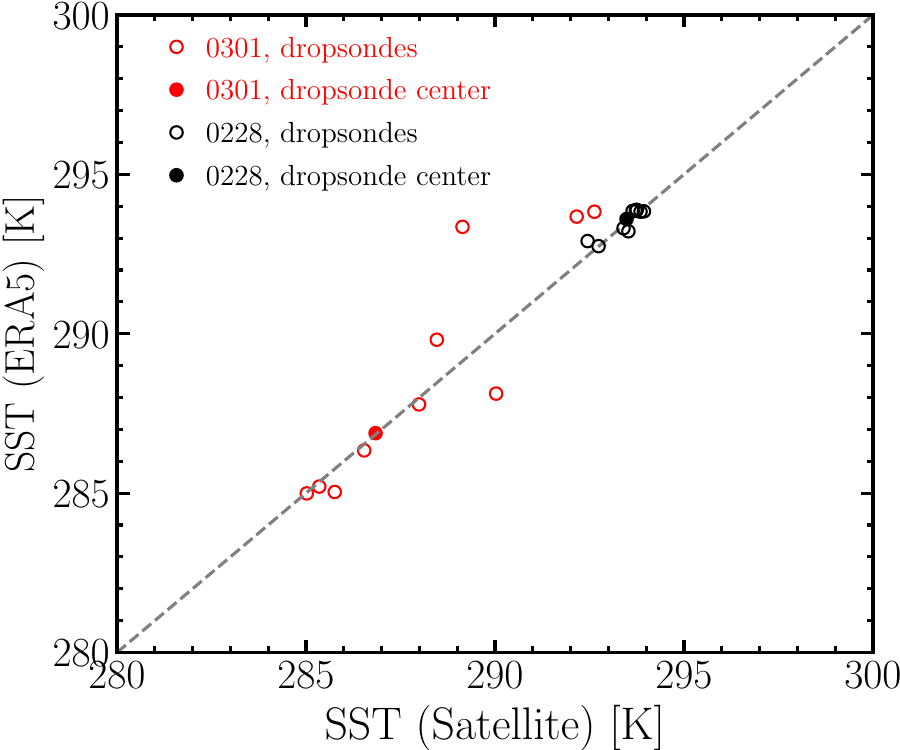}
\end{center}
\caption{
Comparison of SST from satellite retrievals and ERA5 reanalysis data
for both February 28 (black) and March 1 (red) cases .
Open symbols represent SST at the position of each dropsonde.
The two solid dots represent SST at the center of dropsonde circles.
}
\label{sst_IR}
\end{figure}

To derive the surface heat fluxes using
dropsonde measurements and evaluate ERA5 reanalysis data,
we first compare the ERA5 SST to satellite
measurement for the two cases as a quality check. As shown in \Fig{sst_IR},
for both 28 February 2020 (black symbols) and 1 March 2020 cases
(red symbols), SST from ERA5 and
the satellite measurement matches well at the center of
dropsonde circle.  At the location of individual dropsondes
the agreement is reasonable on February 28 while several points
on March 1 are quite off,
which is likely because of the location
mismatch due to the resolution difference and sampling area
being near strong SST gradients (shown in \Fig{drop_map}).
The Normalized Root-Mean-Square Error (NRMSE) is $0.1\%$
for the February 28 case and is $0.6\%$ for the March 1 case.
This comparison suggests that the SST from ERA5 
can also be used as the initial input for our WRF-LES.
ERA5 has assimilated the Operational Sea Surface
Temperature and Sea Ice Analysis (OSTIA) system
for hourly SST starting from 2007 \citep{hirahara201626}.
The OSTIA assimilated the MW-IR measurements.
Thus, the agreement between ERA5 and the satellite retrievals
is expected.

\begin{figure*}[t!]\begin{center}
\includegraphics[width=\textwidth]{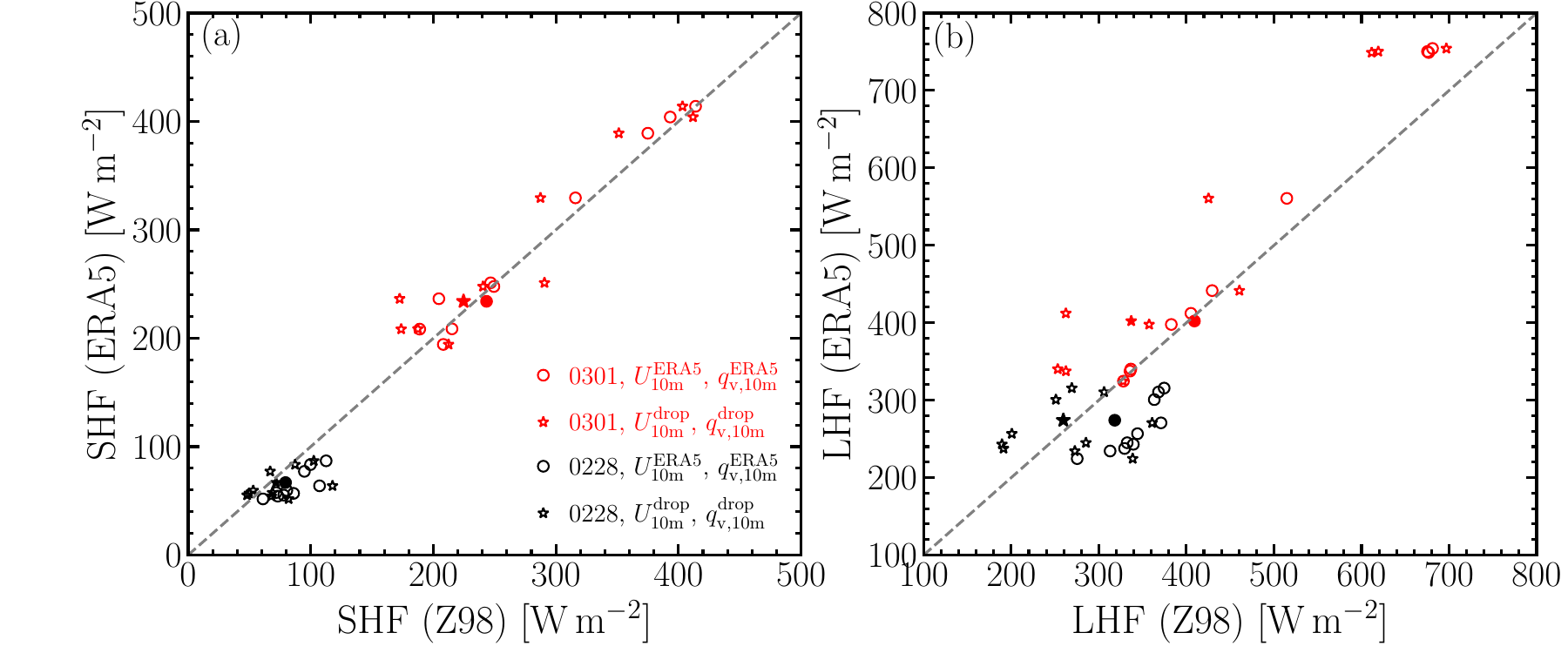}
\end{center}\caption{Comparison between the estimated heat fluxes and the one
from ERA5 reanalysis data at 
16:00 UTC for the February 28 case (black symbols)
and 15:00 UTC for the March 1 case (red symbols).
$T_{\rm s}^{\rm ERA5}$ is adopted for the calculation.
Open symbols represent $T_{\rm s}^{\rm ERA5}$
at the position of individual dropsondes
and solid symbols represent the center of the dropsonde
circle. Two sets of input variables are adopted to calculate the
surface heat fluxes: $U_{10\rm m}^{\rm ERA5}$ and $q_{v,10\rm m}^{\rm ERA5}$ (circles) 
and $U_{10\rm m}^{\rm drop}$ and $q_{v,10\rm m}^{\rm drop}$ (stars). 
}
\label{hf_era5_Z98}
\end{figure*}

Next, we compare surface heat fluxes directly obtained
from ERA5 reanalysis data and the ones estimated from ACTIVATE measurements.
Since the SST obtained from ERA5 agrees
with the satellite measurement,
we try to examine if the ERA5
heat-fluxes can be reproduced from ERA5 SST and dropsonde measurements.
First, we use the Z98 algorithm to calculate
heat fluxes based on $T_{\rm s}^{\rm ERA5}$, $U_{\rm 10m}^{\rm ERA5}$, $q_{v,{\rm 10m}}^{\rm ERA5}$
from ERA5 and $T_{\rm 10m}^{\rm drop}$ from dropsonde.
\Fig{hf_era5_Z98}(a) shows the comparison between the
estimated SHF and the ERA5 reanalysis data
(red and black circles).
The corresponding NRMSE is $5.1\%$
and is $36.4\%$ 
for the March 1 and February 28 case, respectively. The comparison
of LHF is shown in \Fig{hf_era5_Z98}(b) with
NRMSE of $8.5\%$ and
$30.0\%$ for the March 1 (red circles) and
February 28 case (black circles), respectively.
These demonstrate a good agreement between the estimated
heat fluxes and the ERA5 reanalysis data on March 1 case,
given the large spread within the
dropsonde circle. The agreement is particularly good at the circle center (solid symbols).
This suggests that the Z98 algorithm can be used to calculate heat fluxes
and the use of $T_{\rm 10m}^{\rm drop}$ is justified in this study.
It is evident that SHF and LHF calculated using Z98
are underestimated compared to ERA5 for the February 28 case. 
This is because ERA5-$\theta$ is smaller than the dropsonde-$\theta$
within the boundary layers as shown in \Fig{w_comp}(a).
We then use $U_{\rm 10m}^{\rm drop}$ and $q_{v, {\rm 10m}}^{\rm drop}$
obtained from dropsonde measuremens to estimate the fluxes
(stars in \Fig{hf_era5_Z98}), which yields
a NRMSE value of $11.5\%$
for SHF and $19.3\%$
for LHF for the March 1 case, further
indicating that ERA5 gives a good estimate
of turbulent heat fluxes for the March 1 case.
For the February 28 case, NRMSE of SHF and LHF calculated using
$U_{\rm 10m}^{\rm drop}$ and $q_{v, {\rm 10m}}^{\rm drop}$
are $33.2\%$ and $23.1\%$, respectively. This underestimation
is because that $q_v$ from ERA5 is smaller
than the one from dropsonde measurements within the boundary
layer as shown in \Fig{w_comp}(b).
We also compare heat fluxes between the ERA5
and MERRA-2 reanalysis data. Both SHF and LHF agree well
between MERRA-2 and ERA5 for the February 28 case.
However, MERRA-2 underestimates SHF and LHF
compared to ERA5 for the March 1 case (see appendix~\ref{app:shf}).

Overall, by adopting $T_{\rm s}^{\rm ERA5}$, $U_{\rm 10m}$, $q_{\rm v,10m}$,
and $T_{\rm 10m}^{\rm drop}$ from dropsonde measurements to estimate
heat fluxes using Z98 algorithm, we are able to evaluate the heat fluxes
from ERA5. The time-varying ERA5 heat fluxes are
then used in the WRF-LES sensitivity tests.
The method of using dropsonde measurements to estimate
surface heat fluxes was also adopted for studying the
tropical cyclones \citep{powell2003reduced, holthuijsen2012wind, richter2016assessment}.
These studies show that the accuracy of estimated coefficients
based on MO similarity theory decreases with increasing wind speed. In the present study, the wind speed is orders
of magnitude smaller than that of tropical cyclones, which
ensures the accuracy of using dropsonde measurement to
estimate surface heat fluxes.

\section{WRF-LES sensitivities to large-scale forcings and contrast between the two CAO cases}
\label{sec:forcing}

\subsection{Sensitivities to large-scale advective tendencies and relaxation}

In this section, we investigate how to better represent
time-varying meteorological states in idealized WRF-LES applying
either advective tendencies to $\theta$ and $q_v$, relaxation to $u$ and $v$, or both.
Simulations are driven by constant surface fluxes
${\rm SHF} (t_0)$ and ${\rm LHF} (t_0)$.
Here $t_0$ denotes the starting time of simulations.
Since we have shown
in the previous section that ERA5 reanalysis
data agree well with the dropsonde measurements during
the sampling time periods of the two CAO cases,
we adopt hourly $\theta$, $q_v$, $u$, and $v$
vertical profiles from ERA5 reanalysis data and derive the corresponding
vertical profiles of advective tendencies and relaxation adjustments.
The hourly meteorological states simulated in WRF-LES are then
compared to ERA5 reanalysis data that are partly
validated against dropsonde measurements.

\begin{figure*}[t!]\begin{center}
\includegraphics[width=\textwidth]{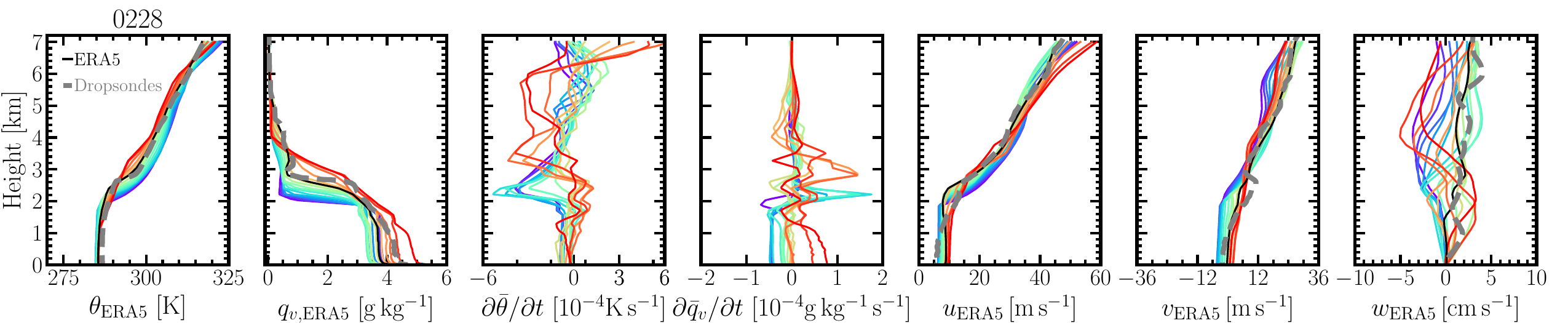}
\includegraphics[width=\textwidth]{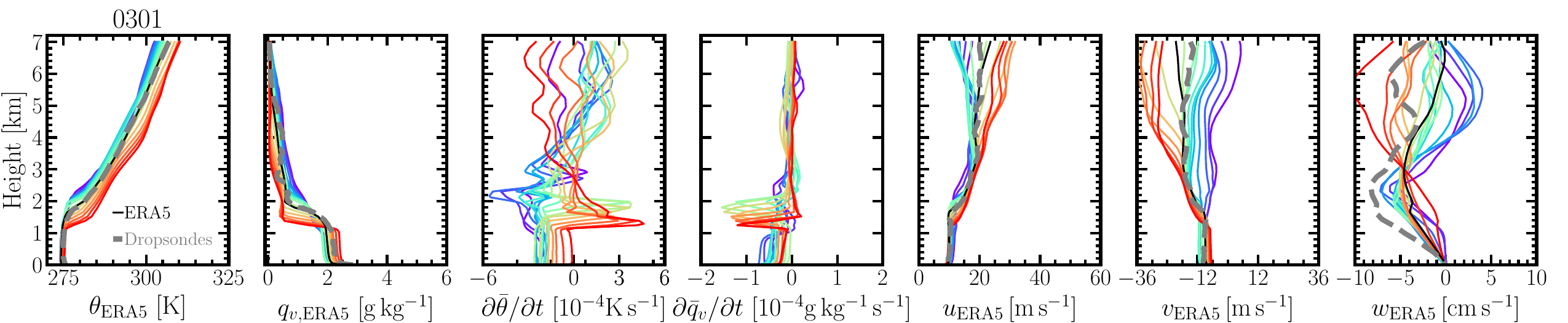}
\end{center}\caption{Hourly meteorological state and forcing profiles
for the February 28 case (simulation 0228A, upper panel) and March 1 case
(simulation 0301A, lower panel) from
ERA5 reanalysis data averaged over a $2^\circ\times2^\circ$ domain. 
The rainbow color scheme represents the time evolution (06:00-21:00 UTC): from purple to red.
The averaged ERA5 reanalysis data over the measurement time period are marked
by black lines, which are compared with the dropsonde measurements
(dashed gray lines).}
\label{input_ls_forcing}
\end{figure*}

\Fig{input_ls_forcing} shows the hourly (rainbow-colored lines) input
meteorological forcing being obtained from ERA5 reanalysis data for the WRF-LES
simulations.  The evolution of vertical profiles are averaged over a
$2^\circ\times2^\circ$ area centered at the middle of the dropsonde circle of
each case. This selected area sufficiently covers the dropsonde circle.
Vertical profiles of $\theta$, $q_v$, $u$, $v$, and $w$ obtained from ERA5
reanalysis data averaged during the measurement time (black solid lines) agree
reasonably well with the dropsonde measurements (gray dashed lines) for the
February 28 case (upper row) and for the March 1 case (lower row).  Vertical
profiles of advective tendencies of $\theta$ and $q_v$ (i.e., $\partial
\bar{\theta}/ \partial t$ and $\partial \bar{q}_v /\partial t$) are calculated
from $\theta_{\rm ERA5}$ and $q_{v,\rm ERA5}$.  Vertical profiles of
$\theta_{\rm ERA5}$, $q_{v,\rm ERA5}$, $u_{\rm ERA5}$, and $v_{\rm ERA5}$ at
06:00 UTC are taken as the input sounding for WRF-LES when the
simulation starts.  We note that $w_{\rm ERA5}$ averaged over the
$2^\circ\times2^\circ$ area at 15:00 UTC differs slightly from the one at the
dropsonde center shown by the red curve in \Fig{w_comp} as expected.  This is
because of the strong spatial variation of $\bar{D}$ as shown in
\Fig{era5_div_0228}.  We have tested the relaxation time scale $\tau$ of $u$ and
$v$ for the February 28 case and found that WRF-LES with $\tau$=30
min, 1 h, and 3 h reveal almost identical vertical profiles and liquid water
path (LWP).  Therefore we adopt $\tau$=1 h for all the simulations as the ERA5
reanalysis data has a time resolution of one hour. The relaxation is applied to
all vertical layers of the LES domain.

\begin{figure*}[t!]\begin{center}
\includegraphics[width=\textwidth]{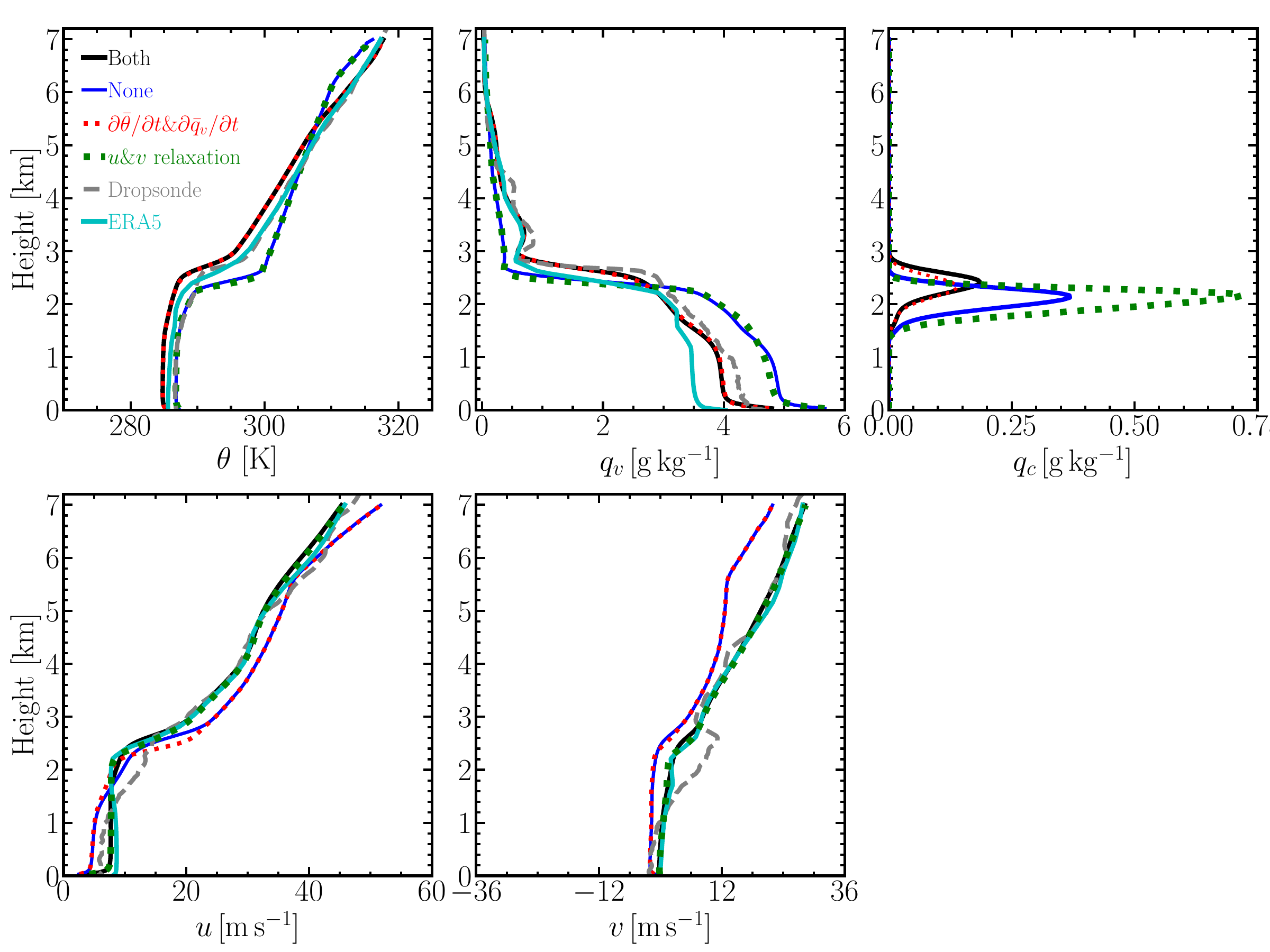}
\end{center}\caption{Domain-averaged vertical profiles for simulation
0228A (black curve), 0228B (green curve), 0228C (red curve), and 0228D (blue
curve) with the corresponding input forcings shown in \Fig{input_ls_forcing} for
the February 28 case during the measurement time listed in \Tab{tab:drop}. The
cyan curve represent profiles from the ERA5 reanalysis data averaged during the
measurement time and the grey curve represent the ones from dropsonde
measurements.
Heat fluxes are from ERA5 reanalysis data:
${\rm SHF}(t_0)=79.91\, \rm{W\, m}^{-2}$ and ${\rm LHF}(t_0)=305.02\, \rm{W\, m}^{-2}$.
}
\label{verticalP_forcing}
\end{figure*}

We first perform a simulation without applying advective tendencies of $\theta$
and $q_v$ and relaxation of $u$ and $v$ (simulation 0228D) for the February 28
case.  It is shown by the blue curves in \Fig{verticalP_forcing} that such a
configuration yields vertical profiles that have a large deviation from the ERA5
reanalysis data (cyan curves) and dropsonde measurements (grey curves).  The
$\theta$ profile from WRF-LES differs considerably from ERA5 above the boundary
layer and the $q_v$ profile shows a more humid boundary layer than the ERA5
(the ratio of $q_v$ from ``Both'' to that from ``ERA5'' is 1.32 averaged within the boundary layer
with a depth of 2.4 km during the measurement time).  The $u$ and $v$
profiles from WRF-LES deviate from the ERA5 and dropsonde measurements.  When
$\partial \bar{\theta}/ \partial t$ and $\partial \bar{q}_v/ \partial t$ are
applied (simulation 0228C), $\theta$ and $q_v$ profiles from WRF-LES agree well
with the dropsonde measurements as shown by the red curves in
\Fig{verticalP_forcing}.  However, $u$ and $v$ profiles still deviate from the
ERA5 reanalysis data.  We then only apply the $u$ and $v$ relaxation to the
WRF-LES (simulation 0228B). As shown by the green curves of
\Fig{verticalP_forcing}, $u$ and $v$ profiles from WRF-LES are in good agreement
with dropsonde measurements even though $\theta$ and $q_v$ profiles differ from
the measurements.  This naturally leads to the configuration of applying
advective tendencies of $\theta$ and $q_v$ together with $u$ and $v$ relaxation
to $u_{\rm ERA5}$ and $v_{\rm ERA5}$ (simulation 0228A).  Such a configuration
leads to vertical profiles ($\theta$, $q_v$, $u$, and $v$) that are comparable
to ERA5 reanalysis data and dropsonde measurements as shown by the black curves
of \Fig{verticalP_forcing}.  Therefore, this combined forcing and relaxation
scheme is justified to simulate the two CAO cases.  Evolution of the vertical
profiles for simulations 0228A, 0228B, 0228C, and 0228D are shown in
section~\ref{app:vp}.  As also shown in \Fig{verticalP_forcing}, the magnitude
and time evolution of cloud water simulated by WRF-LES are sensitive to the
boundary layer meteorological conditions.

\begin{figure}[t!]\begin{center}
\includegraphics[width=0.48\textwidth]{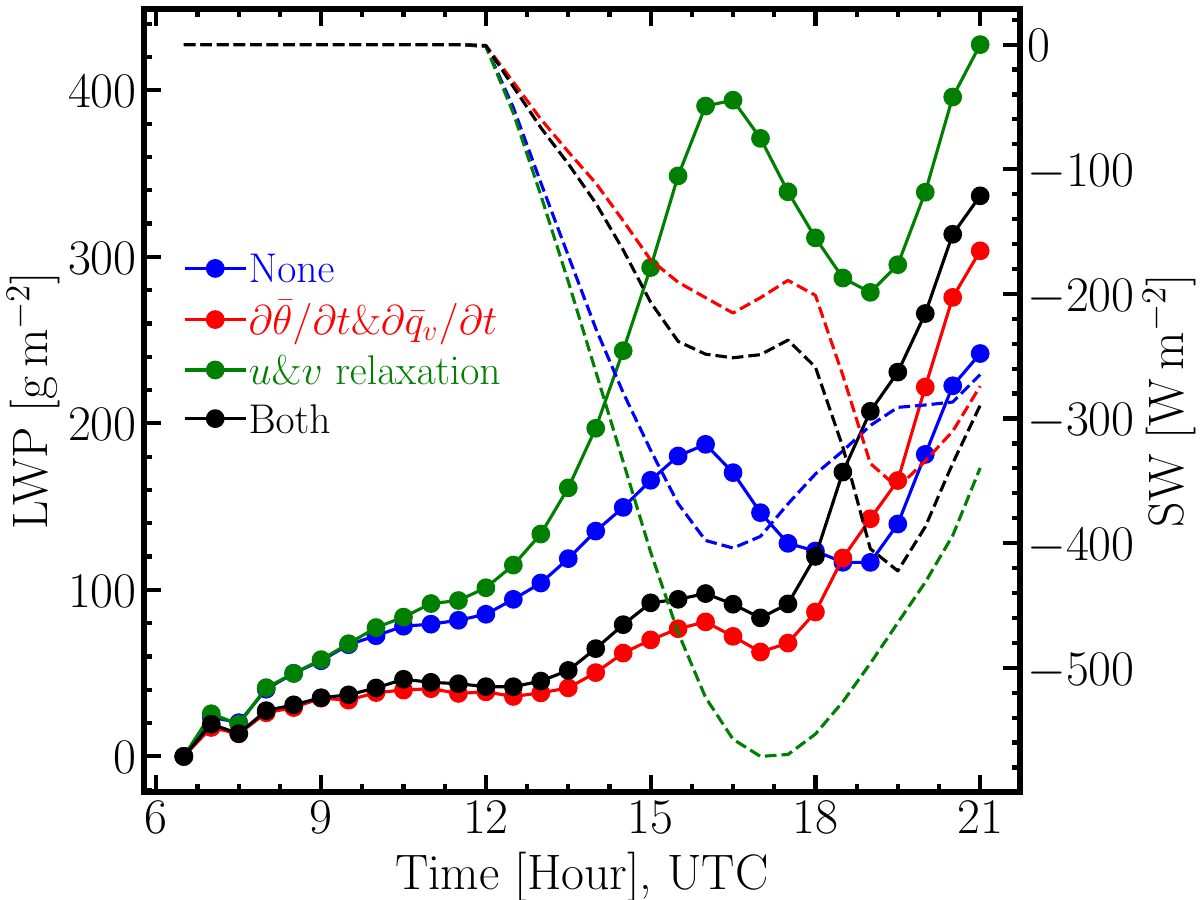}
\end{center}\caption{Time series of domain-averaged LWP
(liquid cloud and rain, solid symbols) and short-wave (SW) cloud forcing at the top of atmosphere
(dashed lines) of simulations (blue:0228D, red:0228B, green:0228C, black:0228A) with different forcing options
as shown in \Fig{verticalP_forcing}.
}
\label{lwp_varyingF_LASSO_0228}
\end{figure}

\begin{figure*}[t!]\begin{center}
\includegraphics[width=\textwidth]{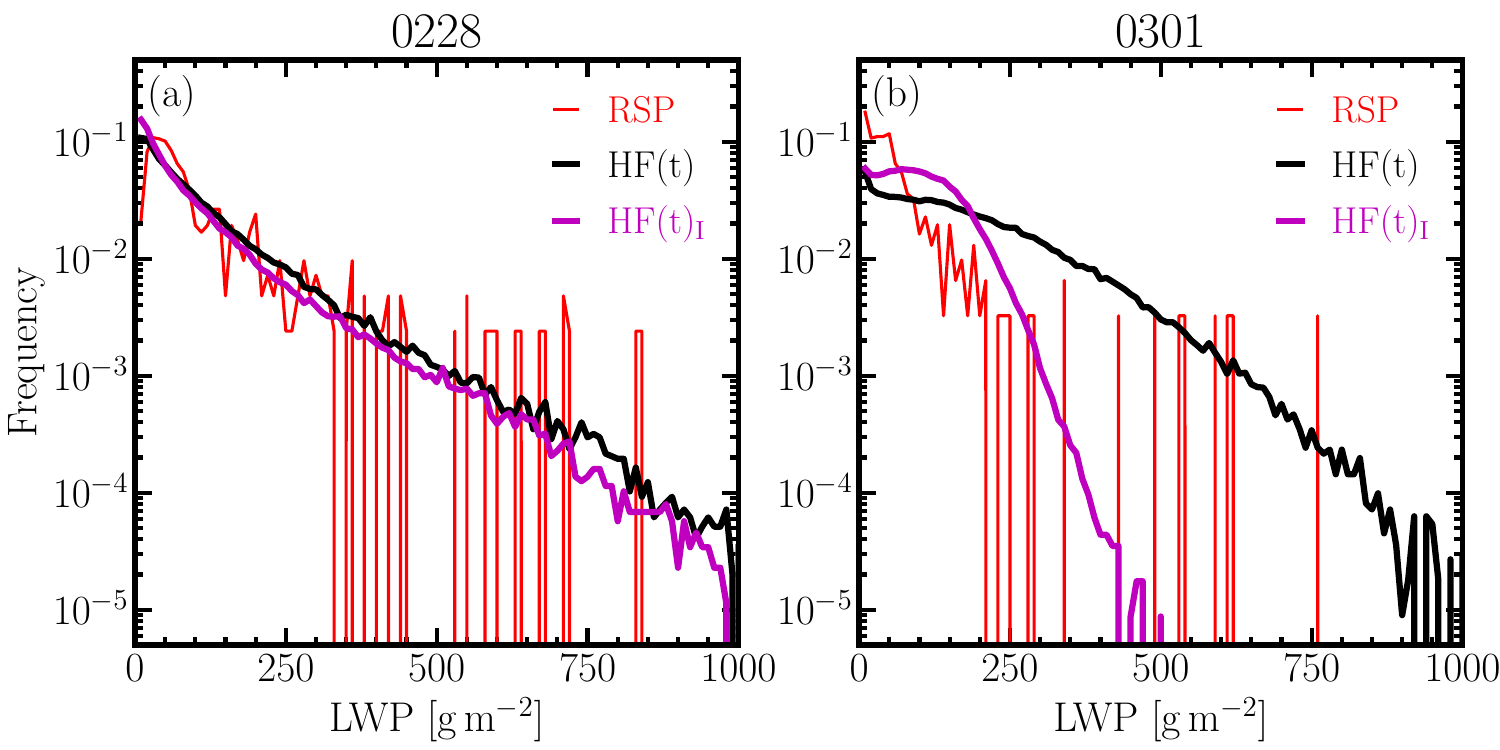}
\end{center}\caption{Comparison of frequency distribution of LWP
between RSP measurements and WRF-LES 
for (a): February 28 (0228E) and (b): March 1 (0301E) cases.
${\rm HF(t)_I}$ denotes heat fluxes calculated interactively (0228F and 0301F)
from WRF-LES.
The frequency for WRF-LES LWP is calculated from 3 snapshots
(every 30 minutes)
during the measurement time.
Note that the boundary layer evolves  as shown
in the last row of \Fig{verticalP_forcing2}.
LWP samples are binned into 100 bins with a uniform width of $10\, \rm{g\, m}^{-2}$.
The minimum value of LWP from the RSP measurement
($2.4\, \rm{g\, m}^{-2}$ and $0.6\, \rm{g\, m}^{-2}$
for the February 28 and March 1 cases, respectively) is
taken as a lower cutoff for the simulated LWP.
}
\label{lwp_pdf_0228Nc}
\end{figure*}

Applying $\partial \bar{\theta}/\partial t$ and $\partial \bar{q}_v/ \partial t$
leads to a colder and less humid boundary layer
(compare simulation 0228C and 0228D)
but allows the boundary layer to grow higher.
This results in  a deeper cloud layer with a reduced
amount of the liquid water content $q_c$,
which is enhanced by a factor of two by applying
$u$ and $v$ relaxation (compare simulation 0228B to 0228D). We then examine the time
evolution of LWP.
LWP peaks around 16:00 UTC and then starts to
decrease as shown in \Fig{lwp_varyingF_LASSO_0228}, which could be due to the solar heating.
The short-wave (SW) cloud forcing at the top of atmosphere increases with increasing LWP
at a fixed time as indicated by the dashed lines in \Fig{lwp_varyingF_LASSO_0228}.
To validate the simulated LWP, we compare it with the
RSP retrievals during the ACTIVATE field campaign.
As shown in \Fig{lwp_pdf_0228Nc}(a), the WRF-LES
(shown as the black line, averaged over the measurement time) agrees
reasonably well with the RSP measurement.
This further illustrates that the WRF-LES is able
to capture the cloud formation and evolution in this case study.
We also tested the tendencies and relaxation forcing for the
March 1 case, which yields the same conclusion as for
the February 28 case.
\Fig{lwp_pdf_0228Nc}(b) shows that
the WRF-LES underestimates the frequency of lower LWP values
(less than $100\, {\rm g\, m}^{-2}$) but overestimates the
frequency between $200-400\, {\rm g\, m}^{-2}$ for the March 1 case.

\subsection{Sensitivities to large-scale divergence $\bar{D}$}

We have shown that applying $\partial \bar{\theta}/\partial t$,
$\partial \bar{q}_v/ \partial t$, $u$ relaxation to $u_{\rm ERA5}$,
and $v$ relaxation to $v_{\rm ERA5}$ is essential
to reproduce the time evolution of meteorological states for the two
CAO cases considered in this study. This configuration
is adopted to further test the sensitivities
of WRF-LES results to $\bar{D}$. We focus on the March 1
since it is more challenging to simulate due to the large surface heat fluxes.

\begin{figure*}[t!]\begin{center}
\includegraphics[width=\textwidth]{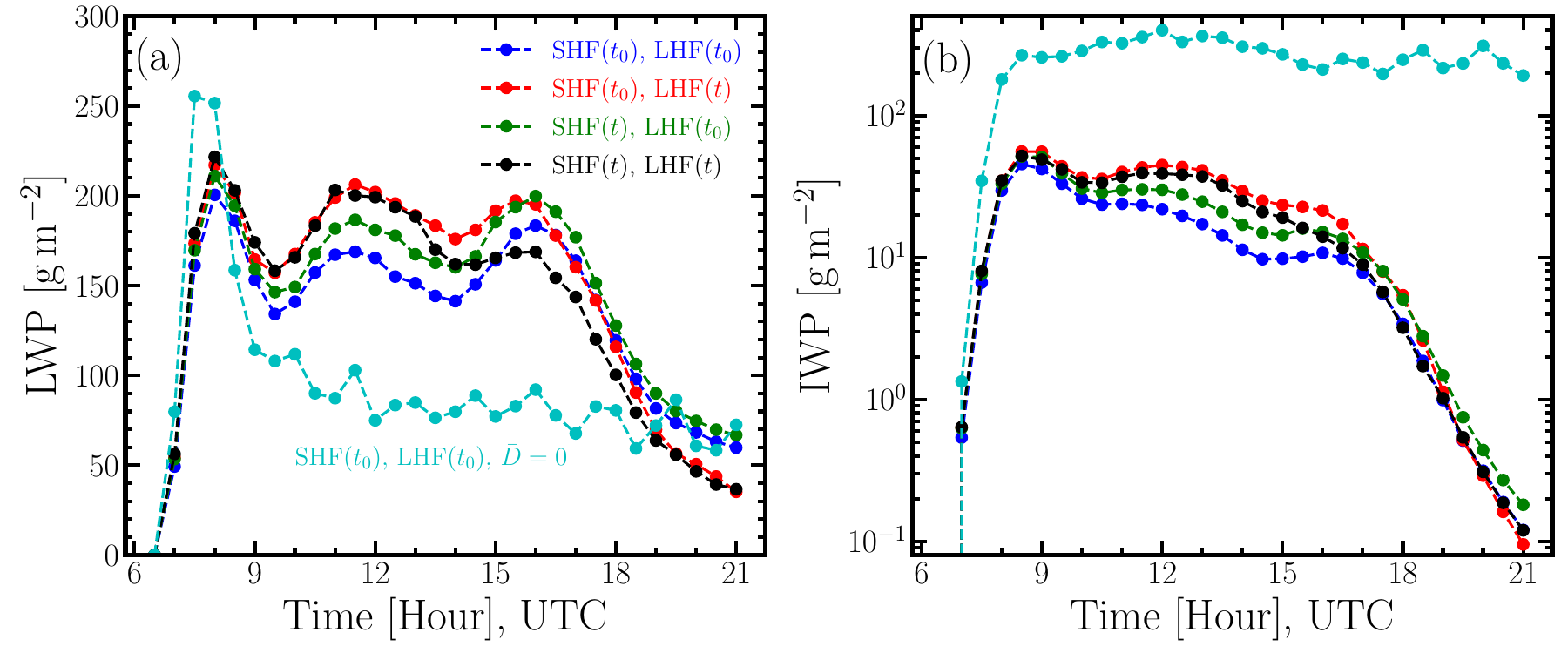}
\end{center}\caption{
Time series of domain-averaged (a): LWP (in-cloud liquid water and rain)
and (b): IWP (ice, graupel, and snow) from WRF-LES (blue: 0301A, red: 0301B, green: 0301C, cyan: 0301D, black: 0301E) with different heat fluxes and large-scale
divergence as indicated by the legends for the March 1 case.
Heat fluxes are from ERA5 reanalysis data, the values of which are
$\rm{SHF}(t_0)=231.76\, \rm{W\, m}^{-2}$ and $\rm{LHF}(t_0)=382.18 \,  \rm{W\,m}^{-2}$.
Here $t_0$ denotes the starting time of the WRF-LES, which is 06:00 UTC.  
}
\label{lwp_varyingF_LASSO_0301}
\end{figure*}

\begin{figure*}[t!]\begin{center}
\includegraphics[width=\textwidth]{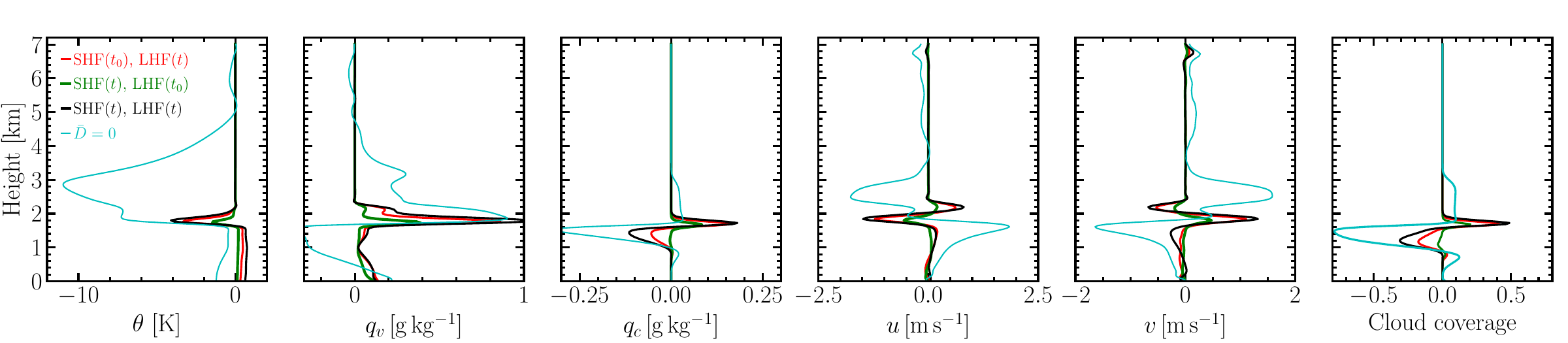}
\end{center}\caption{Deviation of
vertical profiles of simulation 0301B, 0301C,
0301E, and 0301D from the baseline simulation
0301A averaged over the measurement time (3 snapshots over 15:00 to 16:00 UTC).
}
\label{vp_HF}
\end{figure*}

\begin{figure*}[t!]\begin{center}
\includegraphics[width=\textwidth]{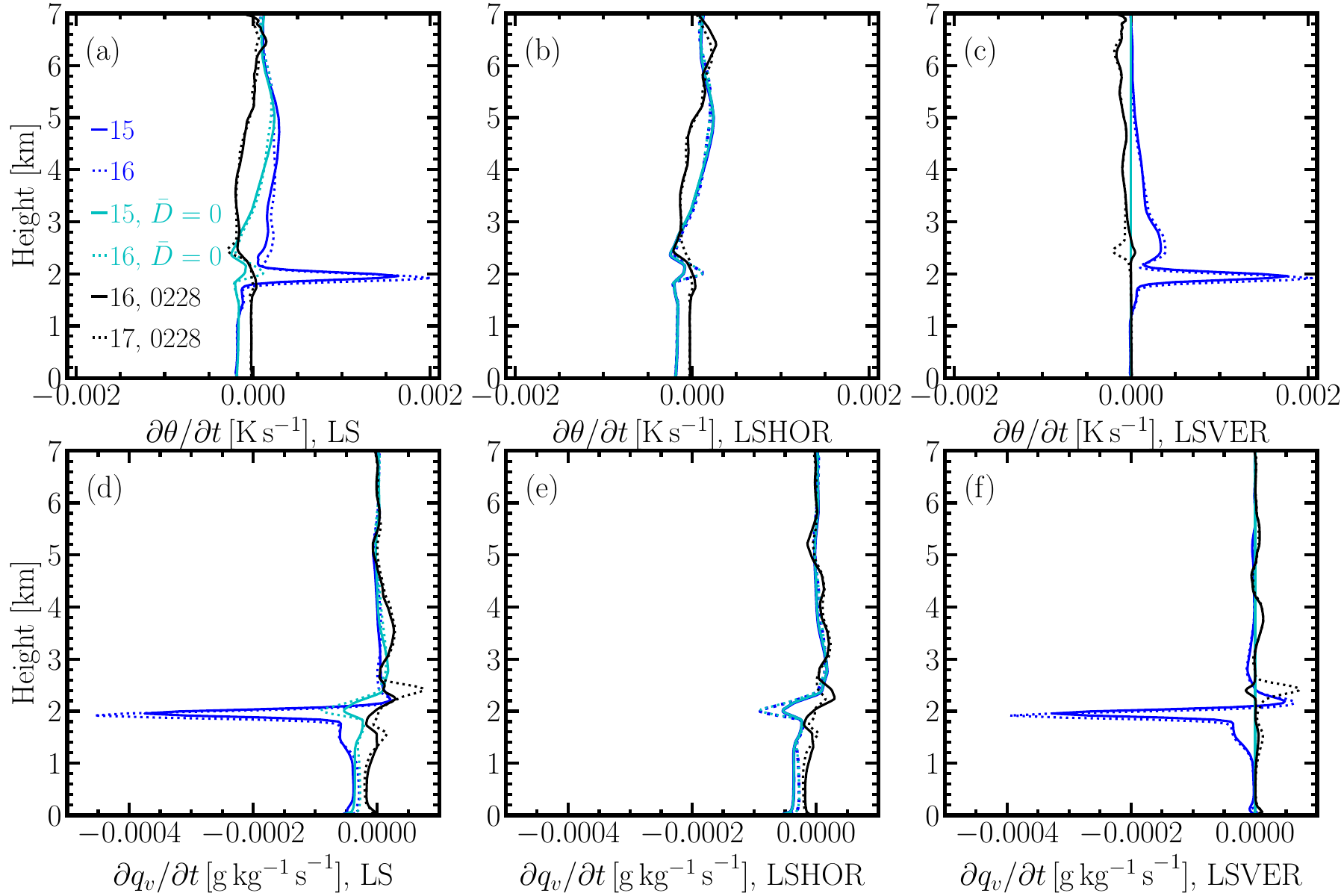}
\end{center}
\caption{Contributions of large-scale forcings to
$\partial \theta/\partial t$ (upper row) and $\partial q_v/\partial t$
(lower row) for simulation
0301A (blue curves) and 0301D (cyan curves)
at 15:00 UTC (solid lines) and 16:00 UTC (dashed lines).
Black lines represent the ones for simulation 0228A
at 16:00 UTC (solid line) and 17:00 UTC (dashed line). 
LS, LSHOR, and LSVER denote large-scale forcing due to
total, horizontal, and vertical advective tendencies, respectively.
}
\label{lsf_0301}
\end{figure*}

We perform two WRF-LES with or without the
large-scale vertical velocity as a forcing (third term
on the r.h.s of \Eqs{eq:dT/dt2}{eq:dqv/dt2}) for the March 1 case.
The forcing configuration for the baseline simulation 0301A
is the same as simulation 0228A.
To examine the impact of large-scale divergence separately,
we conduct a simulation (0301D in
\Tab{tab:param}) that excludes
the forcing term related to $\bar{D}$
(vertical component of the advective tendencies) but
keeps all other forcings the same as the baseline (0301A).
As shown in \Fig{lwp_varyingF_LASSO_0301}, the initial model
spin-up time (06:00-08:00 UTC) is characterized by
a sharp increase of LWP from 0 to about $200\, \rm{g\, m}^{-2}$
for the baseline simulation (blue).
Without the time varying large-scale
divergence (cyan), LWP and Ice Water Path (IWP) experience a larger
increase compared to the baseline (blue) during
the initial time steps because of the lack of subsidence
that tends to suppress the growth of BL.
LWP from simulation
0301D decreases and becomes smaller compared to the one
from simulation 0301A. This is because IWP from 0301D
are about 10 times more than the ones from 0301A.
In the absence of $\bar{D}$, the updraft cooling is more
profound. Thus, the ice formation is enhanced. 
The cyan curve in \Fig{vp_HF} shows the deviation of vertical profiles
of the simulation 0301D to 0301A averaged over the measurement time.
As expected from a positive $w$, a much deeper boundary layer is developed in
the simulation without subsidence.
The moist air is mixed to over 3 km.
This can be explained by the fact that the lack of divergence breaks the balance between the BL
growth driven by surface fluxes and suppression due to subsidence.
\Fig{lsf_0301} shows the contribution of horizontal and vertical
advective tendencies to $\partial \theta/\partial t$ (upper row)
and $\partial q_v/\partial t$ (lower row)
for simulation 0301A and 0301D, respectively.
When $\bar{D}=0$ (simulation 0301D), only horizontal advective
tendencies contribute to $\partial \theta/\partial t$ by comparing the solid cyan curves in \Fig{lsf_0301}(a)--(c).
When $\bar{D}\ne0$, it is evident that the vertical advective term ($\bar{D}$)
dominates the temperature and humidity changes due to large-scale tendencies, especially near the inversion layer.
The contribution from
horizontal terms are small when comparing simulations 0301A
and 0301D. The vertical profiles of
$\partial\theta/\partial t$ and $\partial q_v/\partial t$
only evolve slightly from 15:00 to 16:00 UTC because $\bar{D}$ does
not vary much for the March 1 case.
Recall that for the February 28 case, $\bar{D}$
evolves from convergence to divergence from 16:00 to 17:00 UTC (\Fig{era5_div_0228}(a)).
Consequently, the vertical profiles of $\partial\theta/\partial t$
and $\partial q_v/\partial t$
are nealry zero as shown in \Fig{lsf_0301}.
When comparing vertical profiles of $\partial\theta/\partial t$
and $\partial q_v/\partial t$
for the March 1 (blue lines) and February 28
(black lines) cases, we see that $\bar{D}$ has
a more profound impact for the March 1 case.

To conclude, $\bar{D}$ likely has a strong control
on the time evolution of the boundary
layer in WRF-LES
of the two CAO events we explored here, especially for the March 1 case.
Therefore, including a time-varying $\bar{D}$ profile to 
drive WRF-LES is necessary for simulating the fast-evolving CAO events over WNAO.

\subsection{Sensitivities to surface heat fluxes}
\label{sec:shf}

\begin{figure*}[t!]\begin{center}
\includegraphics[width=\textwidth]{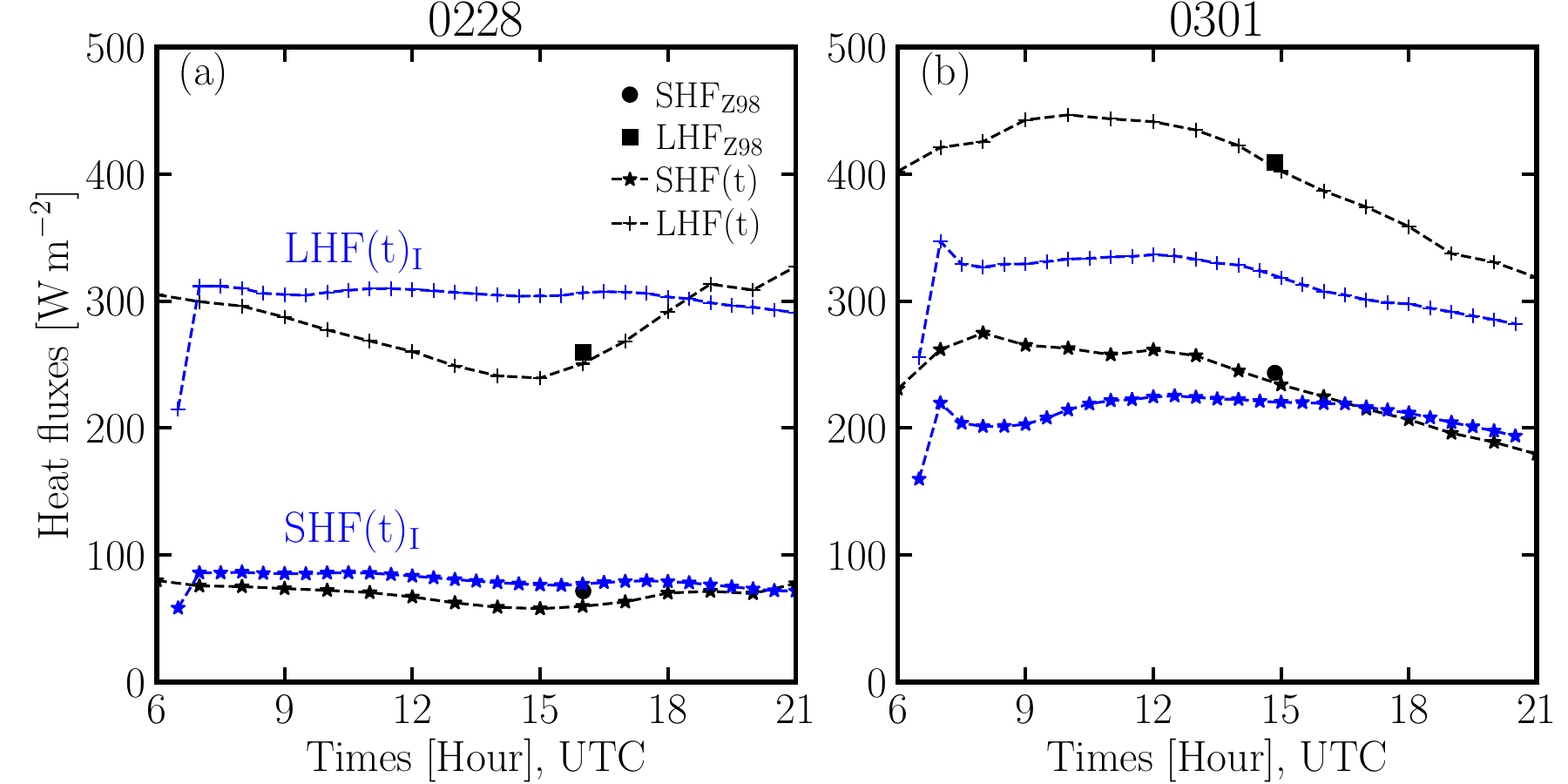}
\end{center}\caption{
Surface heat fluxes from ERA5 (black lines) reanalysis data
at the center of dropsonde circle
on (a) February 28 and (b) March 1.
Solid dots and squares represent heat fluxes at the
center of dropsonde circles calculated based on the Z98 algorithm.
Blue stars and pluses represent surface sensible $\rm{SHF(t)_I}$
and latent heat $\rm{LHF(t)_I}$ fluxes output from WRF-LES (0228F and 0301F), respectively.
}
\label{hf_era5_input}
\end{figure*}

To test the sensitivities to surface fluxes, we perform another three WRF-LES
simulations using the same forcing configuration as the baseline simulation
0301A but with temporally varying and spatially uniform surface heat fluxes for
the March 1 case.  The time series of such surface heat fluxes obtained from
ERA5 reanalysis data at the center of dropsonde circle is shown as the black lines in
\Fig{hf_era5_input}. As shown in \Fig{lwp_varyingF_LASSO_0301}, when the
WRF-LES is forced by ${\rm SHF} (t_0)$ and ${\rm LHF} (t)$ (red
curve), LWP evolves in the same pattern as the baseline but with larger values
between 10:00-16:00 UTC. This is because ${\rm LHF} (t)$ is larger than ${\rm
LHF} (t_0)$ until 15:00 UTC, as shown in \Fig{hf_era5_input}. Overall,
simulations driven by ${\rm LHF} (t)$ result in more LWP compared with the one
by ${\rm LHF} (t_0)$. Simulations forced by ${\rm SHF} (t)$ and ${\rm LHF}
(t_0)$ (green curve) exhibit the same trend as the one by ${\rm SHF} (t_0)$ and
${\rm LHF} (t)$.
When the time-varying ${\rm SHF} (t)$ and ${\rm LHF} (t)$ are both applied to
the WRF-LES (black curve), the initial increase in ${\rm SHF} (t) \, \& \, {\rm LHF} (t)$,
as compared to ${\rm SHF} (t_0)\, \& \, {\rm LHF} (t)$ (red), does not have an impact on the LWP.
Since the forcing ${\rm SHF} (t)$ and ${\rm LHF} (t)$ only vary slightly, the
mean LWP values do not show a significant difference when comparing the four
WRF-LES.  We also compare the IWP as shown in
\Fig{lwp_varyingF_LASSO_0301}(b).  The evolution of these quantities follow the
same trend as LWP.  \Fig{vp_HF} shows the corresponding deviations of vertical
profiles of simulation 0301B, 0301C, 0301E, and 0301D from the baseline
simulation 0301A.  These profiles are averaged over the measurement time (3
snapshots over 15:00 to 16:00 UTC).  Differences at the inversion layer (about 2
km) are the most pronounced.  The green curves (${\rm SHF}(t)$, ${\rm
LHF}(t_0)$) deviate the least from the blue curves (baseline simulation) while
the red (${\rm SHF}(t_0)$, ${\rm LHF}(t)$) and black (${\rm SHF}(t)$, ${\rm
LHF}(t)$) curves diverge the most within the boundary layer. The red and black
curves are almost identical except for the slight difference in $q_c$.

We also perform WRF-LES with interactive surface heat fluxes
estimated from a prescribed constant SST from ERA5 and model simulated
atmospheric states for both cases. A constant ERA5-SST is used
here because ERA5-SST does not vary at the location of dropsonde center from
06:00 UTC to 21:00 UTC. \Fig{hf_era5_input}(a) shows that surface heat fluxes
($\rm{SHF_{\rm I}}$ and $\rm{LHF_{\rm I}}$) calculated within the WRF-LES
surface scheme \citep{beljaars1995parametrization, chen2001coupling} are close
to the ones from ERA5, leading to a similar LWP (\Fig{lwp_0301_dx}) and
meteorological states (\Fig{vp_HFt_dx}) for the February 28 case.  The frequency of
LWP from simulation 0228E (prescribed $\rm{HF}(t)$ from ERA5) and 0228F
($\rm{HF}(t)$ calculated interactively within WRF-LES) agree excellently with the
RSP measurement as shown in \Fig{lwp_pdf_0228Nc}(a). For the March 1 case, the
surface latent heat flux from WRF-LES is substantially weaker than the one from ERA5
(\Fig{hf_era5_input}(b)), resulting in a drier BL (\Fig{vp_HFt_dx})
and smaller LWP (\Fig{lwp_0301_dx}). The frequency of LWP
from simulation 0301F agrees better with RSP than that from 0301E.
Nevertheless, we use prescribed surface heat fluxes from ERA5 in our LES
because there is no direct measurement of surface heat fluxes from the ACTIVATE
campaign. We aim to unravel aerosol-meteorology-cloud interactions and to
improve its parameterizations in the Earth System Models by using
LES constrained by ACTIVATE measurements and reanalysis data. 

Simulations with finer horizontal resolution
($dx=100\, \rm{m}$) yields similar LWP (\Fig{lwp_0301_dx})
and almost identical vertical profiles (\Fig{vp_HFt_dx}) as the ones
with $dx=300\, \rm{m}$ for both cases.
The energy power spectra at 1 km height during
the measurement time is shown in \Fig{ps_0228}. As expected,
a larger inertial range is observed for simulation with
$dx=100\, \rm{m}$.
Nevertheless, this does not affect the simulated LWP and
BL thermodynamics, which
justifies our use of $dx=300\, \rm{m}$.

\Fig{horizC_0228} shows the instantaneous field of $\theta$,
$q_v$, $q_c$, and TKE at UTC 16:00 and 2.5 km (near cloud top)
for the February 28 case (simulation 0228G with dx=100 m). The thermodynamics
fields exhibits same spatial patterns as TKE. Same for 
the March 1 case (simulation 0301G with dx=100 m) at 1.5 km
as shown in \Fig{horizC_0301}.

\begin{table}[t!]
\caption{Surface heat fluxes during the dropsonde
measurement time. ``Flux'' represents moisture and
heat fluxes calculated from LES $\overline{w^\prime q_v^\prime}$
and $\overline{w^\prime \theta^\prime}$ (0228E and 0301E) at the bottom model layer, respectively.
} 
\centering
\setlength{\tabcolsep}{1pt}
\begin{tabular}{|c|c|c|c|c|c|c|c|c|}
\hline
\multirow{2}{*}{Case} & \multicolumn{4}{c|}{SHF [$\rm{W\, m}^{-2}$]} & \multicolumn{4}{c|}{LHF [$\rm{W\, m}^{-2}$]} \\ 
\cline{2-9}
& ERA5 & Z98 & I & Flux & ERA5 & Z98 & I & Flux \\ 
\cline{2-9}
\hline
0228 & 59.7 & 71.7 & 77.2 & 64.4   & 250.7 & 259.5 & 306.6& 273.1 \\ 
0301 & 234.1 & 243.4 &220.3 & 232.2  & 402.3 & 409.2& 318.4 & 406.7 \\ 
\hline
\end{tabular}
\label{tab:HF}
\end{table}

\section{Turbulent fluxes: validating LES against aircraft in-situ measurements}

\begin{figure*}[t!]\begin{center}
\includegraphics[width=0.95\textwidth]{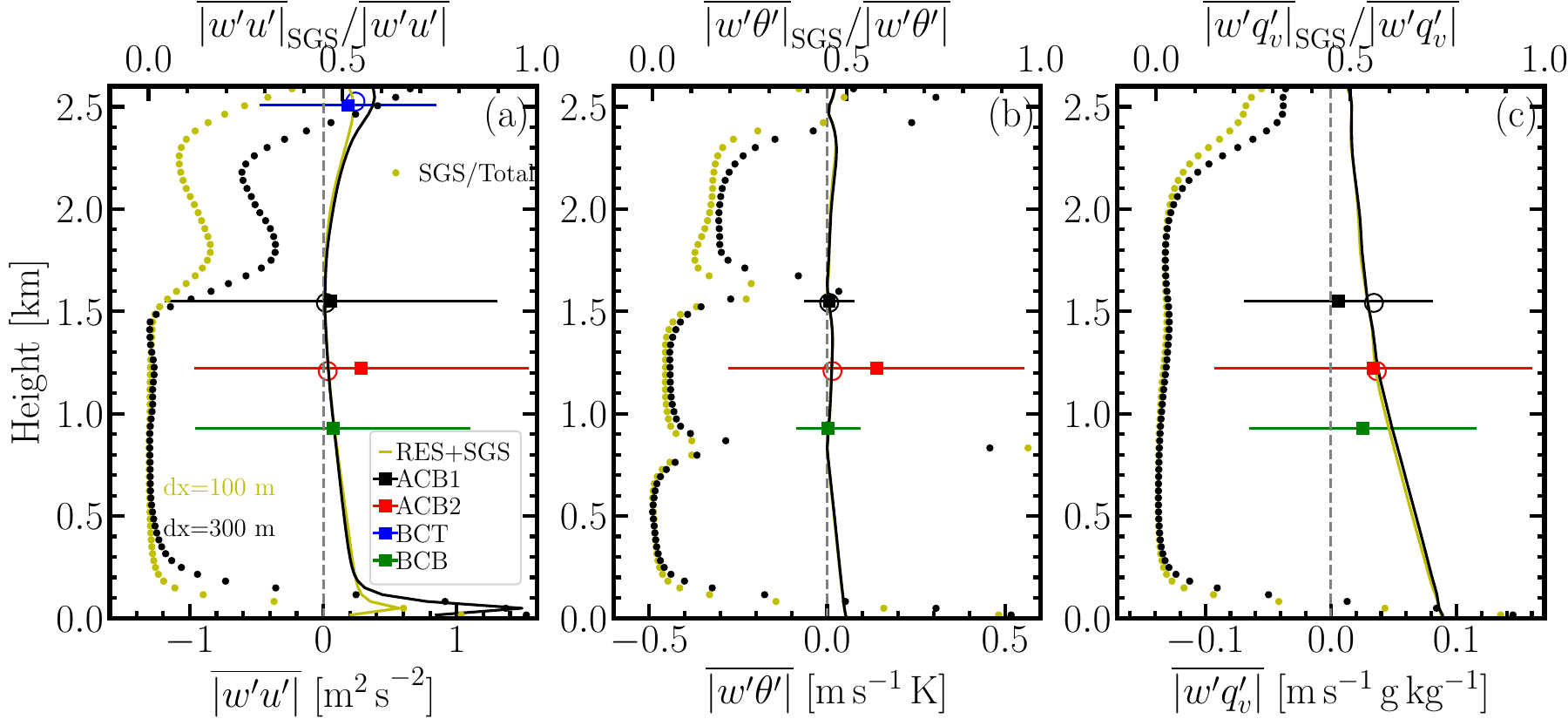}
\end{center}\caption{(a) momentum, (b) heat, and (c) moisture
fluxes within BL during dropsonde measurement time for the February 28 case.
Open circles represent total fluxes, i.e., resolved (RES) plus sub-grid scale
(SGS) fluxes, from simulation 0228G. Solid squares represent the ones from different
flight legs (``ACB'', ``BCT'' and ``BCB'' denotes above cloud-base,
below cloud-top,and below cloud-base, respectively). The
error bars represent one standard deviation of fluxes.
Solid and dotted lines represent the total fluxes and the ratio between SGS and the total fluxes from LES, respectively.
The closest snapshots in both time and height from simulation 0228G are used to
compare to the measurements.  $\overline{w^\prime \theta^\prime}$ and
$\overline{w^\prime q_v^\prime}$ from Falcon measurements are calculated by
matching the sampling time of $q_v$ and $\theta$ to the averaged $w$,
respectively.  The time series and vertical profiles of $w^\prime$, $q_v^\prime$
and $\theta^\prime$ from the measurements are shown in \Fig{fluc_0228}.  Flight
time and height of the four flight legs are listed in \Tab{tab:legs}.
$\overline{w^\prime q_v^\prime}$ and $\overline{w^\prime \theta^\prime}$ from
the BCT leg is not shown due to limited sampling. Vertical profiles from LES are averaged during the dropsonde measurement time. Moisture and heat fluxes
calculated from the bottom layer of LES $\overline{w^\prime q_v^\prime}$ and $\overline{w^\prime
\theta^\prime}$ (0228E and 0301E), respectively, are listed in
\Tab{tab:HF}.
}
\label{tur_flux_0228}
\end{figure*}

\begin{figure*}[t!]\begin{center}
\includegraphics[width=0.95\textwidth]{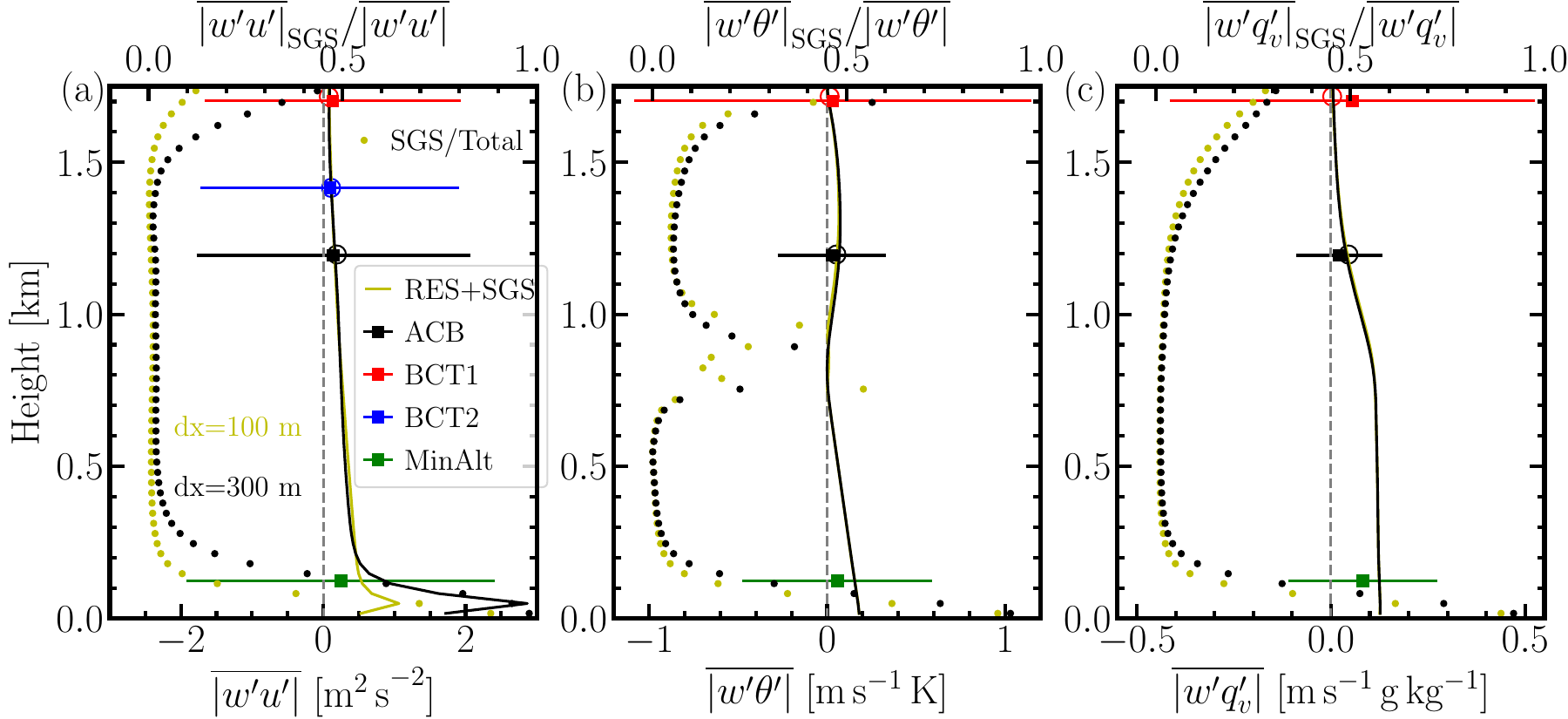}
\end{center}\caption{Same as \Fig{tur_flux_0228} but for the March 1 case (simulation 0301G and 0301E). $\overline{w^\prime q_v^\prime}$ is not shown for the flight leg
BCT2 because $q_v$ was not measured. ``MinAlt'' denotes minimum altitude ($\sim 150\,  {\rm m}$).
The corresponding time series and
vertical profiles of the measured $w^\prime$, $\theta^\prime$, and $q_v^\prime$ are shown in \Fig{fluc_0301}.
}
\label{tur_flux_0301}
\end{figure*}

To validate LES against in-situ measurements during the ACTIVATE
campaign, we compare the measured turbulent fluxes
from the Falcon aircraft flying in the BL to the ones from LES.
We select two above cloud-base (ACB), one below cloud-top (BCT),
and one below cloud-base (BCB) flight legs
during the dropsonde measurement time (16:00-17:00 UTC) on February 28. The time
series and vertical profiles of $w^\prime$, $q_v^\prime$ and $\theta^\prime$
from the four flight legs (ACB1, ACB2, BCT, BCB) are shown in \Fig{fluc_0228}.  The
sampling time and altitude variation of each flight leg is about 10 minutes or
less and about 17 m (\Tab{tab:legs}), respectively.  Since the vertical layer
thickness of LES is about 33 m within BL, we compare turbulent fluxes at the LES
layer center that is closest to the height of each flight leg. The closest LES
snapshot (every 30 minutes) to the flight sampling time is used for comparison.
To calculate turbulent fluxes from measurements, the sampling time of $T$ and
$q_v$ are mapped to that of averaged wind speed, which has 20 times higher
sampling frequency.  \Fig{tur_flux_0228} shows the comparison of turbulent
fluxes between the Falcon measurements and LES. The sampling frequency of $T$
and $q_v$ is 1 Hz, which is equivalent to a spatial distance of $100\, \rm{m}$
given that the flight speed is about $100\, {\rm m\, s}^{-1}$.  Such a spatial
distance is comparable to the mesh size of LES.  The LES is able to reproduce
$\overline{w^\prime u^\prime}$ and $\overline{w^\prime \theta^\prime}$ measured
during flight legs ACB1, BCT, and BCB. It captures the measured $\overline{w^\prime
q_v^\prime}$ at flight leg ACB2.  The measured $\overline{w^\prime u^\prime}$,
$\overline{w^\prime \theta^\prime}$ and $\overline{w^\prime q_v^\prime}$ agree
well with the ones from LES for the March 1 case, as shown in
\Fig{tur_flux_0301}.

Comparison of turbulent fluxes between simulation 0228E ($dx=300\,
\rm{m}$) and 0228G ($dx=100\, \rm{m}$) is also shown in \Fig{tur_flux_0228}.
The parameterized subgrid-scale (SGS) turbulent fluxes are strong
within the surface layer for both simulations,
above which all the eddies are resolved by LES as suggested by
the ratio between SGS and the total fluxes (i.e. yellow and black dots). $\overline{w^\prime u^\prime}$
within the surface layer is strongly mesh-size dependent, which is not the case
for $\overline{w^\prime \theta^\prime}$ and $\overline{w^\prime q_v^\prime}$.
Therefore, the simulation with dx=300 m yields the same LWP as the one with
dx=100 m as discussed in section~\ref{sec:forcing}.\ref{sec:shf}. The same conclusion can be drawn for
the March 1 case as shown in \Fig{tur_flux_0301}.

\section{Discussion and conclusions}

We have reported two contrasting cold air outbreak (CAO) cases
observed during the ACTIVATE field campaign and the
corresponding WRF-LES modeling of them. The February 28 case is
characterized by weaker turbulent surface heat fluxes
(${\rm SHF=79.91\, W\, m}^{-2}$ and ${\rm LHF=305.02\, W\, m}^{-2}$)
than those of the March 1 case
(${\rm SHF=231.76\, W\, m}^{-2}$ and ${\rm LHF=382.18\, W\, m}^{-2}$).
The divergence is on the order of $10^{-5}\, {\rm s}^{-1}$ for both
cases, which is about 10 times larger than common marine cases
(e.g., $\bar{D} \approx 10^{-6}\, {\rm s}^{-1}$ in the DYCOMS-II case simulated
by \citet{wang2009modeling}) and about two times larger than the CAO case in
\citet{Roode2019}.  A deeper, warmer, and more humid boundary layer was observed
for the February 28 event than the one on March 1.

To examine and validate different prescribed forcing options to drive WRF-LES,
we first evaluate divergence obtained from the ERA5 reanalysis data against the
one derived from dropsonde measurements for the two CAO cases. The divergence
profile and the corresponding vertical velocity obtained from ERA5 reanalysis
data at the center of dropsonde circle are able to capture the structure of the
ones estimated from dropsonde measurements for the March 1 case.  This gives us
the confidence to adopt the time-varying divergence profiles from ERA5 to drive
our WRF-LES.

Since the surface turbulent heat fluxes are partly determined by SST, we compare
SST from ERA5 to the one from satellite retrievals. They agree very well for
both February 28 and March 1 cases.  Therefore, SST from ERA5 together with the
10 m temperature, water vapor mixing ratio, and wind speed from dropsonde
measurements are used to calculate heat fluxes for the March 1 case and those
from ERA5 for the February 28 case using the bulk aerodynamic algorithms from
\citet{zeng1998intercomparison}.  The estimated sensible and latent heat fluxes
agree well with the ones directly obtained from ERA5 reanalysis data for the
March 1 case case. They are underestimated by about $30\%$ compared to the ERA5
heat fluxes for the February 28 case.

By applying the surface heat fluxes, large-scale temperature
and moisture advective tendencies, and wind relaxation adjustments
from ERA5 to the WRF-LES,
the simulated meteorological states for both CAO cases match the
ERA5 reanalysis data and the ACTIVATE field campaign measurements.
We also conduct WRF-LES sensitivity simulations on the surface fluxes
and divergence and find that the divergence is important in
suppressing the evolution of the boundary layer and achieves the observed states
of the boundary layer for this case,
while surface heat fluxes are more influential for the simulated LWP.
The frequency of LWP produced from our WRF-LES agrees
reasonably well with the measured ones from the ACTIVATE
campaign  for  both  the  February  28  case.
Since the large-scale tendencies profiles vary with
time for the two CAO cases, it is important to apply
time-varying tendencies to the WRF-LES instead  of constant ones.

In summary, with initial conditions, large-scale forcings, and turbulent surface
heat fluxes obtained from ERA5 and validated by ACTIVATE airborne measurements,
WRF-LES is able to reproduce the observed boundary-layer meteorological states
and LWP for two contrasting CAO cases.  This manifests the meteorological impact
on marine boundary layer and clouds associated with CAO over WNAO.  This study
(Part 1) paves the path to further investigation of aerosol effects on cloud
microphysics during the CAO events to be reported in the forthcoming companion
paper (Part 2).

\acknowledgments

This work was supported through the ACTIVATE Earth Venture Suborbital-3 (EVS-3)
investigation, which is funded by NASA’s Earth Science Division and managed
through the Earth System Science Pathfinder Program Office.  The Pacific
Northwest National Laboratory (PNNL) is operated for the U.S. Department of
Energy by Battelle Memorial Institute under contract DE-AC05-76RLO1830.  We wish
to thank the pilots and aircraft maintenance personnel of NASA Langley Research
Services Directorate for their work in conducting the ACTIVATE flights.  We
thank Andrew S Ackerman for discussions.  The source code used for the
simulations of this study, the Weather Research and Forecasting (WRF) model, is
freely available on \url{https://github.com/wrf-model/WRF}.  The simulations
were performed using resources available through Research Computing at PNNL.

\datastatement

ACTIVATE data are publicly available at: \url{https://www-air.larc.nasa.gov/cgi- bin/ArcView/activate.2019}

MW-IR SST are produced by Remote Sensing Systems and sponsored by NASA. Data are
available at \url{www.remss.com}

\appendix

\section{Dropsonde measurements}
\label{app:D}

This appendix is to review the method being adopted to
calculate divergence $D$ from the dropsonde measurements and to
test statistical convergence of $D$ to the number of
dropsondes used in the calculation. 

The integral form of \Eq{eq:con} is
\EQ
\int {\bm \nabla}\cdot {\bm u} {\rm d}S=0.
\EN
Thus, 
\EQ
\int \left(\frac{\partial u}{\partial x}+\frac{\partial v}{\partial y}\right) {\rm d}S=
-\int \frac{\partial w}{\partial z} {\rm d}S=-A\frac{\partial w}{\partial z}.
\label{eq:integral}
\EN
According to the Stokes theorem, \Eq{eq:integral} can be written as
\EQ
\frac{\partial w}{\partial z}=-D=-\frac{1}{A}\oint_l v_n {\rm d}l.
\label{eq:D2}
\EN

Based on \Eq{eq:D2}, \citet{lenschow2007divergence} noted
that the most efficient flight track is a circle-like shape since the
circle has the largest enclosed area of any closed curve
and the turning rate of flight is slow. Such a method to
calculate $D$ is called the ``linear integral method''.
This method requires a closed and circular flight track and
a linearly evolving wind speed, none of which can be
satisfied. Therefore, \citet{lenschow2007divergence} developed
the ``regression method'', which can alleviate the requirements
of the linear integral method. The first order Taylor expansion
of horizontal wind velocity ${\bm v}={\bm v}(u,v)$ at the dropsondes center is
\EQ
{\bm v} = {\bm v_0}+\frac{\partial {\bm v}}{\partial x}\Delta x
+ \frac{\partial {\bm v}}{\partial y}\Delta y+
\frac{\partial {\bm v}}{\partial t}\Delta t,
\EN
where $\Delta x$ and  $\Delta y$ are the eastward and
northwest displacements from the center of dropsondes.
Assuming a stationary state, the term $\frac{\partial {\bm v}}{\partial t}\Delta
t$ can be neglected.  This assumption suggests that all the dropsondes should be
released simultaneously, which is not feasible experimentally.  The sampling lag
in space and time between different dropsondes may cause error of calculating
$\bar{D}$. However, \citet{bony2019measuring} demonstrated that the stationarity
assumption is not bad.

\begin{figure*}[t!]\begin{center}
\includegraphics[width=\textwidth]{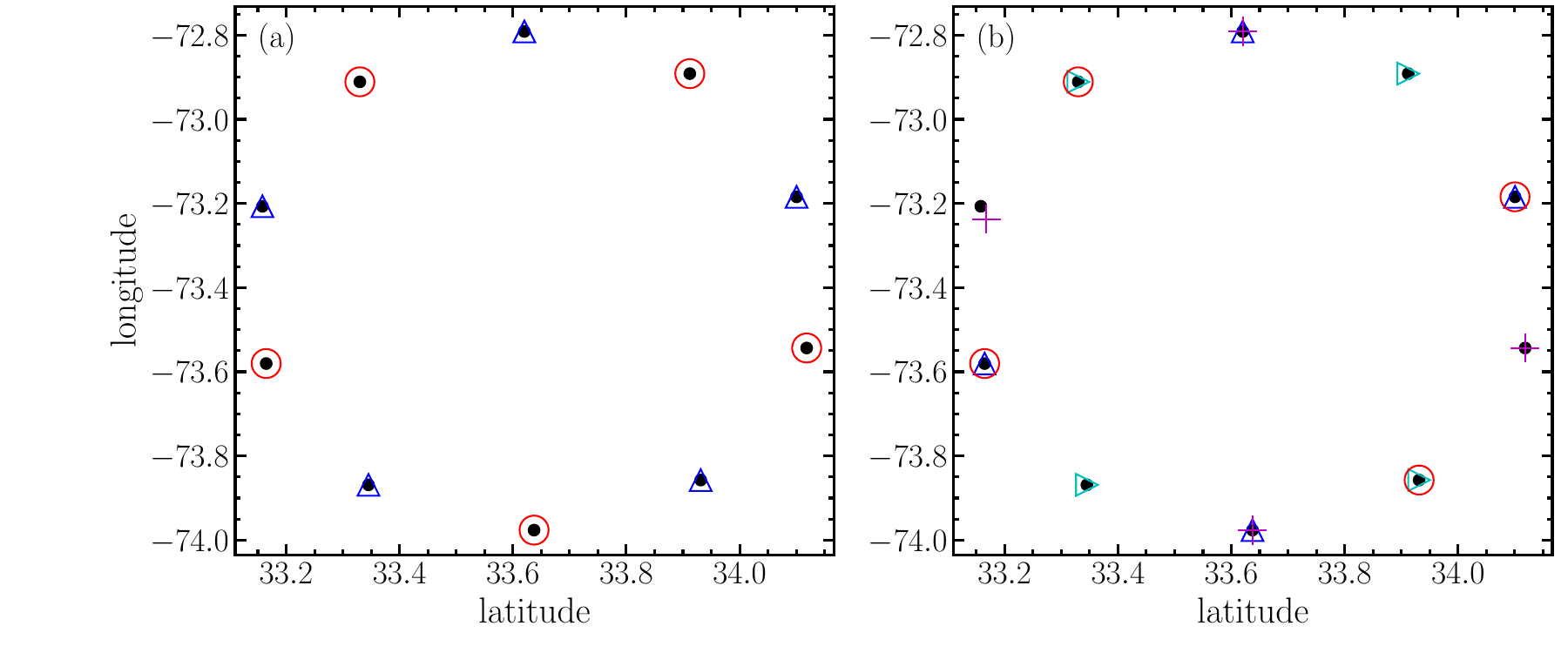}
\end{center}
\caption{Latitude and longitude coordinates of dropsondes released on February 28 case.
Black dots represent the 10 dropsondes.  (a): Red circles and blue triangles
represent two subsets of 5 dropsondes, respectively.
(b): Red circles, blue triangles, cyan triangles, and red crosses
represent four subsets of 4 dropsondes, respectively.
}
\label{drop_select0228}
\end{figure*}

\begin{figure*}[t!]\begin{center}
\includegraphics[width=\textwidth]{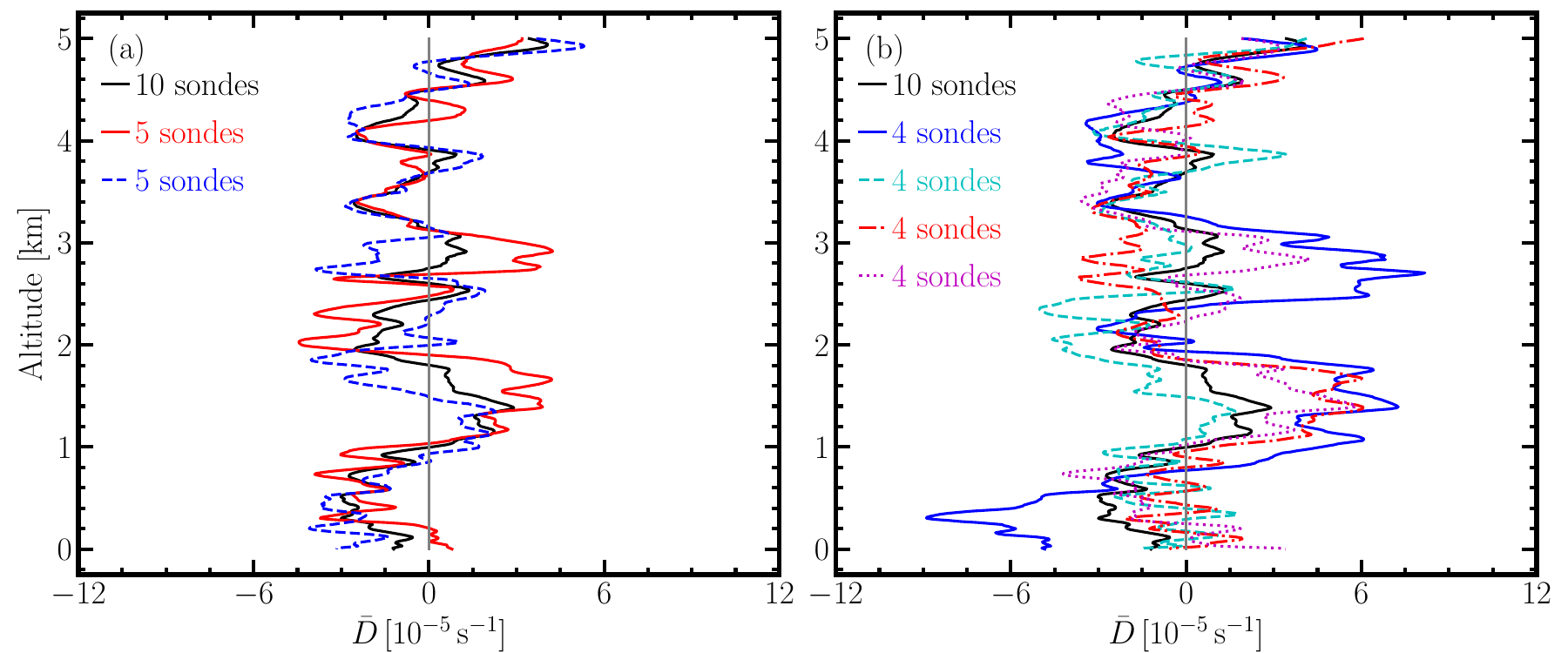}
\end{center}
\caption{Corresponding $\bar{D}$ profiles derived from dropsonde measurements shown in
\Fig{drop_select0228}.}
\label{drop_div0228}
\end{figure*}

We then test the sensitivity of $\bar{D}$ to
the number of dropsondes used in the calculation.
\Fig{drop_select0228} shows the circular distribution of
10 dropsondes (black solid dots) for the February 28 case.
Here, only 10 dropsondes are used to calculate the divergence
because two (out of the 11 dropsondes)
were released at the same location.
We select two subsets as shown in \Fig{drop_select0228}(a).
The corresponding $\bar{D}$ is shown in \Fig{drop_div0228}(a).
$\bar{D}$ derived from
the 5-dropsonde circle agrees with that from the 10-dropsonde circle. 
We also test four subsets of 4-dropsonde circles,
as shown in \Fig{drop_select0228}(b), and the corresponding $\bar{D}$ is
shown in \Fig{drop_div0228}(b). Same as the 5-dropsonde circles, 
4-dropsonde subset agree with the 10-dropsonde circle
even though the individual subsets exhibit differences.

\begin{figure*}[t!]\begin{center}
\includegraphics[width=\textwidth]{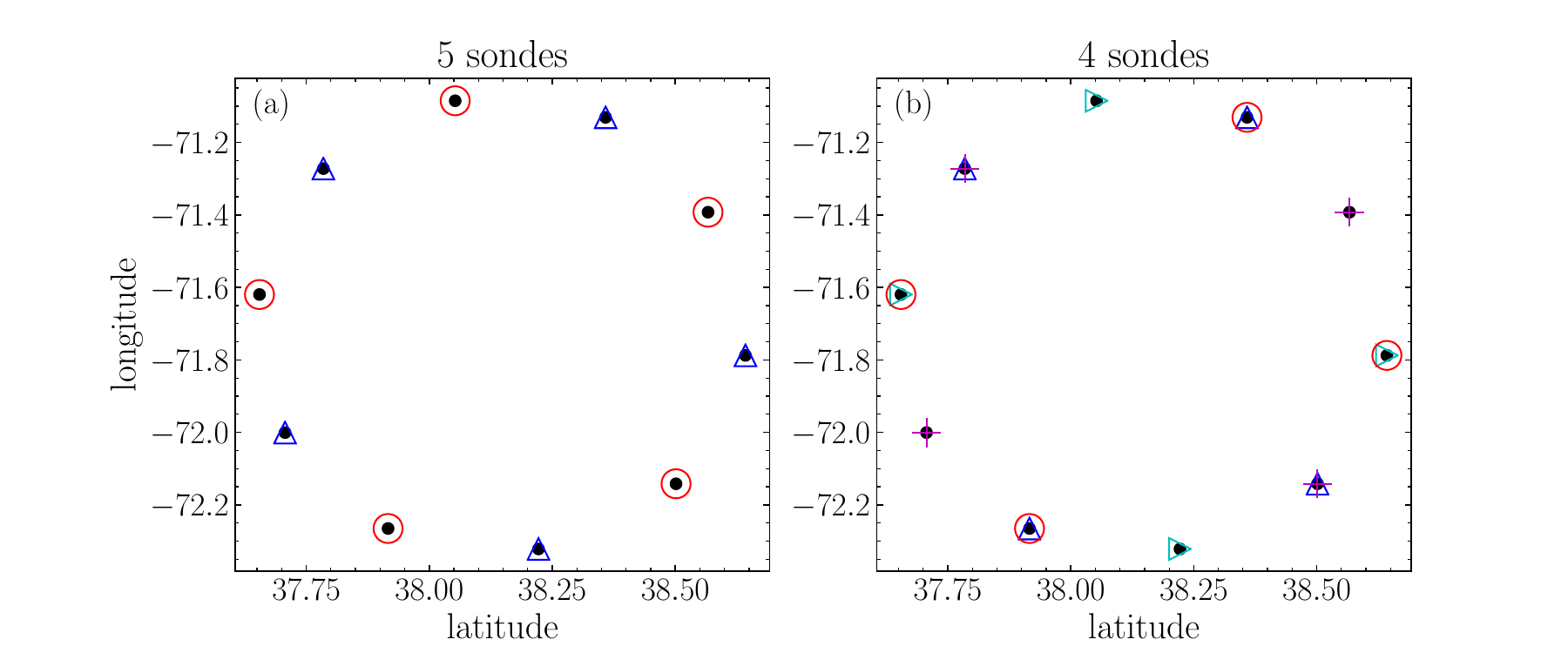}
\end{center}
\caption{Same as \Fig{drop_select0228} but for the March 1 case.}
\label{drop_select}
\end{figure*}

\begin{figure*}[t!]\begin{center}
\includegraphics[width=\textwidth]{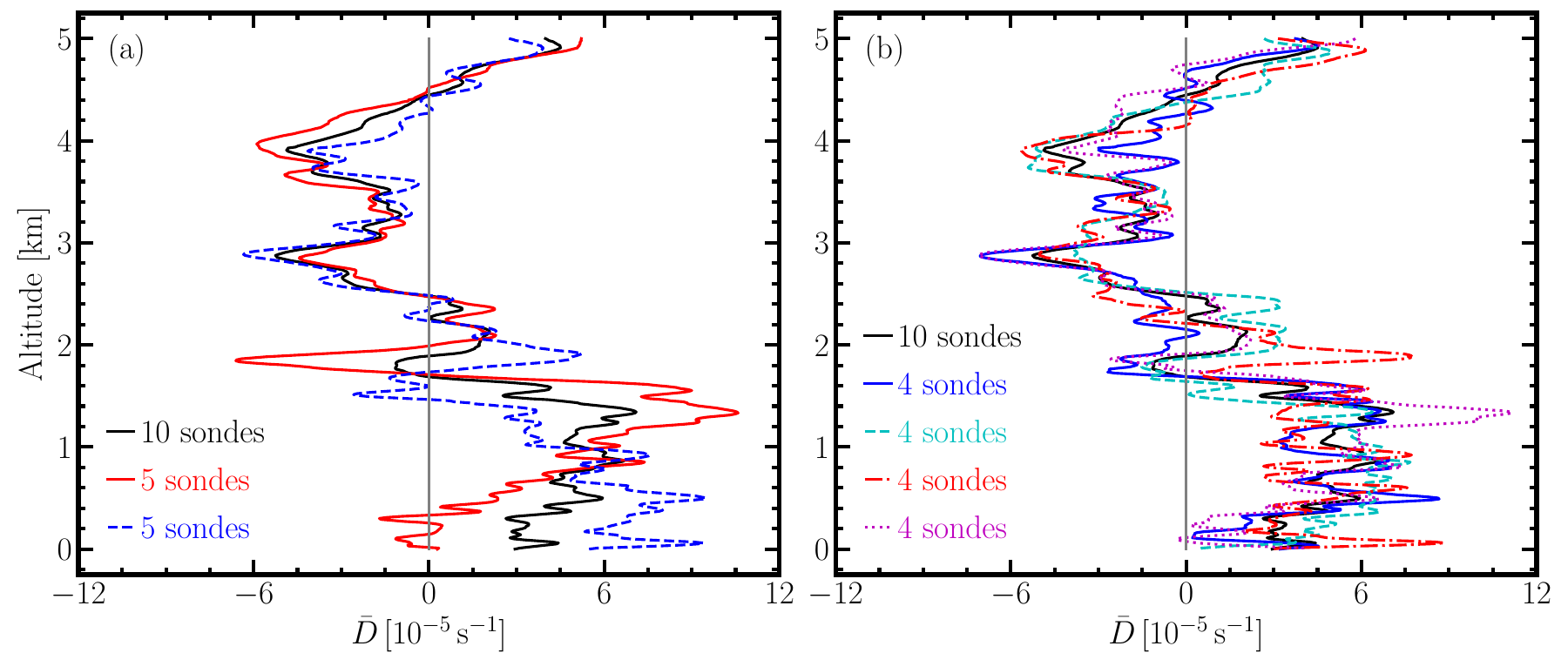}
\end{center}
\caption{Corresponding $\bar{D}$ profiles derived from groups of dropsondes shown in
\Fig{drop_select}.}
\label{drop_div}
\end{figure*}

We also apply the same analysis to dropsonde measurements being carried out
on the March 1 case as shown in \Fig{drop_select} and \Fig{drop_div}.
$\bar{D}$ derived from
the 5-dropsonde circle agrees with that from the 10-dropsonde circle
above the inversion layer but differs 
within the boundary layer. 
The difference is even larger between the two different sets of
5-dropsonde circles (red and blue curves in \Fig{drop_div}(a))
in the boundary layer. Interestingly, $\bar{D}$ from the 
4-dropsonde subset (\Fig{drop_div}(b))
is closer to the one derived from the 10-dropsonde.
Therefore, reducing the total number of dropsondes in
a circle results in statistical uncertainties.
\citet{bony2019measuring} suggested that at least 12 dropsondes are needed to
estimate $\bar{D}$.  More dropsondes can indeed improve the accuracy of the
estimation in the tropics as shown in Figure 5 of \citet{bony2019measuring}.
However, the additional two dropsondes are not expected to make a big
difference.

\section{Surface heat fluxes: ERA5 versus MERRA-2}
\label{app:shf}

\begin{figure*}[t!]\begin{center}
\includegraphics[width=\textwidth]{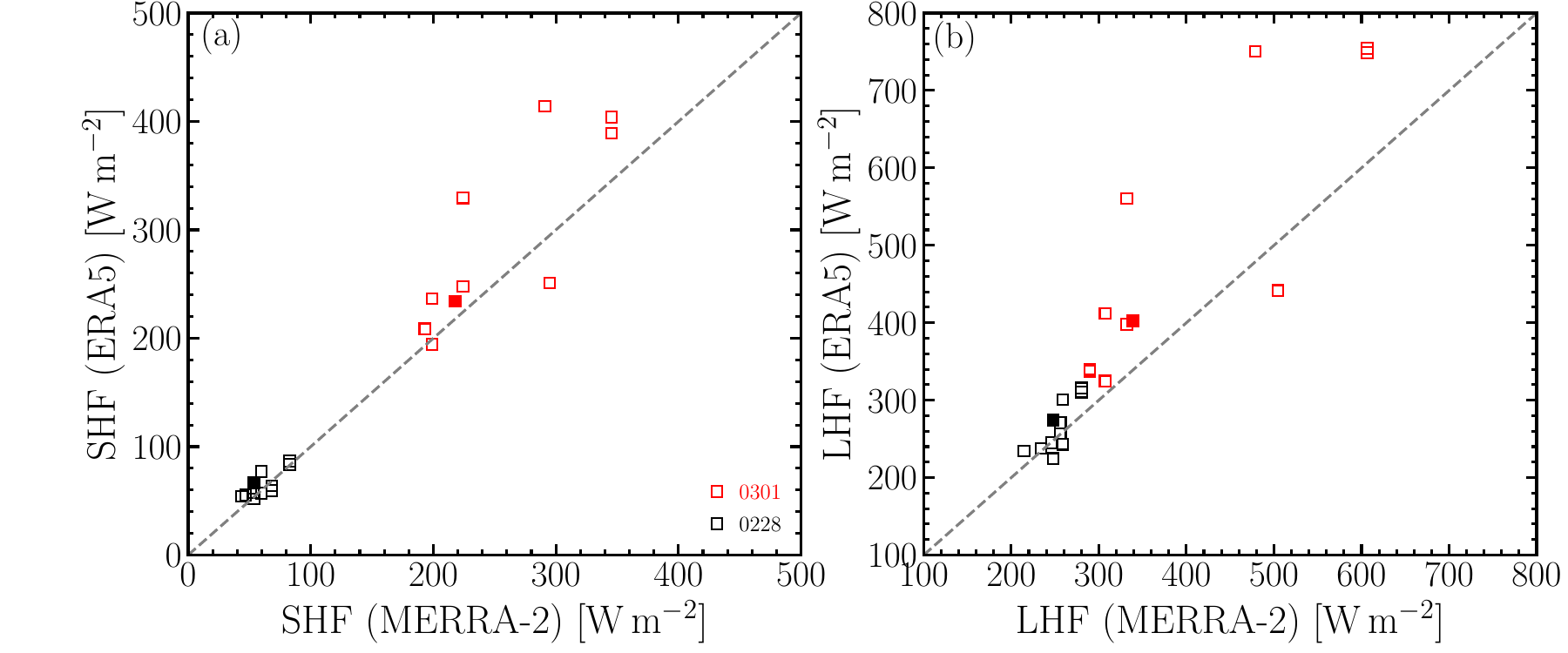}
\end{center}
\caption{Comparison of heat fluxes between ERA5
and MERRA-2 reanalysis data at 15:00 UTC for the February 28
case (black squares) anad 16:00 UTC for the March 1 case (red squares).
Heat fluxes from the MERRA-2 reanalysis data are averaged
between 16:30 and 17:30 for the February
28 case and between 14:30 and 15:30 for the March 1 case.
}
\label{hf_era5_merra2}
\end{figure*}

\Fig{hf_era5_merra2} shows the comparison of heat fluxes
between ERA5 and MERRA-2 reanalysis data.
MERRA-2 underestimates the heat fluxes compared to ERA5. 

\section{Horizontal resolution and interactive surface heat fluxes}

\begin{figure*}[t!]\begin{center}
\includegraphics[width=0.48\textwidth]{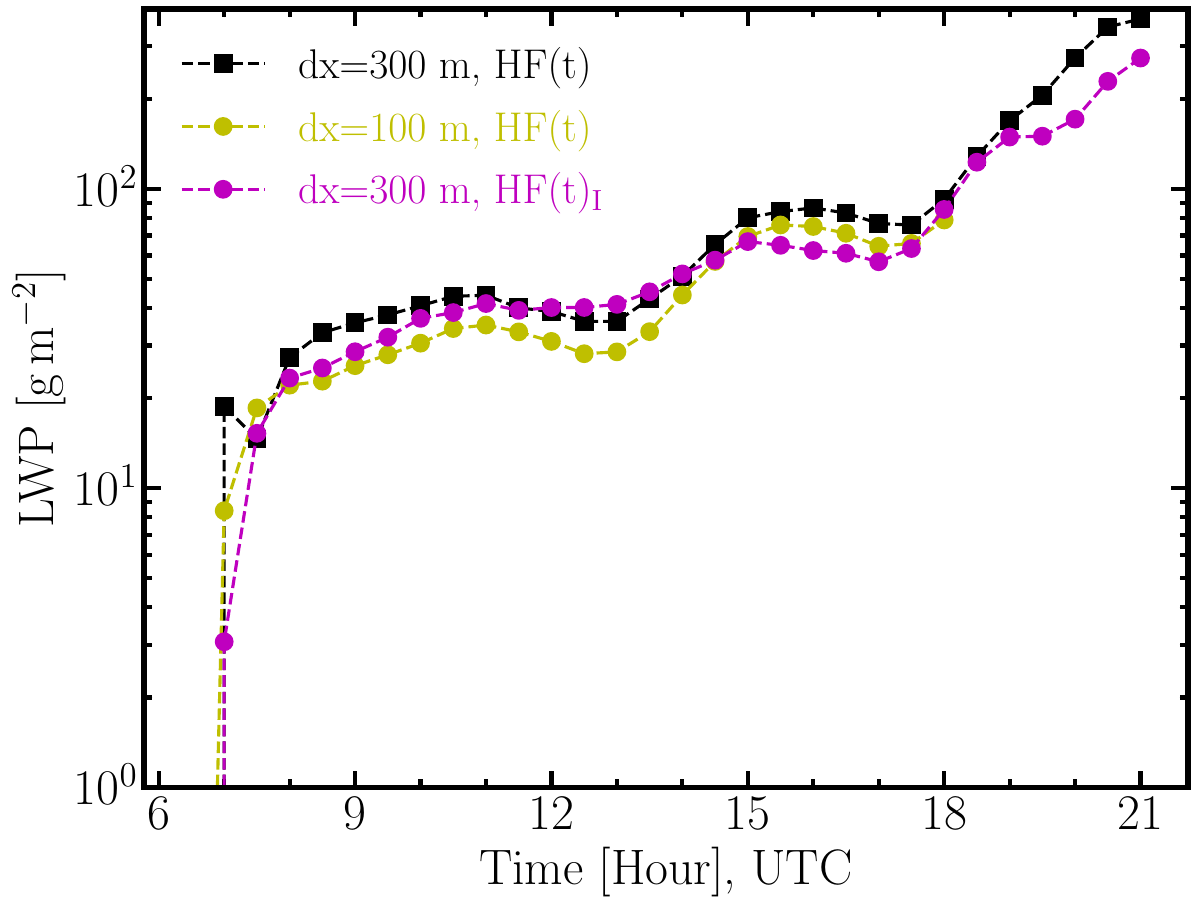}
\includegraphics[width=0.48\textwidth]{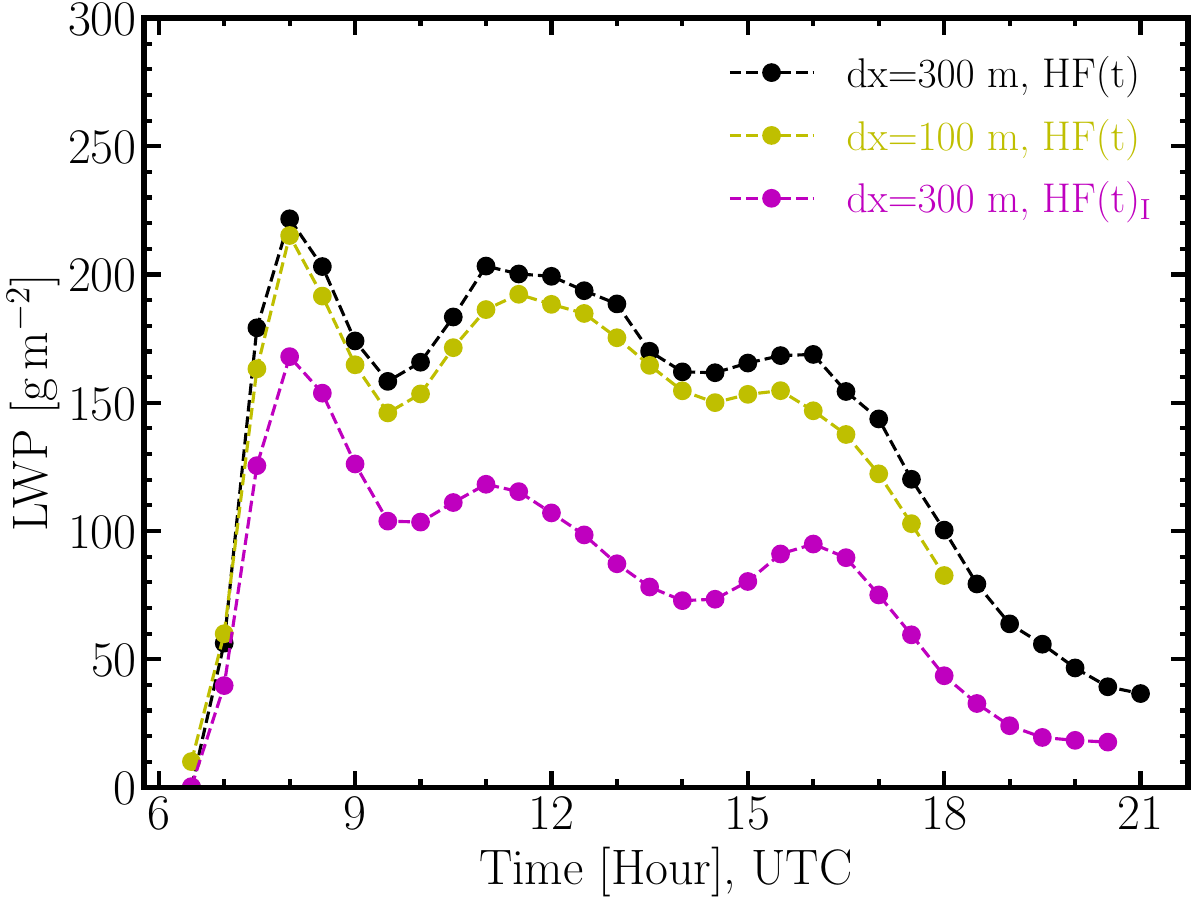}
\end{center}
\caption{Horizontal resolution and interactive surface heat fluxes
dependency for the February 28 (left-hand side) and March 01 cases (right-hand
side). $\rm{SHF} (t)$ and $\rm{LHF} (t)$ from ERA5 are adopted.
$\rm{HF(t)_I}$ denotes the heat fluxes
calculated interactively with ERA5-SST as input.
Black, yellow, and magenta lines represent simulation 0228E (0301E), 0228G (0301G), and
0228F (0301F), respectively.
}
\label{lwp_0301_dx}
\end{figure*}

\begin{figure*}[t!]\begin{center}
\includegraphics[width=\textwidth]{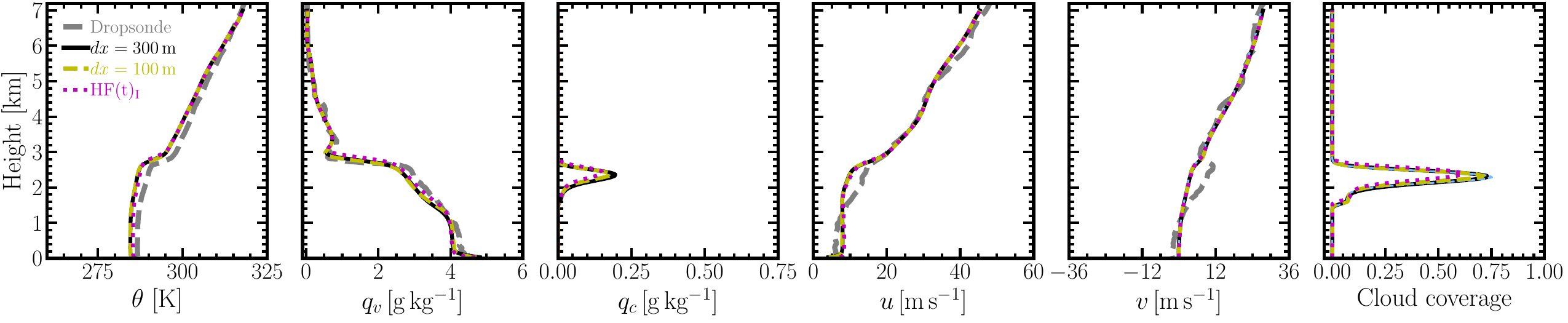}
\includegraphics[width=\textwidth]{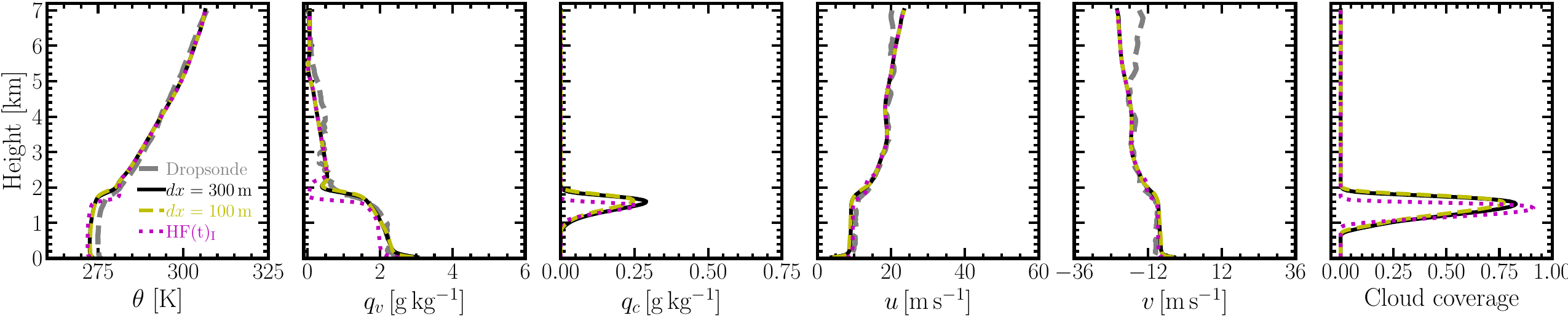}
\end{center}
\caption{Vertical profiles for the February 28 (upper row)
and March 1 (lower row) case during the dropsonde measurement. Same simulations as in \Fig{lwp_0301_dx}.
}
\label{vp_HFt_dx}
\end{figure*}

\Fig{lwp_0301_dx} and \Fig{vp_HFt_dx} shows the horizontal resolution and
interactive surface heat fluxes dependency for both cases. The interactive
heat fluxes result in smaller LWP and IWP. This is due to smaller heat fluxes
shown in \Fig{hf_era5_input}.

\begin{figure*}[t!]\begin{center}
\includegraphics[width=0.48\textwidth]{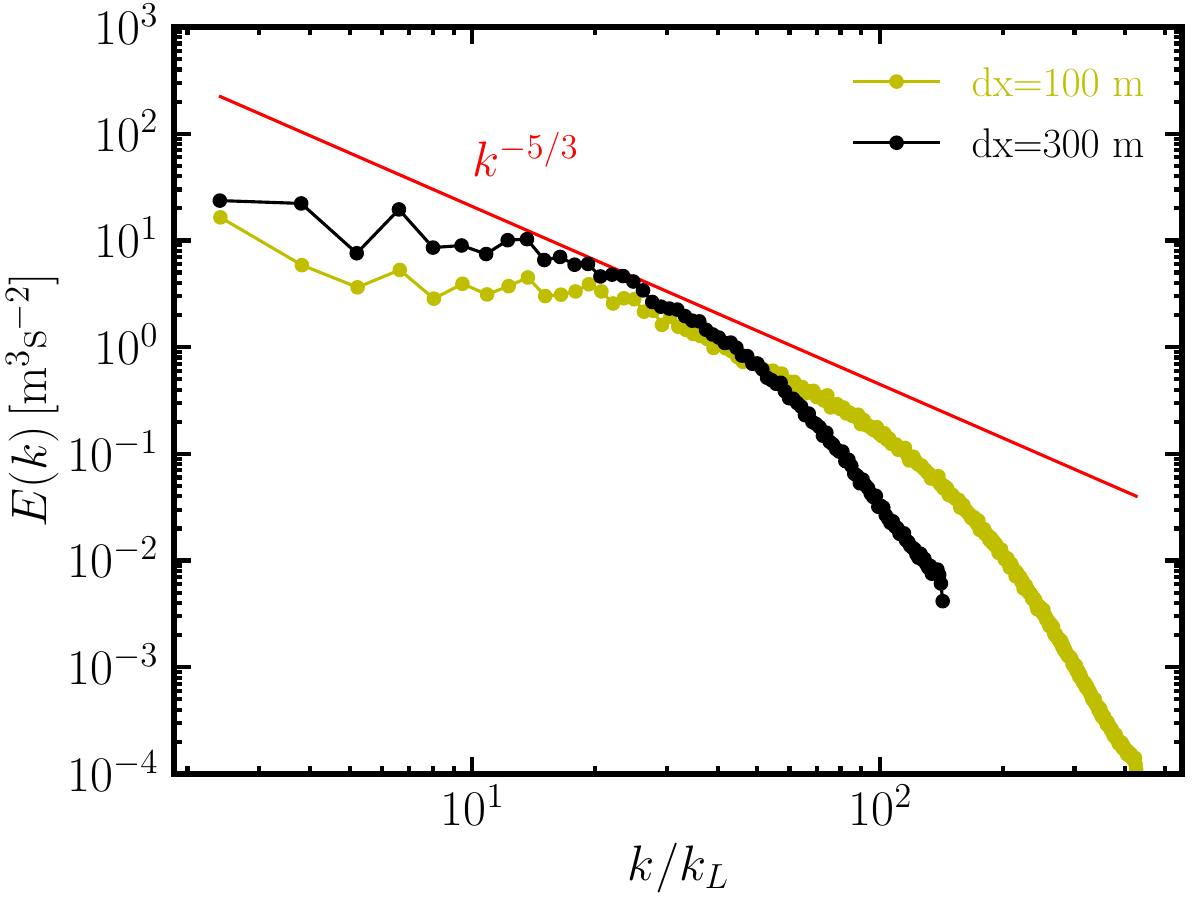}
\includegraphics[width=0.48\textwidth]{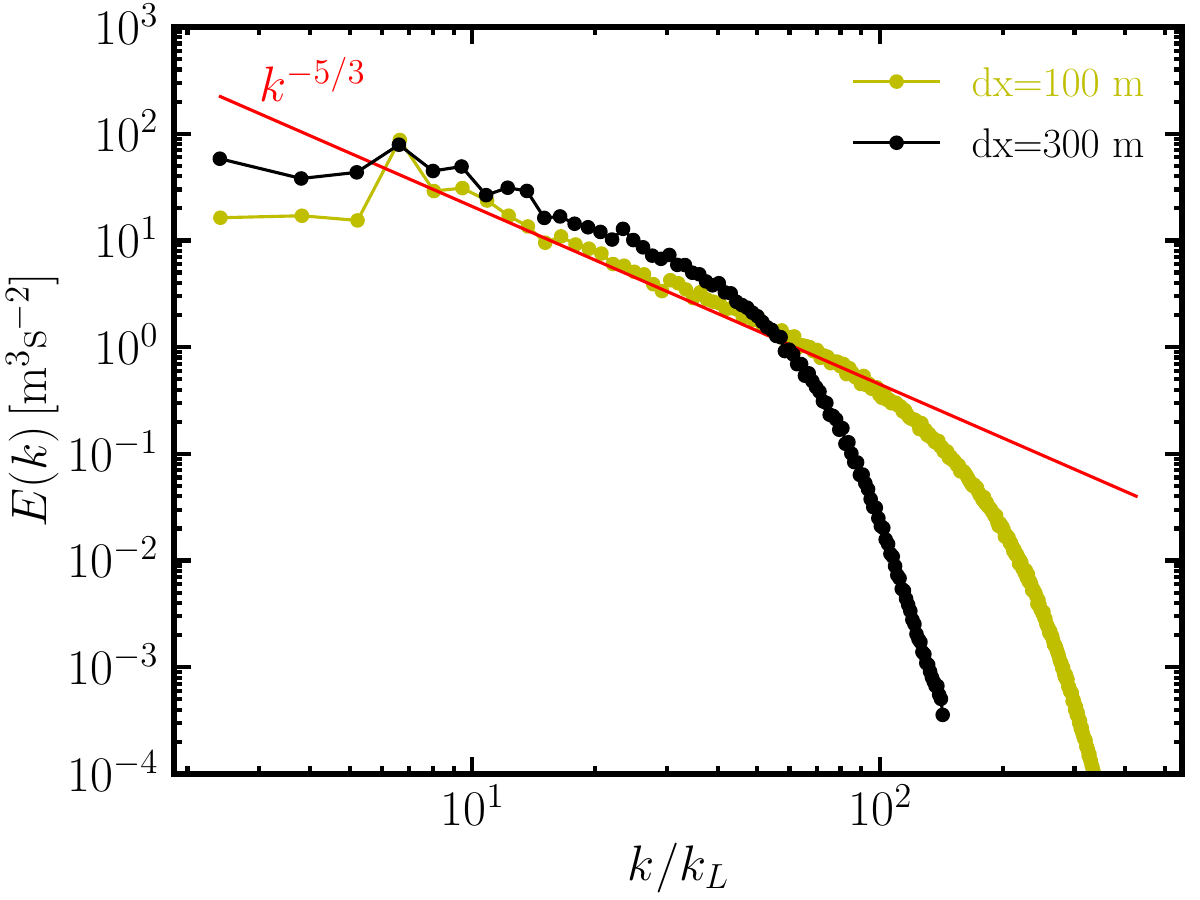}
\end{center}
\caption{Energy power spectra at $H=1 {\rm km}$ averaged during the
measurement time for the February 28 (left-hand side, 16:00-17:00 UTC)
and March 1 (right-hand side, 15:00-16:00 UTC) case.
The abscissa is normalized by $k_L=2\pi/L_x$ to
demonstrate at which length scale the eddies are
not resolved, i.e., $E(k)$ deviates away from the
Kolmogorov scaling $k^{-5/3}$ (red curve). 
Black and red yellow lines represent simulation
0228E (0301E) and 0228G (0301G), respectively.
}
\label{ps_0228}
\end{figure*}

\section{Vertical profiles for LES with different forcing}
\label{app:vp}

\begin{figure*}[t!]\begin{center}
\includegraphics[width=\textwidth]{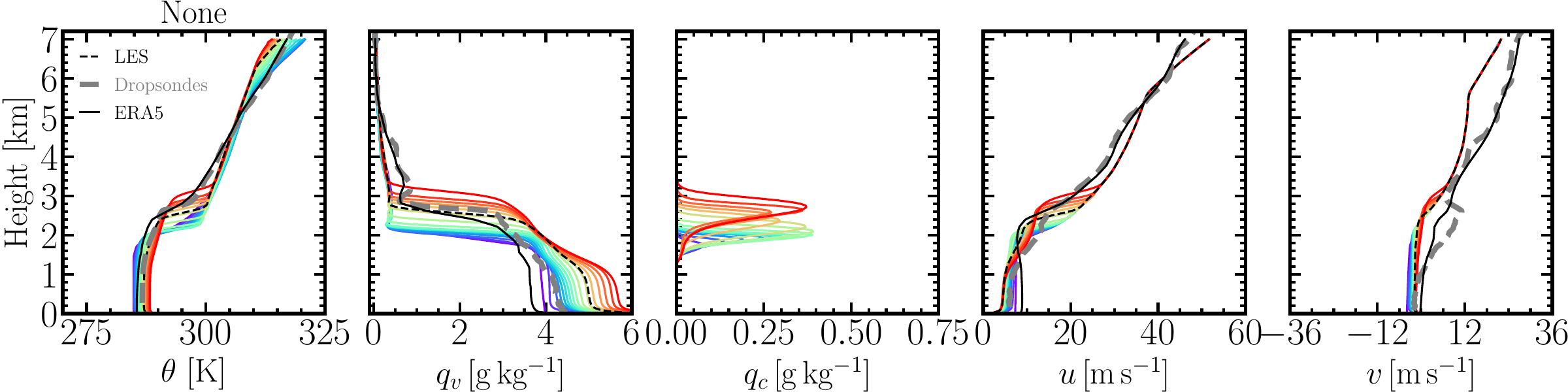}
\includegraphics[width=\textwidth]{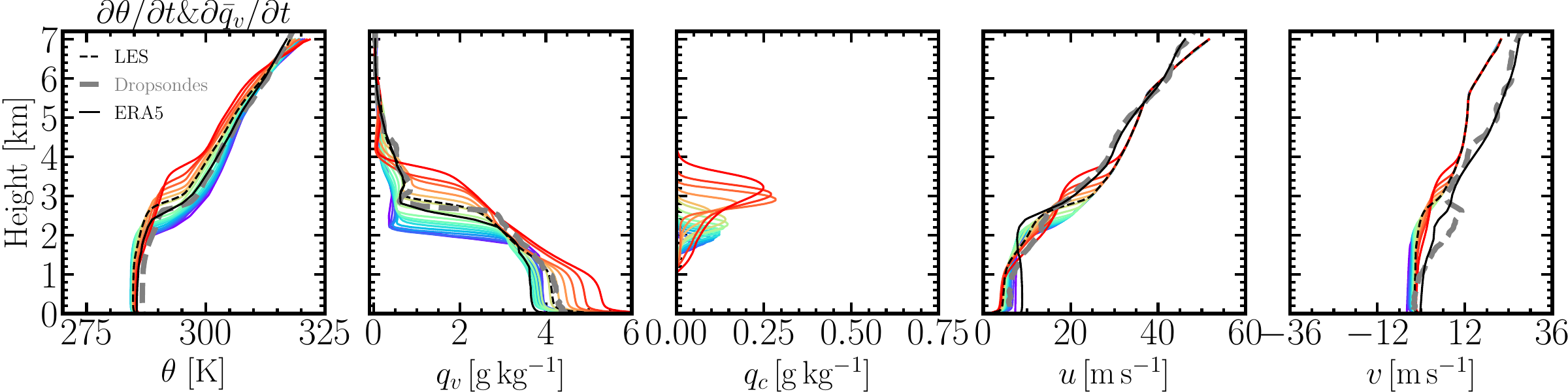}
\includegraphics[width=\textwidth]{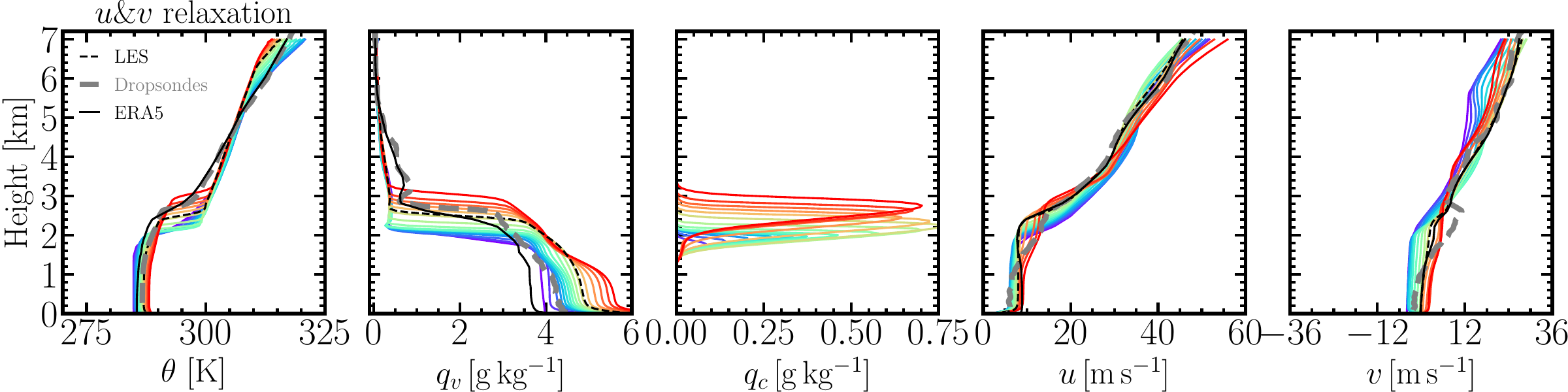}
\includegraphics[width=\textwidth]{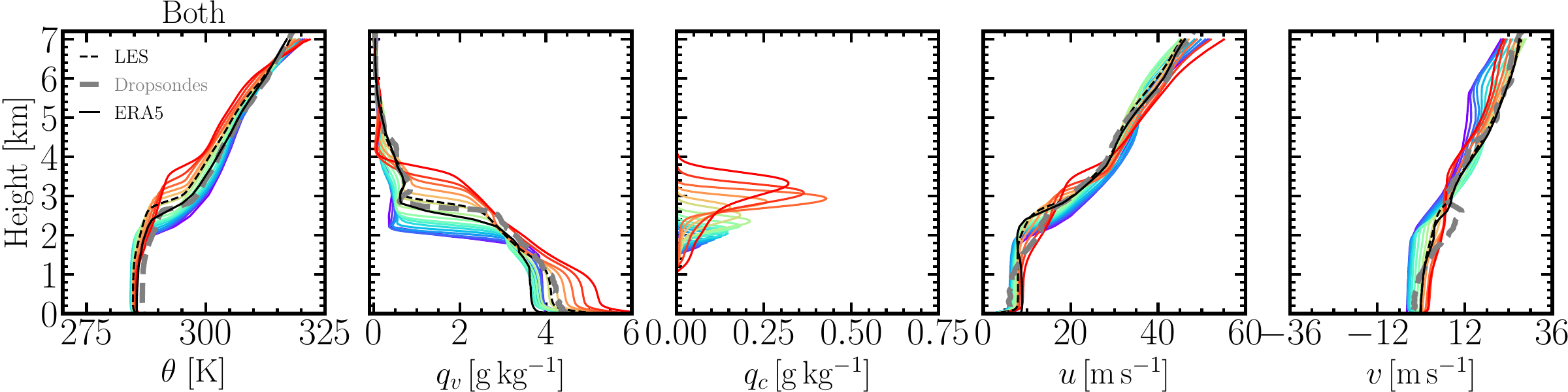}
\end{center}
\caption{Evolution of domain-averaged vertical profiles from the WRF-LES
simulation with the corresponding input forcings shown in \Fig{input_ls_forcing}
for the February 28 case.  The rainbow color scheme represents the time
evolution (06:00-21:00 UTC): from purple to red. The solid black line represents
the ERA5 reanalysis data and the dashed one represent the WRF-LES
averaged during the measurement time. The grey curve represents the dropsonde
measurement.  From the top to the bottom, rows represent the simulation with no
forcing, only advective tendencies ($\partial \bar{\theta}/\partial t$ and
$\partial \bar{q}_v/ \partial t$), only relaxation of $u$ and $v$, and advective
tendencies ($\partial \bar{\theta}/\partial t$ $\partial \bar{q}_v/ \partial t$)
plus $\bar{u}$ and $\bar{v}$ relaxation to $u_{\rm ERA5}$ and $v_{\rm ERA5}$,
respectively.
}
\label{verticalP_forcing2}
\end{figure*}

\Fig{verticalP_forcing2} shows the evolution of vertical
profiles of the meteorological state for simulations
shown in \Fig{verticalP_forcing}.

\Fig{horizC_0228} shows horizontal cross-section of $\theta$,
$q_v$, $q_c$, and TKE at UTC 16:00 and a height of 2.5 km
for the February 28 case (simulation 0228G). Those for the
March 1 case (simulation 0301G) are shown in \Fig{horizC_0301}.

\section{Instantaneous fields and in-situ measurements}

\begin{figure*}[t!]\begin{center}
\includegraphics[width=\textwidth]{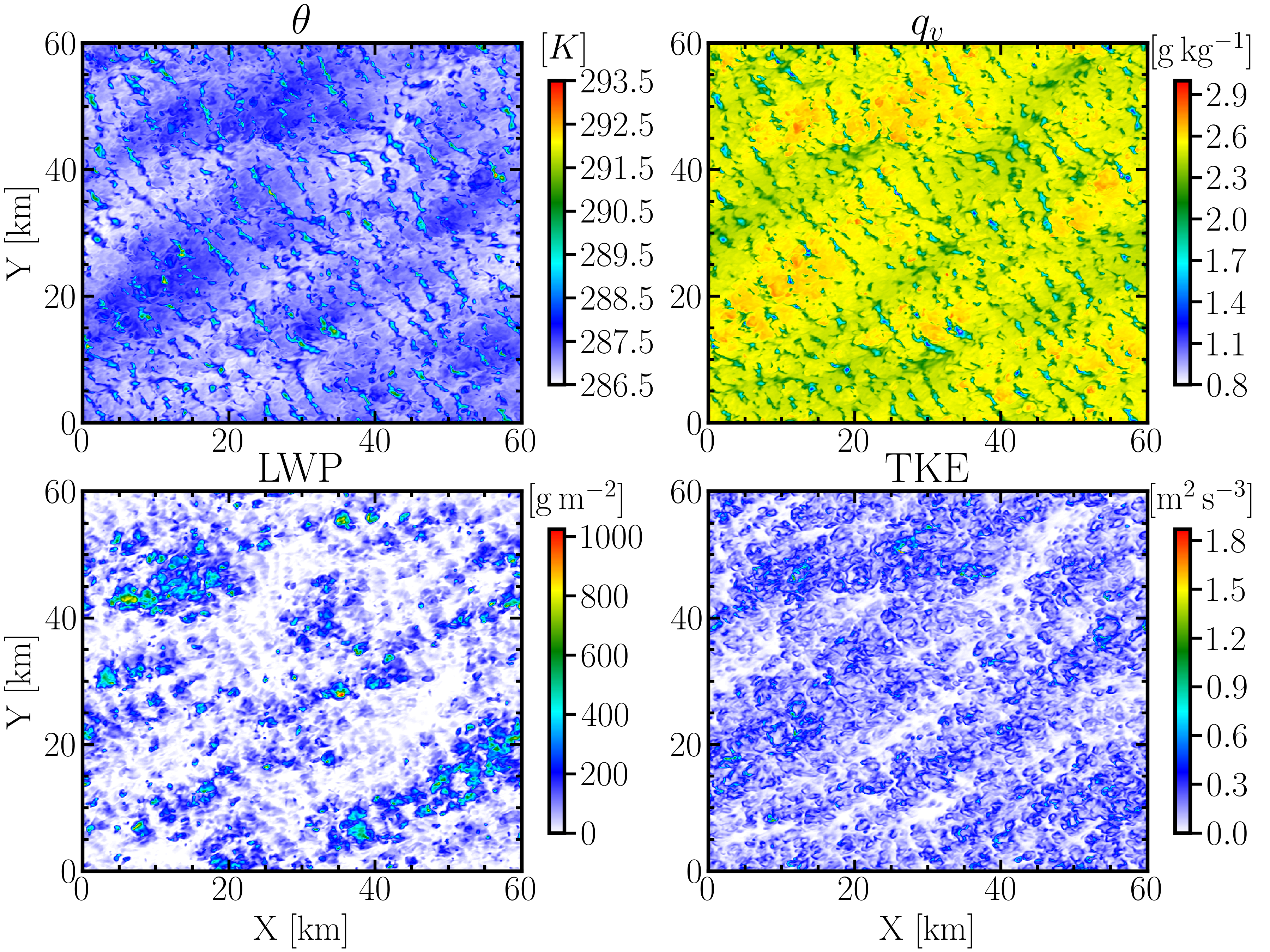}
\end{center}\caption{Horizontal cross-section of $\theta$,
$q_v$, $q_c$, and TKE at UTC 16:00 and 2.5 km (near cloud top)
for the February 28 case (simulation 0228G).
}
\label{horizC_0228}
\end{figure*}

\begin{figure*}[t!]\begin{center}
\includegraphics[width=\textwidth]{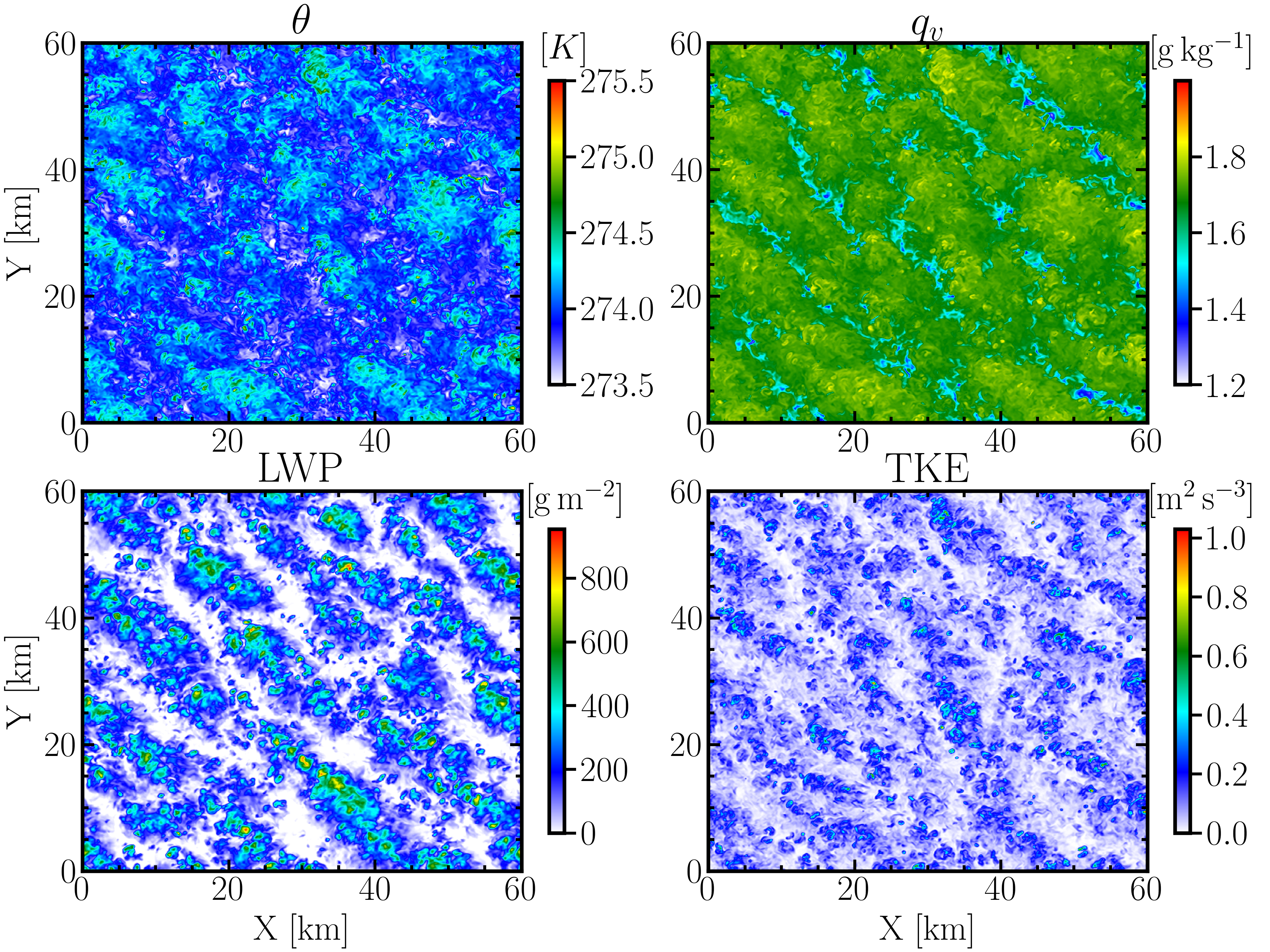}
\end{center}\caption{Same as \Fig{horizC_0228}
but for the March 1 case (simulation 0301G) at 1.5 km.
}
\label{horizC_0301}
\end{figure*}

\begin{figure*}[t!]\begin{center}
\includegraphics[width=\textwidth]{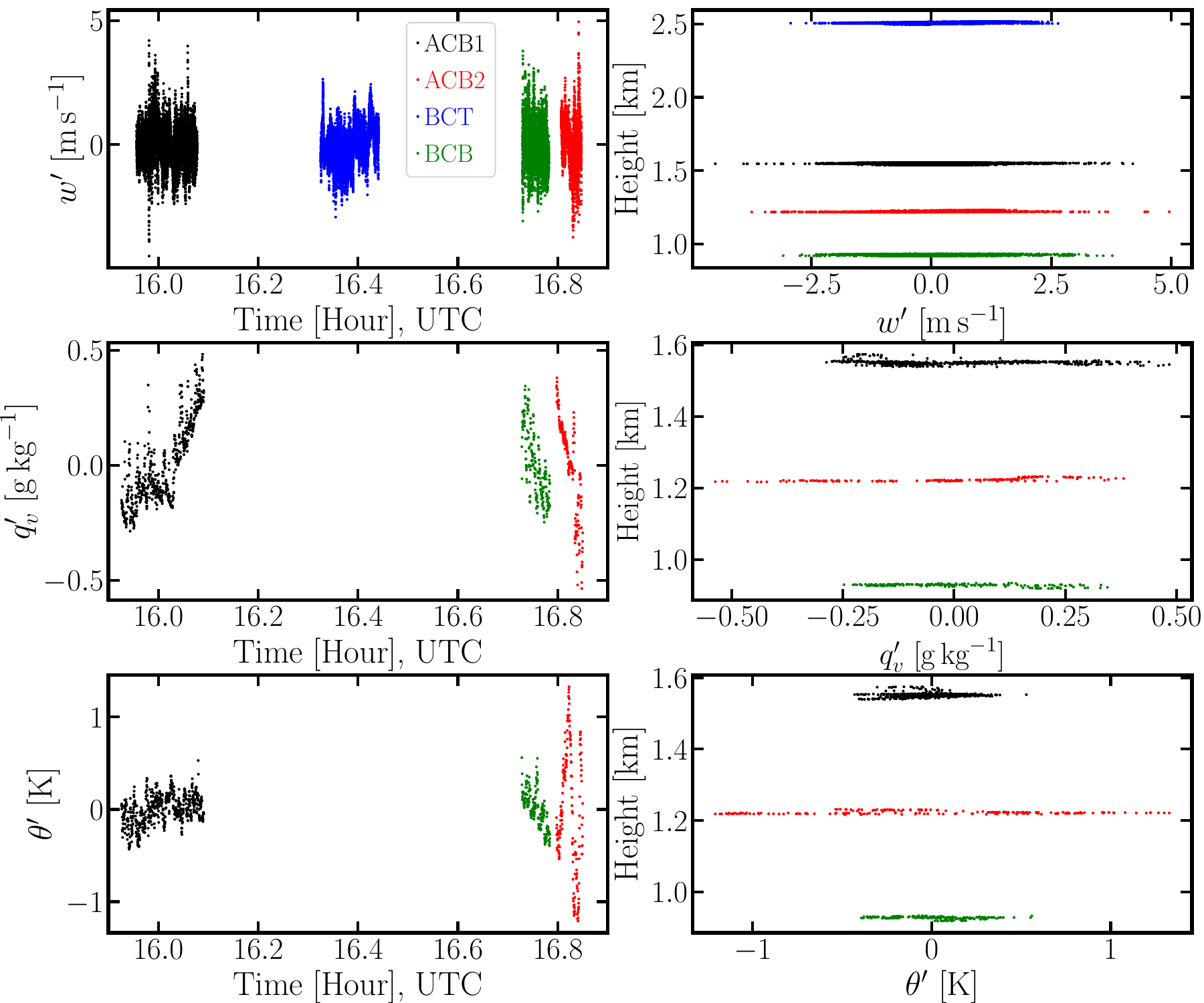}
\end{center}\caption{
Time series and vertical profiles of $w^\prime$, $q_v^\prime$,
and $\theta^\prime$ from the Falcon measurements for the February 28 case.
Flight time and height of each leg are listed in \Tab{tab:legs}.
}
\label{fluc_0228}
\end{figure*}

\begin{figure*}[t!]\begin{center}
\includegraphics[width=\textwidth]{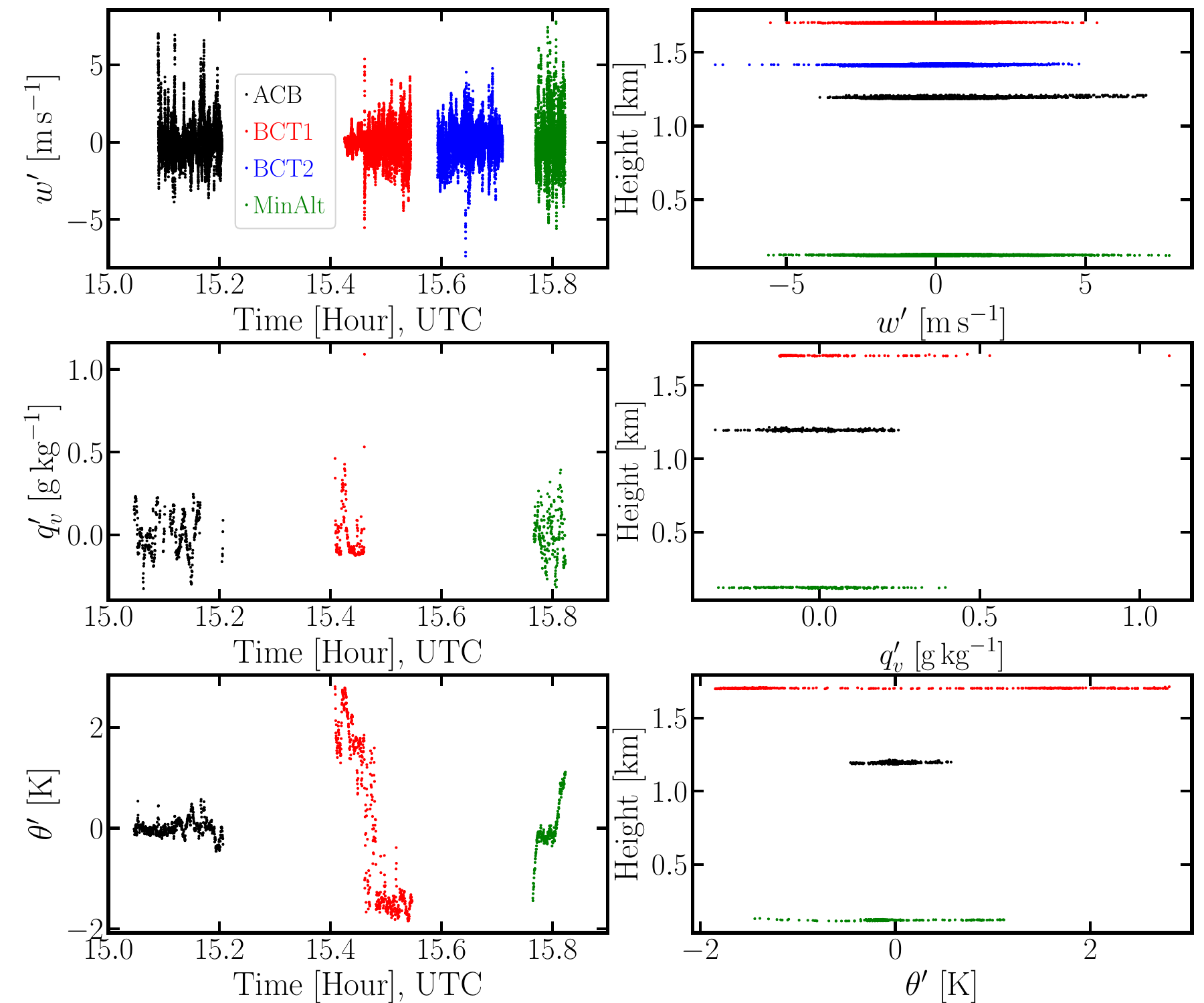}
\end{center}\caption{
Same as \Fig{fluc_0228} but for the March 1 case.
Flight time and height of each leg are listed in \Tab{tab:legs}.
}
\label{fluc_0301}
\end{figure*}

\Fig{horizC_0228} and \Fig{horizC_0301} show
instantaneous fields for the February 28 and March 1 cases,
respectively. Time series and vertical profiles of $w^\prime$, $q_v^\prime$, and $\theta^\prime$ from the Falcon measurements are
shown in \Fig{fluc_0228} and \Fig{fluc_0301} for the two cases.

\begin{table}[t!]
\caption{Falcon flight legs for the February 28 and
March 1 cases.
``ACB'', ``BCT'', ``BCB'', ``MinAlt'' denotes above cloud-base,
below cloud-top, below cloud-base, and minimum altitude, respectively.}
\centering
\setlength{\tabcolsep}{1pt}
\begin{tabular}{|c|c|c|c|}
\hline
Case & Flight legs  & Time [s], UTC & Height [m] \\ 
\hline
     & ABC1 & 15:55:35-16:05:28 & 1538.37-1555.35  \\ 
0228 & ABC2 & 16:47:51-16:51:00 & 1217.93-1230.76  \\ 
     & BCT & 16:18:41-16:26:56 & 2496.92-2514.76  \\ 
     & BCB & 16:43:41-16:47:03 & 918.58-934.12  \\ 
\hline
     & ABC & 15:02:47-15:12:24 & 1185.53-1210.00  \\ 
0301 & BCT1 & 15:24:31-15:32:51 & 1697.50-1708.11  \\ 
     & BCT2 & 15:34:07-15:42:44 & 1405.56-1423.46  \\ 
     & MinAlt & 15:45:54-15:49:26 & 116.56-128.13  \\ 
\hline
\multicolumn{4}{p{0.4\textwidth}}{}
\end{tabular}
\label{tab:legs}
\end{table}

\clearpage
\clearpage

\end{document}